\documentclass{aa} 

\usepackage{graphicx}
\usepackage{mathtools}
\usepackage{comment}
\usepackage{amsmath}	% Advanced maths commands
\usepackage{amssymb}	% Extra maths symbols
\usepackage[dvipsnames]{xcolor}
\usepackage[squaren, Gray, cdot]{SIunits}
\usepackage{txfonts}
\usepackage{subfigure}
\usepackage{lscape}
\usepackage{pifont}
\usepackage{comment}
\usepackage{natbib}

\usepackage{placeins}

\usepackage{caption}
\DeclareCaptionFormat{cont}{#1-continued#2#3\par}

\usepackage[pdfencoding=auto, psdextra, unicode]{hyperref}
\definecolor{ballblue}{rgb}{0.13, 0.67, 0.8}

\hypersetup{
    colorlinks=true,
    linkcolor=RubineRed,
    filecolor=NavyBlue,
    urlcolor=NavyBlue,
    citecolor=NavyBlue
}

\makeatletter
\renewcommand*\aa@pageof{, page \thepage{} of \pageref*{LastPage}}
\makeatother

\begin{document} 
   \title{Gas condensation in Brightest Group Galaxies unveiled with MUSE.}\subtitle{Morphology and kinematics of the ionized gas.}
   \author{V.~Olivares\inst{1,2}, P.~Salom\'e\inst{1}, S.~L.~Hamer\inst{3}, F.~Combes\inst{1,4}, M.~Gaspari\inst{5,6}, K.~Kolokythas\inst{7}, E.~O'Sullivan\inst{8}, R.~S.~Beckmann\inst{9,10}, A.~Babul\inst{11}, F.~L.~Polles\inst{1,12}, M.~Lehnert\inst{13,9}, S.~I.~Loubser\inst{7}, M.~Donahue\inst{14}, M.-L.~Gendron-Marsolais\inst{15}, P.~Lagos\inst{16}, G.~Pineau~des~Forets\inst{1,17}, B.~Godard\inst{1,18}, T.~Rose\inst{19}, G.~Tremblay\inst{8}, G.~Ferland\inst{2}, P.~Guillard\inst{9}}

    \institute{
            LERMA, Observatoire de Paris, PSL Research Univ., CNRS, Sorbonne Univ., 75014 Paris, France % 1
            \and
            Department of Physics and Astronomy, University of Kentucky, 505 Rose Street, Lexington, KY 40506, USA % 2
            \and
            Department of Physics, University of Bath, Claverton Down, BA2 7AY, UK % 3
            \and
            Coll\'ege de France, 11 Place Marcelin Berthelot, 75005 Paris, France % 4
            \and
            INAF, Osservatorio di Astrofisica e Scienza dello Spazio, via P. Gobetti 93/3, 40129 Bologna, Italy % 5
            \and
            Department of Astrophysical Sciences, Princeton University, Princeton, NJ 08544, USA  % 6
            \and
            Centre for Space Research, North-West University, Potchefstroom 2520, South Africa % 7
            \and
            Harvard-Smithsonian Center for Astrophysics, 60 Garden Street, Cambridge, MA 02138, USA % 8
            \and
            Sorbonne Université, CNRS, UMR 7095, Institut d’Astrophysique de Paris, 98bis bd Arago, 75014 Paris, France % 9
            \and
            Institute of Astronomy and Kavli Institute for Cosmology, University of Cambridge, Madingley Road, Cambridge CB3 0HA, UK % 10
            \and
            Department of Physics and Astronomy, University of Victoria, Victoria, BC, V8W 2Y2, Canada % 11
            \and
    	    SOFIA Science Center, USRA, NASA Ames Research Center, M.S. N232-12 Moffett Field, CA 94035, USA % 12
            \and
            Universit\'e Lyon 1, ENS de Lyon, CNRS UMR5574, Centre de Recherche Astrophysique de Lyon, F-69230 Saint-Genis-Laval, France % 13
            \and 
            Department of Physics and Astronomy, Michigan State University, East Lansing, MI 48824, USA  %14
            \and
            European Southern Observatory, Alonso de C\'oordova 3107, Vitacura, Casilla 19001, Santiago de Chile % 15
            \and 
            Instituto de Astrof\'isica e Ci\^encias do Espa\c{c}o, Universidade do Porto, CAUP, Rua das Estrelas, 4150-762 Porto, Portugal % 16
            \and
            Universit\'e Paris-Saclay, CNRS, Institut d’Astrophysique Spatiale, 91405, Orsay, France % 17
            \and
            École normale supérieure, Université PSL, Sorbonne Université, CNRS, LERMA, 75005, Paris, France % 18
    	    \and
    	    Department of Physics and Astronomy, University of Waterloo, Waterloo, ON N2L 3G1, Canada % 19
            }

   \date{Received XX; Accepted XX}
   \abstract{
    The origin of the cold gas in central galaxies in groups is still a matter of debate. We present Multi-Unit Spectroscopic Explorer (MUSE) observations of 18 optically selected local ($z\leq$0.017) Brightest Group Galaxies (BGGs) to study the kinematics and distribution of the optical emission-line gas. MUSE observations reveal a distribution of gas morphologies including ten complex networks of filaments extending up to $\sim$10~kpc to two compact ($<$3~kpc) and five extended ($>$5~kpc) disk-dominated structures. Some rotating disks show rings and elongated structures arising from the central disk. The kinematics of the stellar component is mainly rotation-dominated, which is very different from the disturbed kinematics and distribution found in the filamentary sources. The ionized gas is kinematically decoupled from the stellar component for most systems, suggesting an external origin for the gas. We find also that the H$\alpha$ luminosity correlates with the cold molecular gas mass. By exploring the thermodynamical properties of the X-ray atmospheres, we find that the filamentary structures and compact disks are found in systems with small central entropy values, $K$, and $t_{\rm cool}/t_{\rm eddy}$ ratios. This suggests that, like for Brightest Cluster Galaxies (BCGs) in cool core clusters, the ionized filaments and the cold gas associated are likely formed from hot halo gas condensations via thermal instabilities, consistently with the Chaotic Cold Accretion simulations (as shown via the C-ratio, ${\rm Ta_t}$, and k-plot). We note that the presence of gaseous rotating disks is more frequent than in BCGs. An explanation for the origin of the gas in those objects is a contribution to gas fueling by wet-mergers or group satellites, as qualitatively hinted by some sources of the present sample. Nonetheless, we discuss the possibility that some extended disks could also be a transition stage in an evolutionary sequence including filaments, extended disks and compact disks, as described by hot gas condensation models of cooling flows.
   }
   
    \authorrunning{V. Olivares et al.}
    \maketitle

\section{Introduction}
\label{sec:introduction}

Elliptical galaxies, particularly those sitting at the core of galaxy clusters, were thought to be red and dead, lacking cold gas reservoirs and containing mainly old stars. Notwithstanding, in the last decade, new surveys have dramatically changed this picture with the arrival of state of the art telescopes. In particular, studies of the nearby Universe have found that many elliptical galaxies can harbor a complex multiphase medium, containing a reservoir of cold molecular ($<100~K$) gas traced by CO emission lines \citep[e.g.,][]{edge01,salome03,lim08,temi18,tremblay18,olivares19,russell19,rose19b,rose20}, {warm ($\sim$10$^{3}$~K) molecular hydrogen  \citep[e.g.,][]{edge+02}, a warm ionized ($\sim$10$^{4}$~K) gas traced mainly through optical H$\alpha$+[NII] emission lines  \citep[e.g.,][]{heckman89,hatch07,werner14,mcdonald10,mcdonald12,hamer16}, OVI emitting gas ($\sim$10$^{5.5}$~K, e.g., \citealt{bregman05})}, and a very hot ($\sim10^{7}$~K) atmosphere \citep{sun09a,osullivan11b,Su2015,gaspari19,werner19}. However, the origin of the cold gas is a matter of discussion.

Most of the studies related to the origin of the cold gas have focused principally on the most massive and brightest elliptical galaxies located at the core of galaxy clusters, so-called brightest cluster galaxies. Observations of the ionized gas in galaxy clusters, traced with the H$\alpha$ emission line, show spectacular filamentary structures extending up to $\sim$70~kpc from the core of the central galaxy \citep[e.g.,][]{heckman89,hatch07,mcdonald10,mcdonald11b,hamer16,hamer19,tremblay18,olivares19,ciocan21}. Systems with nebular emission, cold molecular gas, and ongoing star formation are preferentially found in galaxy clusters with short central cooling times, $\lesssim$1~Gyr and low central entropy values, $\lesssim$30~keV~cm$^{2}$ \citep[e.g.,][]{cavagnolo08,bildfell08,pipino09,Rafferty_2008,pulido18,loubser16}. These correlations have been interpreted as an indication of ICM cooling, with the gas becoming thermally unstable and condensing to form multiphase filaments  \citep[e.g.,][]{gaspari12,prasad15,Voit_2017,beckmann19,li20}. Early theoretical studies suggested that the ICM condenses when the cooling time over the free-fall time, $t_{\rm cool}/t_{\rm ff}$, is below 10 \citep[e.g.,][]{sharma12,mccourt12,li14,prasad15}. Other simulations and analytical studies predict the onset of condensation when $t_{\rm cool}/t_{\rm ff}$ is between 5--20, as an outcome of precipitation-regulated feedback \citep[e.g.,][]{gaspari12,prasad18,prasad20,Voit_2017,beckmann19}. In a similar vein, simulations by \citet{Gaspari_2018} predict that the ICM condenses through turbulent non-linear instabilities triggered by the AGN feedback when the cooling time over the eddy-turnover time is close to unity, generating extended H$\alpha$ nebulae (see also \citealt{prasad18}). On the other hand, some studies have proposed that filaments form out of the low-entropy gas lifted by AGN bubbles \citep[e.g.,][]{Revaz_2008,pope10,li14,mcnamara16}. Furthermore, simulations carried out by \citet{qiu19, qiu20} show that multiphase filaments can form from warm AGN-driven outflows when the cooling time is shorter than rising timescales. In particular, the condensed multiphase gas plays an essential role in the self-regulation of the central Super Massive Black Hole (SMBH), in particular by feeding and triggering the AGN via Chaotic Cold Accretion (CCA, \citealt{Gaspari_2013,gaspari20} for a review; see also \citealt{pizzolato05} and \citealt{king06} for related papers but different modeling).

%. https://academic.oup.com/mnras/article/421/4/2809/1084198

While a consensus has emerged that cold gas in BCGs forms through the hot atmosphere's cooling, less attention has been paid to the lower-mass counterparts, galaxy groups (see \citealt{oppenheimer21} for a recent review). Some galaxy groups show remarkable similarities to galaxy clusters. Many of them are observed to possess a hot intragroup medium (IGrM) \citep[e.g.,][]{dong10, osullivan18a}, that cools through X-ray emission in a similar form to the ICM in galaxy clusters. Of galaxy groups in the local Universe with detected IGrM, about $\sim$80\% of them have a central temperature decline indicating a rapid radiative cooling \citep{dong10,osullivan17}. 

Some of the brightest X-ray galaxy groups have short central cooling times, H$\alpha$ emitting gas, and cold molecular clouds around their central galaxies \citep[e.g.,][]{werner14, david14, david17, temi18, schellenberger20}, similar to what is found within the massive BCGs. In particular, \citet{werner14} presented H$\alpha$+[NII] imaging of warm ionized gas of 8 BGGs, using narrow-band imaging from the \textit{Southern Astrophysical Research} (SOAR) telescope. These observations reveal extended filamentary emission-line nebulae in some sources, suggesting that the cold gas is produced by cooling from the hot phase, but disks, rings, and compact gas distribution are also present. In the case of a galaxy merger (or interaction) event, a fraction of the gas is likely to evade the shock process and fall to the center of the galaxy, often producing a central disk or rings \citep{mazzuca06, Eliche-Moral10}, but can also form gaseous streams of low entropy gas (see \citealt{poole06}). An internal origin for the cold gas has also been suggested, through the cooling of material injected from the stellar population (stellar-mass loss, \citealt{davis11, Voit_2011, bregman-parriott09}).

Recently, \citet{osullivan15,osullivan18b} reported molecular gas detection in 36/53 central group galaxies from the Complete Local Volume Groups Sample (CLoGS, \citealt{osullivan17}), using Institut de Radioastronomie Millimetrique (IRAM) 30m and Atacama Pathfinder Experiment telescope (APEX) observations. CLoGS\footnote{\url{http://www.sr.bham.ac.uk/~ejos/CLoGS.html}} is a statistically complete optically-selected survey of 53 nearby (D$<$80\,Mpc)  galaxy groups, observed in X-ray, optical, and radio wavebands. The CLoGS survey provides comprehensive coverage of the galaxy groups to study the origin of the cold gas in BGGs. Our goal here, is to determine the optical emitting gas properties of these central group galaxies to explore the origin and life-cycle of the cold material. Additionally, galaxy groups are a critical and diverse environment to study the life-cycle of gas, and a great laboratory to investigate turbulence, non-gravitational and environmental processes, such as AGN-feedback, stellar feedback, and mergers, due to the group's shallower gravitational potential and the smaller separation between galaxy members \citep{sun09a}.

In this paper, we present optical observations of 18 BGGs, which are part of the CLoGS, with the MUSE instrument on the Very Large Telescope (VLT). The present paper focuses on the distribution and kinematics of the ionized gas. A dedicated analysis of the line ratios, the excitation of the gas, and and chemical abundances will be presented in a future paper. This paper is laid out as follows. In section~\ref{sec:sample} we present the description of the sample. In Section~\ref{sec:muse}, the MUSE observations and data analysis are described. The description of the distribution and kinematics of gas and stars is presented in Section~\ref{sec:results}. The origin of the warm ionized gas and a possible evolution sequence is discussed in Section~\ref{sec:discussion}. Finally, the conclusions are summarized in Section~\ref{sec:conclusion}. 

Throughout this paper, we assume a standard cosmology with H$_{\rm 0}$=70\,km s$^{-1}$\,Mpc$^{-1}$, and $\Omega_{\rm m}$=0.3.

\section{Sample}\label{sec:sample}
To investigate the various properties and nature of the optical line-emitting gas in nearby BGGs, we observed with MUSE 18 sources drawn from the 26-member CLoGS high-richness sub-sample (R\footnote{Richness (R) is defined as the number of galaxies within the group with B-band luminosity log L$_{\rm B}$ $>$ 10.2 \citep{osullivan17}.}$=$4-8) \citep{osullivan15, osullivan17, osullivan18b}. The CLoGS sample consists of 53 galaxy groups selected via friends-of-friends and hierarchical clustering algorithms from the LEDA (Lyon-Meudon Extragalactic Database; \citealt{garcia93}) galaxy catalog \citep{makarov14}. Worth noting that the groups were chosen to own an elliptical as a group-dominant galaxy. The CLoGS provides an ideal sample in which to study the origin of the gas within lower-mass systems. For the interested reader, a detailed description of the CLoGS sample selection criteria can be found in \citet{osullivan17}. In terms of stellar velocity dispersion (a proxy for gravitational potential), the MUSE sources show no significant difference from the other CLoGS group dominant galaxies in the high or low-richness sub-samples. Table~\ref{tab:sample2} and Table~\ref{tab:prop_multiwave}, summarize the MUSE observations and the main properties for each target, respectively. 

Our MUSE sample spans a wide range of stellar masses, SFRs, X-ray and radio properties. The stellar masses span a factor of approximately 5, from 4$\times$10$^{10}$ to 26$\times$10$^{10}$~M$_{\odot}$, With NGC\,1060 being the most luminous, and NGC\,584 the dimmest. For comparison, BCGs have typical stellar masses higher than $>$10$^{11}$~M$_{\odot}$. The stellar masses were determined using a mass-to-light ratio model, and using galaxy colors from SDSS \textit{ugriz} and 2MASS (2 Micron All Sky Survey) \textit{JHK} magnitudes \citep{osullivan18b}.
 
Optical emission line narrow-band spectroscopy had already been carried out for a couple of sources (NGC\,584, NGC\,5846, NGC\,7619), by using narrow-band filter centered on the H$\alpha$+[NII] emission lines from \textit{New Technology Telescope} (NTT) \citep{macchetto96}. The authors identify different morphologies from compact up to filamentary as we also found in galaxy clusters. The observations reveal H$\alpha$+[NII] luminosities of 18--40$\times$10$^{39}$\,erg~s$^{-1}$. Detection of warm ionized gas through H$\alpha$+[NII] imaging and spectroscopy has been presented only in one source of our MUSE sample (NGC\,5846), using SOAR observations \citep{werner14}. The SOAR observations show optical emission concentrated in the innermost 1.5~kpc of the galaxy, but the distribution of the gas is filamentary, with some sign of interaction with the central AGN.

Based on their central cooling time classification presented by \citet{hudson10} some groups have either weak-cool-core (WCC), or strong-cool-core (SCC) \citep{osullivan17}. Galaxy groups have lower central temperature than galaxy clusters, and given that emission lines are more effective at radiating away thermal emission at $\leq $1~keV than the continuum, that leads to shorter central cooling times in galaxy groups. In \citet{osullivan17} has been discussed an alternative approach to classifying groups in cool-core (CC) and non-cool-core (NCC) based on the temperature profiles of the X-ray emitting gas, although adiabatic heating can flatten the temperature profiles. \citet{osullivan17} found that NCC and CC groups have nearly identical entropy and $t_{\rm cool}/t_{\rm ff}$ profiles. In addition, the amplitude of the entropy profiles are almost 10 times higher than the self-similar values inside R$_{\rm 500}$, indicating that the entropy may highly affected by AGN and SN feedback, cooling, and star formation \citep{oppenheimer21}.

Extended X-ray emission ($>$65~kpc) with luminosities ($L_{\rm x-ray} >$ 10$^{41}$~erg~s$^{-1}$) and temperatures typical of group-scale haloes are seen in a few of our sources. Some other systems have an X-ray distribution with galaxy-scale sizes of 10--65~kpc with X-ray luminosities on the order of 10$^{40}$~erg~s$^{-1}$, or a point-like appearance (with an extension smaller than the \textit{XMM} PSF). In these groups, a variety of X-ray morphologies are observed, including galaxy mergers (e.g., NGC\,1060 and NGC\,7619), and one sloshing-disturbed system (NGC\,5846). In a few cases, disturbances in the X-ray emitting gas are suspected to be caused by a central radio AGN (NGC\,193, NGC\,4261; \citealt{osullivan17}). See Table~\ref{tab:prop_multiwave} for more details on each source.

Radio-jets have been identified in some sources of our MUSE sample, with a range of physical scale, 10--80~kpc, and radio power, 10$^{20}$ to 10$^{24}$ W~Hz~$^{-1}$, using the VLA and the GMRT, at 1.4~GHz, 610 and 235~MHz \citep{kolokythas18}. However, most of the sources lack radio jets. Instead, the radio emission show a diffuse or point-like distribution.

Several sources have confirmed molecular gas detection through CO(1-0) and CO(2-1) emission lines using IRAM 30m and APEX observations. Those observations reveal a broad range of molecular gas masses of 10$^{7}$ up to 10$^{9}$~M$_{\odot}$ \citep{osullivan15,osullivan18b}, a few orders of magnitude lower than it is usually found in the BCGs (10$^{10}$--10$^{11}$~M$_{\odot}$; \citealt{edge01,salome03}). From double-horned HI and CO profiles, \citet{osullivan18b} reported that some sources (e.g., NGC\,924, NGC\,940 and ESO\,507-25) likely host molecular disks. ALMA observations of one of our targets, NGC\,5846 \citep{temi18}, reveals 10$^{5}$~M$_{\odot}$ of giant molecular clouds associated with some filaments and dust structures.

Our sources also have very low SFRs, 0.01--5~M$\odot$~yr$^{-1}$ (SFR=L$_{\rm FIR}$/(5.8$\times$10$^{9}$~M$_{\odot}$); \citealt{kennicutt98}). The SFRs were derived from FIR-based observations with The Infrared Astronomical Satellite (IRAS) or Spitzer Space Telescope (Spitzer). They also present short depletion times $<$1~Gyr ($\tau$=5.8 M$_{\rm mol}$/L$_{\rm FIR}$~Gyr) indicating a fast replenishment of their gas reservoirs \citep{osullivan15,osullivan18b}.

\begin{table*}[htb!]
\caption{The MUSE observational parameters listed for each object in the BGG sample.}
\setlength{\tabcolsep}{3.5pt}
\begin{center}
\scalebox{0.9}{
\begin{tabular}{llccccccccccc}
    \noalign{\smallskip} \hline \hline \noalign{\smallskip}
    \small Group & BGG & $z$ & Scale & Exp. time & Seeing & $\alpha_{J2000}$ & $\delta_{J2000}$  & $R$ &  G. & H$\alpha$ info & Emission lines \\
    & &  & (kpc/$\arcsec$) & (s) & ($\arcsec$) & & &  &Type & & detected\\
    \noalign{\smallskip}
     (1) & (2) & (3) & (4) & (5) & (6) & (7) & (8) & (9) & (10) & (11) & (12)\\
    \noalign{\smallskip} \hline \noalign{\smallskip}
    LGG\,9 & NGC\,193 	& 0.014723 & 0.359 & 3$\times$900 & 1.98 & 00:39:18.6 & $+$03:19:52 & 7 &E & Emission &  {\scriptsize H$\alpha$, [NII], [OIII], H$\beta$, [SII]}\\
    
    LGG\,18 & NGC\,410 	& 0.017659 & 0.373 & 3$\times$900 & 1.10 & 01:10:58.9 & $+$33:09:07 & 6 & E2 & Emission & {\scriptsize H$\alpha$, [NII], [OIII], H$\beta$, [SII]}\\
    
    LGG\,27 &NGC\,584 	& 0.006011 & 0.121 & 3$\times$900 & 1.59 & 01:31:20.7 & $-$06:52:05 & 4 & S0 & Abs/Emis & {\scriptsize H$\alpha$, [NII], [OIII], H$\beta$, [SII]}\\
    
    LGG\,31 & NGC\,677 	& 0.017012 & 0.378 & 3$\times$900 & 0.65 & 01:49:14.0 & $+$13:03:19 &7& E3 & Emission & {\scriptsize H$\alpha$, [NII], [OIII], H$\beta$, [SII], [OI]}\\
    
    LGG\,42 &NGC\,777 	& 0.016728 & 0.354 & 3$\times$900 & 0.86 & 02:00:14.9 & $+$31:25:46 & 5 & E1 &Abs/Emis &  {\scriptsize H$\alpha$, [NII]} \\
    
    LGG\,58 &NGC\,940 	& 0.017075 & 0.359 & 3$\times$900 & 0.77 & 02:29:27.5 & $+$31:38:27 & 3 & S0 & Abs/Emis & {\scriptsize H$\alpha$, [NII], [OIII], H$\beta$, [SII] } \\
    
    LGG\,61 &NGC\,924 	& 0.014880 & 0.310 & 3$\times$900 & 0.73 & 02:26:46.8 & $+$20:29:51 & 3 & S0 & Emission & {\scriptsize H$\alpha$, [NII], [OIII], H$\beta$, [SII], [OI]}\\
    
    LGG\,66 & NGC\,978 	& 0.015794 & 0.334 & 3$\times$900 & 1.16 & 02:34:47.6 & $+$32:50:37 & 7 & E/S0 &Abs/Emis  & {\scriptsize H$\alpha$, [NII], [OIII], H$\beta$, [SII], [OI]}\\
    
    LGG\,72 &NGC\,1060 	& 0.017312 & 0.368 & 3$\times$900 & 0.43 & 02:43:15.0 & $+$32:25:30 &8 & E/S0 & Abs/Emis &
    {\scriptsize H$\alpha$, [NII], [OIII], [SII]} \\
    
    LGG\,103 &NGC\,1453 & 0.012962 & 0.305 & 3$\times$900 & 0.80 & 03:46:27.2 & $-$03:58:08 & 4 & E3 & Emission & {\scriptsize H$\alpha$, [NII], [OIII], H$\beta$, [SII], [OI]}\\
    
    LGG\,117 &NGC\,1587 & 0.012322 & 0.247 & 3$\times$900 & 0.97 & 04:30:39.9 & $+$00:39:42 & 4 &E1  & Abs/Emis & {\scriptsize H$\alpha$, [NII], [OIII], H$\beta$, [SII], [OI]}\\
    
    LGG\,262 &NGC\,4008 & 0.012075 & 0.262 & 3$\times$900 & 0.79 & 11:58:17.0 & $+$28:11:33 & 4 &E5  & Abs/Emis & {\scriptsize H$\alpha$, [NII],  H$\beta$}\\
    
    LGG\,276 &NGC\,4169 & 0.012622 & 0.218 & 3$\times$900 & 0.68 & 12:12:18.8 & $+$29:10:46 & 4 &S0a & Emission & {\scriptsize H$\alpha$, [NII], [OIII], H$\beta$, [SII], [OI]}\\
    
    LGG\,278 &NGC\,4261 & 0.007378 & 0.155 & 3$\times$880 & 0.78 & 12:19:23.2 & $+$05:49:31 & 7& E2 & Emission & {\scriptsize H$\alpha$, [NII], [OIII], H$\beta$, [SII], [OI]}\\
    
    LGG\,393 &NGC\,5846 & 0.005711 & 0.126 & 7$\times$880 & 0.90 & 15:06:29.3 & $+$01:36:20 & 5 &E0 & Emission & {\scriptsize H$\alpha$, [NII], [OIII], H$\beta$, [SII], [OI]}\\
    
    LGG\,421 &NGC\,6658 & 0.014243 & 0.305 & 3$\times$870 & 1.82 & 18:33:55.6 & $+$22:53:18 & 4 &E2 & Absorption & -- \\
    
    LGG\,473 & NGC\,7619 & 0.012549 & 0.262 & 3$\times$870 & 0.47 & 23:20:14.5 & $+$08:12:22 & 8 &E3 & Abs/Emis & {\scriptsize H$\alpha$, [NII], [SII], [OI]}\\
    
    LGG\,310 &ESO\,507-25 & 0.010788 & 0.218 & 3$\times$900 & 1.90 & 12:51:31.8 & $-$26:27:07 & 4 & S0 & Emission & {\scriptsize H$\alpha$, [NII], [OIII], H$\beta$, [SII], [OI]}\\
    
    \noalign{\smallskip} \hline \noalign{\smallskip}
\label{tab:sample2}
\end{tabular}}
\end{center}
\tablefoot{
(1) Group name (LGG: Lyon Groups
of Galaxies). (2) BGG name. (3) Redshifts of each group, taken from \citet{osullivan15} and \citet{osullivan18b}. (4) Scale. (5) MUSE exposure time on source.
(6) Average DIMM seeing of the MUSE observations. 
(7) and (8) Right Ascension and Declination of each group. 
(9) Richness $R$ values for each source \citet{osullivan17}. 
(10) Morphological classification of the galaxy based on \citet{deVaucouleurs91}. 
(11) In some cases, the H$\alpha$ line appears very weak or even completely absent from some parts of the cube where it is otherwise expected to exist (such as regions where the other ionized lines are stronger, e.g. [NII]$\lambda$6583). This is because the stellar continuum is very high, and the H$\alpha$ flux is low. Those systems are being labeled with H$\alpha$ in absorption. Some of the H$\alpha$ flux is recovered after the stellar continuum subtraction (see Fig.~\ref{fig:spectrum_examples} for an example){, however, it is possible that we are not able to recover entirely the underlying H$\alpha$ emission in some cases.} (12) Emission lines detected on the MUSE data cube. 
}\\ 
\end{table*}

\begin{table*}
\caption{Multi-wavelength properties of the BGG sample.}
\setlength{\tabcolsep}{2.5pt}
\begin{center}
\centering
\renewcommand{\arraystretch}{1.0}
\scalebox{0.72}{
\begin{tabular}{lccccccccccccccccc}
    \noalign{\smallskip} \hline \hline \noalign{\smallskip}
    \rm BGG & 
    M(H$_{2}$) &
    log(M$_{\star}$) &
    SFR &
    CCT &
    min($t_{\rm cool}/t_{\rm ff}$) &
    $t_{\rm cool}/t_{\rm eddy}$ &
    \multicolumn{2}{c}{Core type} &
    T$_{\rm sys}$ &
    K$_{0}$ &
    K$_{10}$ &
    R$_{500}$ &
    M$_{500}$ &
    Radio &
    LLS &
    X-ray &
    Notes \\
    \rm &
    (10$^{8}$ M$_{\odot}$)  &
    (M$_{\odot}$) &
    (M$_{\odot}$\,yr$^{-1}$)    &
    (Gyr)&
         &
         &
    T profile &
    Hudson &
    (keV) &
    (keV\,cm$^{-2}$) &
    (keV\,cm$^{-2}$) &
    (kpc) &
    (10$^{13}$\,M$_{\odot}$) &
    &
    (kpc)&
    & \\
    \noalign{\smallskip}
     (1) & (2) & (3) & (4) & (5) & (6) & (7) & (8) & (9) & (10) & (11) & (12) & (13) & (14) & (15) & (16) & (17) & (18) \\
    \noalign{\smallskip} \hline \noalign{\smallskip}
    NGC\,193 & <1.35 & 10.92 & -- & -- & -- & -- & CC & -- & 0.88$^{+0.02}_{-0.03}$ & 11.31 &24.3 & 432 & 2.33  & JET & 80 & GRP & shock \\%/X-cavity/ \\
    
    NGC\,410 & <1.17 & 11.37 & -- & 0.102 & 18.2  & 0.5 & CC & SCC &0.98$\pm$0.02 & 4.50 & 34.5 & 458 & 2.78  & pnt & $\leq$11& GRP & \\
    
    NGC\,584 & <0.11 & 10.62 & 0.024 & -- & --  & -- & -- & -- & -- &  <270.62 &>272.7 & --& -- & pnt & $\leq$3 & -- & \\
    
    NGC\,677 & <2.25 & 10.91 & --& 2.042 & 47.9 & 1.69 & CC & WCC & 0.79$\pm$0.01& 5.82 &31.9 &  406 & 1.94 & diffuse & 30 & GRP &\\
    
    NGC\,777 & <2.08 & 11.32 & 0.014 & 0.314 & 39.2 & 1.0 & NCC & SCC &  0.89$\pm$0.02& 23.84 & 34.8 & 434 & 2.37 &  pnt & $\leq$8 & GRP &\\
    
    NGC\,924 & 0.52$\pm$0.10 &  10.76 &0.545  & -- & -- & --& -- &-- & -- & <403.08 &>406.1 & -- & -- &  pnt & $\leq$4 & -- &\\
    
    NGC\,940 & 61.0$\pm$2.15 & 10.94 & 2.0 & -- & -- & -- &-- & -- & 0.41$^{+0.04}_{-0.03}$ & <548.74 & >552.9 &  282 & 0.65 & pnt &$\leq$6 & pnt &\\
    
    NGC\,978 & <0.70 & 11.08 & 0.157 &1.421 & 80.6* & -- &  (NCC) & WCC & 0.49$^{+0.15}_{-0.11}$ & 7.29 & 23.6 & 312 & 0.87 &  pnt &$\leq$3 & gal&\\
    
    NGC\,1060 & <0.78 & 11.42 & 5.6 & 0.132 & 10.4 & 0.6 & CC & SCC & 1.02$\pm$0.01 & 11.20 & 41.8 & 468 & 2.97 &  jet & 14 & GRP & merger\\%/Merger \\
    
    NGC\,1453 & <0.78 & 11.16 & 0.217 & 0.390 & 64.4 & 1.6  & NCC & SCC & 0.74$\pm$0.03 & 14.08 & 40.4 & 392 & 1.74 & pnt & $\leq$11& GRP &\\
    
    NGC\,1587 & 2.30$\pm$0.48 & 11.01 & -- & 0.101 &  31.6* & 0.5 & NCC & SCC & 0.37$^{+0.08}_{-0.06}$ & 1.22 & 15.3 & 267 & 0.55 &diffuse & 22 & GRP &\\
    
    NGC\,4008 & <0.73 & 10.90 & 0.128 & 0.199 &16.9* & 0.7 &  (NCC) & SCC & 0.56$^{+0.04}_{-0.07}$ & 9.98 & 20.5 & 336 & 1.09 & pnt &$\leq$7 & gal & \\
    
    NGC\,4169 & 1.44$\pm$0.34 & 10.89 & 4.972 & -- & -- & -- & -- & -- & 0.71$^{+0.08}_{-0.10}$ & <744.33 & >750.0 & 384 & 1.63 &  pnt & $\leq$3 & pnt &\\
    
    NGC\,4261 & 0.112$\pm$0.005{$^{\star}$} & 11.05 & 0.034 & 0.097 & 15.0 & 1.2 & CC & SCC & 1.36$^{+0.03}_{-0.02}$ & 3.91 & 39.1 & 552 & 4.83 & JET& 80 & GRP &\\
    
    NGC\,5846 & 0.14$\pm$0.06 & 10.83 & 0.019 &0.138 & 14.7 &  0.8 &CC & SCC & 0.95$\pm$0.01 & 7.07 & 25.6 &  452 & 2.65 & jet & 12 & GRP & sloshing\\%/Sloshing \\
    
    NGC\,6658 & <0.71 & 10.76 & -- & -- & -- & -- & -- & --  & 0.29$^{+0.08}_{-0.06}$ & <586.82 & >591.3 & 233 & 0.36 &  -- & -- & pnt &\\
    
    NGC\,7619 & <0.33 & 11.21 & 0.255 & 0.135 &24.5 &  0.7 & CC & SCC & 1.00$\pm$0.01 & 6.62 & 36.4 &  464 & 2.88 & pnt &$\leq$6 &  GRP & merger \\%\\/Merger \\
    
    ESO\,507-25 & 4.23$\pm$0.56 & 10.95 & 0.497 & -- & -- & -- & -- & -- & -- & <331.42 & >333.9 & -- & -- &  diffuse & 11 & -- \\ 
    \noalign{\smallskip} \hline \noalign{\smallskip}
\label{tab:prop_multiwave}
\end{tabular}}
\end{center}
\tablefoot{(1) BGG name.\\
(2) Molecular gas mass from single-dish IRAM and APEX observations \citep{osullivan15,osullivan18b}. $\star$ From ALMA observations \citep{boizelle21}.\\
(3) Stellar mass \citep{osullivan05,osullivan18b}. (4) Star formation rate  \citep{osullivan05,osullivan18b}.\\
(5) Central cooling time at 10 kpc. (6). min t$_{\rm cool}$/t$_{\rm ff}$ values. Systems where no deprojected profile was available are marked with an asterisk and were calculated at 10~kpc \citep{osullivan17}.\\
(7) Values of t$_{\rm cool}$/t$_{\rm eddy}$, where t$_{\rm cool}$ is measure at 10~kpc \citep{osullivan17}. The eddy turn-over timescale, t$_{\rm eddy}$, has been measured following \citet{olivares19}.\\
(8) and (9) Type core indicates the classification of groups as cool-core/non-cool-core based on their temperature profiles, and based on the classification of \citet{hudson10} as strong (SCC), weak (WCC), and non-cool-core (NCC). Taken from \citet{osullivan17}.
Entries in brackets indicate systems where only a projected temperature profile with $<$3 bins is available.\\
(10) System temperature \citet{osullivan17}.\\
(11) Entropy measured at 1~kpc based on X-ray profiles from \citet{osullivan17}.\\ 
(12) Entropy measured at 10~kpc \citet{osullivan17}.\\
(13) R$_{500}$ refer to the radii at which the cluster mass density is 500 times the critical density of the universe.\\
(14) Total mass within R$_{500}$.\\
(15) Radio morphology from \citet{kolokythas18}. The physical scale of the radio emission moves from few kpcs with an (i) point-like (pnt) radio source with sizes  $\leq$11~kpc, (ii) diffuse emission (diffuse) with no clear jet or lobe structure, (iii) small-scale ($<$20~kpc) jets (jet) confined within the stellar body of the host galaxy, and (iv) large-scale ($>$20~kpc) jets (JET)  extending beyond the host galaxy and into the IGrM.\\
(16) The largest linear size (LLS) of the radio source, taken from \citet{kolokythas18}.\\
(17) X-ray morphology classification and notes from \citet{osullivan17}. For systems where thermal emission was detected, we classified the extent of the gas halo as either group-like (GRP) (extent $>$65~kpc), galaxy-like (Gal) (extent $\sim$10-65~kpc) or point-like (point) (unresolved, extent smaller than the XMM PSF). Although somewhat arbitrary, these classifications give a simple picture of the scale of the emission.\\
(18) X-ray features.
}
\end{table*}

\section{MUSE Observations and Analysis}\label{sec:muse}
\begin{figure}[htb!]
    \centering
    \includegraphics[width=0.5\textwidth]{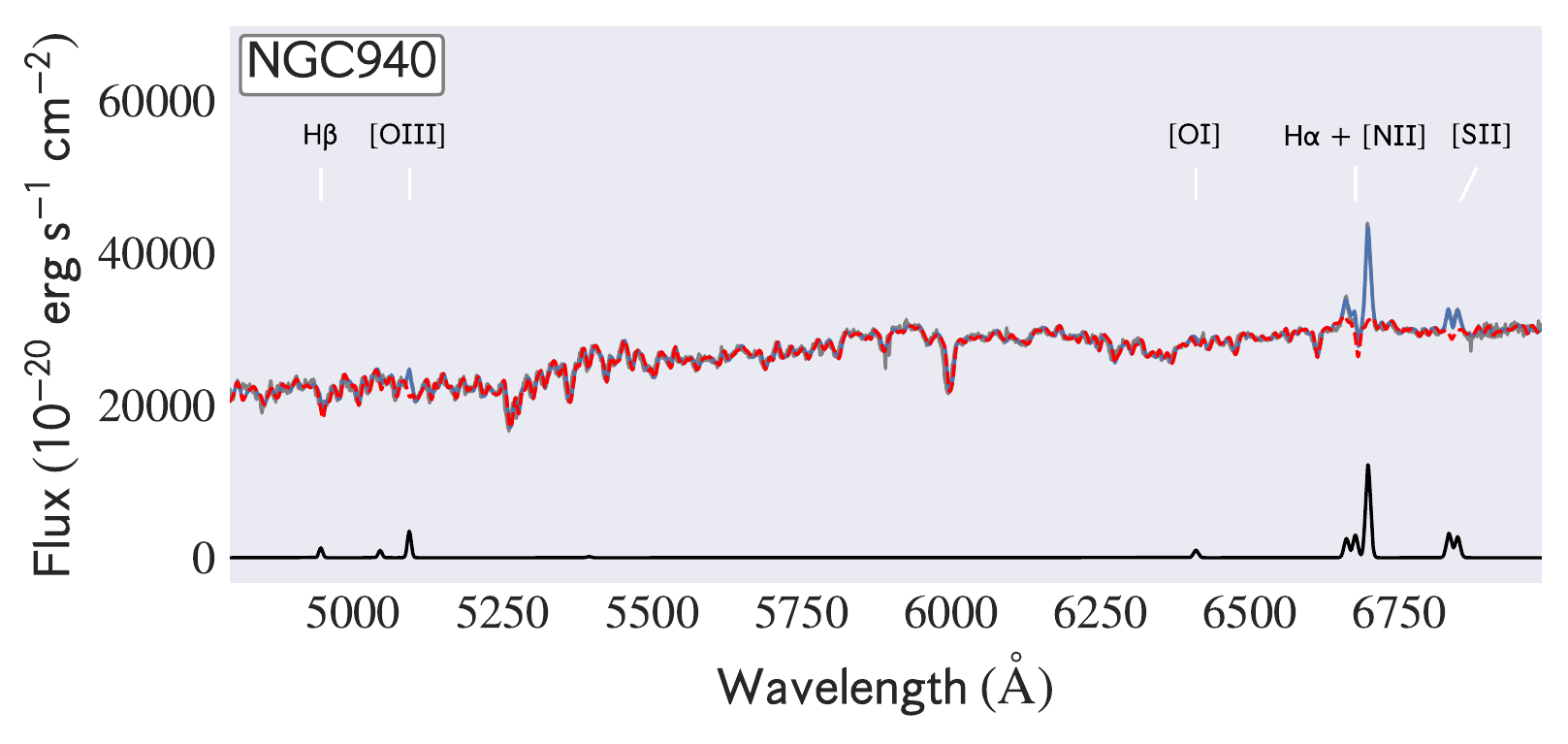}
    \caption{The MUSE optical spectrum extracted from a 1$\arcsec$ square aperture from the center of NGC\,940 is shown in grey. The nebular (blue solid-line) emission and stellar (red dashed-line) components are shown. The spectrum after stellar continuum subtraction is shown with a solid black line, revealing that some of the H$\alpha$ and H$\beta$ flux is recovered due to the underlying absorption features in the stellar atmospheres. The detected emission lines are labeled at the top of the spectrum. The MUSE coverage extends to around 9300\,\mbox{\AA}, but we have fitted and truncated it at 7500\,\mbox{\AA}.}
    \label{fig:spectrum_examples}
\end{figure}

The observations of the central group galaxies were obtained using the MUSE instrument during period 97 (097.A-0366(A) program; PI: S.~Hamer) between April and July of 2016. The MUSE IFU on the ESO VLT is a large-FOV (1$\arcmin$ $\times$ 1$\arcmin$) optical ($\sim$0.46--0.93~$\mu$m) instrument with a resolving power of R$\sim$3,000 at the center of the band. Two BGGs of 097.A-0366(A) program were not observed, NGC\,2563 and NGC\,1167, and will not be discussed in this work. The observations were carried out in the wide-field mode (WFM) of the instrument, which provided a spatial sampling of 0.2$\arcsec$ over a FOV that was 1$\times$1 arcmin. The observations consisted of four pointings. Three of 870s to 900s, were positioned at the center of each galaxy group, and a fourth targeted a nearby empty region for 120s to provide an accurate measure of the sky spectrum, uncontaminated by emission lines from the central galaxy. The observations were taken with seeing in the range of 0.5$\arcsec$ to 1.9$\arcsec$. See Table~\ref{tab:sample2} for more details for each source.

The data were reduced using the MUSE data reduction pipeline (version 2.6) provided by the ESO with the individual recipes executed from the European Southern Observatory Recipe Execution Tool (ESOREX v.3.13.1) command-line interface. After combining the science frames, the sky subtraction was done using the \texttt{ZAP} (version 2.1; \citealt{soto16}), a high precision sky subtraction tool. We included a sky-cube in order to improve the process in addition to the sky subtraction of the MUSE pipeline.

The data were corrected for galactic extinction using an O'Donnell extinction curve \citep{odonnell94} and the dust maps of \citet{schlegel98}. The continuum and nebular emission were fitted simultaneously, using the \texttt{Platefit} code\footnote{http://userpages.planetefl.fr/flamareille/galaxie/}, described in \citet{tremonti04}. The MUSE coverage extends to 9,300\,\mbox{\AA}, but we have fitted and truncated it at 7,500\,\mbox{\AA}. Within this range, the important optical emission lines remain. In a few sources with weak optical emission lines but a strong continuum, the Balmer absorption features from the stellar atmospheres dominate over the emission lines, making it challenging to accurately fit the Balmer emission features. Those sources were labeled as Absorption in Table~\ref{tab:sample2}. In such cases, the H$\alpha$ line often appears very weak or even completely absent from some parts of the cube, usually at the central position of the galaxy (see Fig.~\ref{fig:spectrum_examples} for an example), where the continuum is strong. Therefore, for the purpose of this work, we present the [NII]$\lambda$6583 as a reliable tracer of the ionized gas distribution for all the sources. To tackle this issue, a stellar model was reproduced using the latest version of the \texttt{pPXF} (Penalized Pixel-Fitting) code \citep{cappellari17}, taking stellar templates from the Indo-US spectral library, with T$_{eff}$ $\sim$3000-30000~K and [Fe/H]$\sim$3--1 \citep{valdes04}, and fitting continuum and nebular emission, simultaneously. We chose this library because of its spectral resolution of 1.35\AA. The fit of the emission lines was carried out by using Gaussians in velocity space to an adjustable list of lines (e.g., H$\alpha$, [NII]$\lambda$6548, [OI]$\lambda$6300, [OIII]$\lambda$5003, H$\beta$, [SII]$\lambda\lambda$6717, and [SII]$\lambda\lambda$6732). All Forbidden and Balmer emission lines were tied to have the same velocity and velocity dispersion.

\subsection{Stellar Kinematics}
To extract the stellar kinematics from the MUSE spectrum, we used the \texttt{pPXF} code, fitting the observed spectra with the stellar population templates from the Indo-US spectral library, as described above. Then we performed the fit to spatially binned spectra based on S/N ratio (with a minimum S/N of 50), with the \texttt{Voronoi} tessellation techniques in the line-free stellar continuum \citep{cappellari03}. This allows us to derive for each \texttt{Voronoi} bin a rotational velocity and velocity dispersion estimate. The final maps of the kinematics of the stellar component are presented in Section~\ref{sec:results} for each source. Note that all the MUSE velocity maps shown in this paper are projected at a zero-point set to the stellar systemic velocity of each source. This one is measured through the different absorption features (such as the sodium D absorptions, NaD\,$\lambda$5895.9 and NaD\,$\lambda$5889.9), that trace the galaxy's stellar component by taking the median value of all spaxels.

\subsection{Double Velocity Components} We note that some sources revealed the presence of second velocity components, with double optical emission lines per spaxel. Some examples of those components are the off-nuclear region seen in NGC\,940 and ESO\,507-25 (see Fig.~\ref{fig:double_components} for some examples of the integrated spectra over those regions). Our fitting procedure was performed with a single-gaussian per spaxel per optical emission line and tied to have the same velocity and velocity dispersion. Consequently, some of the optical emission coming from the second velocity components was lost from our fitting procedure. To quantify the flux coming from those components, we created a residual map by subtracting the H$\alpha$ integrated emission-line map from the observed continuum-subtracted cube to the fitted cube. Then we measured the residual flux coming from the pixels with a high $>$3-5 S/N. The second component is a small fraction of the first one. {We found that although 25--75\% of the total flux originates from the areas in which a second component is observed, the flux of the second component makes up only a small fraction ($\sim$2-4\%) of the total flux of each source.} Double components thus remain rare and represent a very small fraction of the total emission. Therefore, we do not discuss the double-components here and leave them for a future study. An interesting value is the velocity shift between the two components $+$200 -- $+$350~km~s$^{-1}$ and $-$300~km~s$^{-1}$ depending on the source. We suspect that those double components trace scattered gas threads aligned along the line of sight but with different orientations. Similar findings have been reported in a few cool-core clusters of galaxies, like Abell\,1664 \citep{russell13, russell19, olivares19}. \citet{hamer16} also found nine multiple velocity component sources within the sample of 73 BCGs/BGGs using VIMOS observations.

\section{Results}\label{sec:results}
\subsection{Stellar Distribution and Kinematics}

\def\setwithtotal{0.8}
\def\setwidthsmall{0.8}
\begin{figure*}[htbp!]
    { \noindent \bf \large Extended Rotating Disks}\\
        \begin{center}
            \vspace{-0.7cm}
        \subfigure{\includegraphics[width=\setwidthsmall\textwidth]{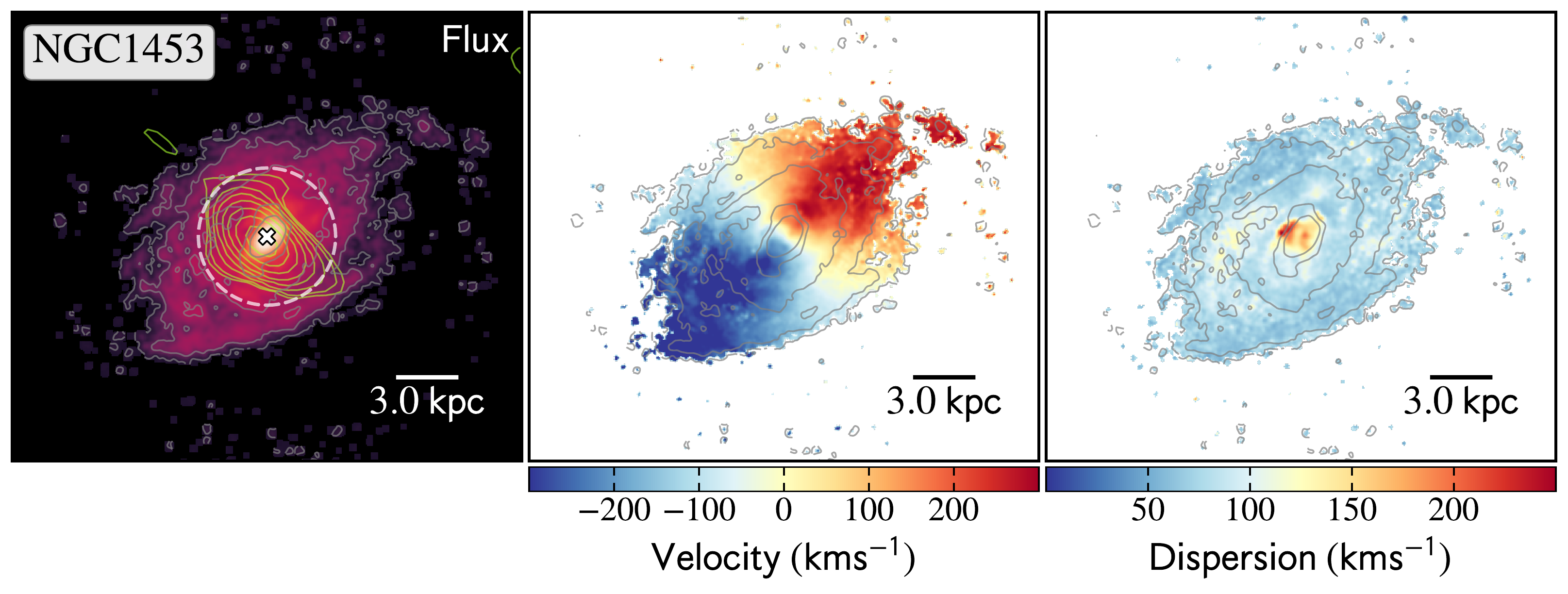}}\\
            \vspace{-0.5cm}
        \subfigure{\includegraphics[width=\setwidthsmall\textwidth]{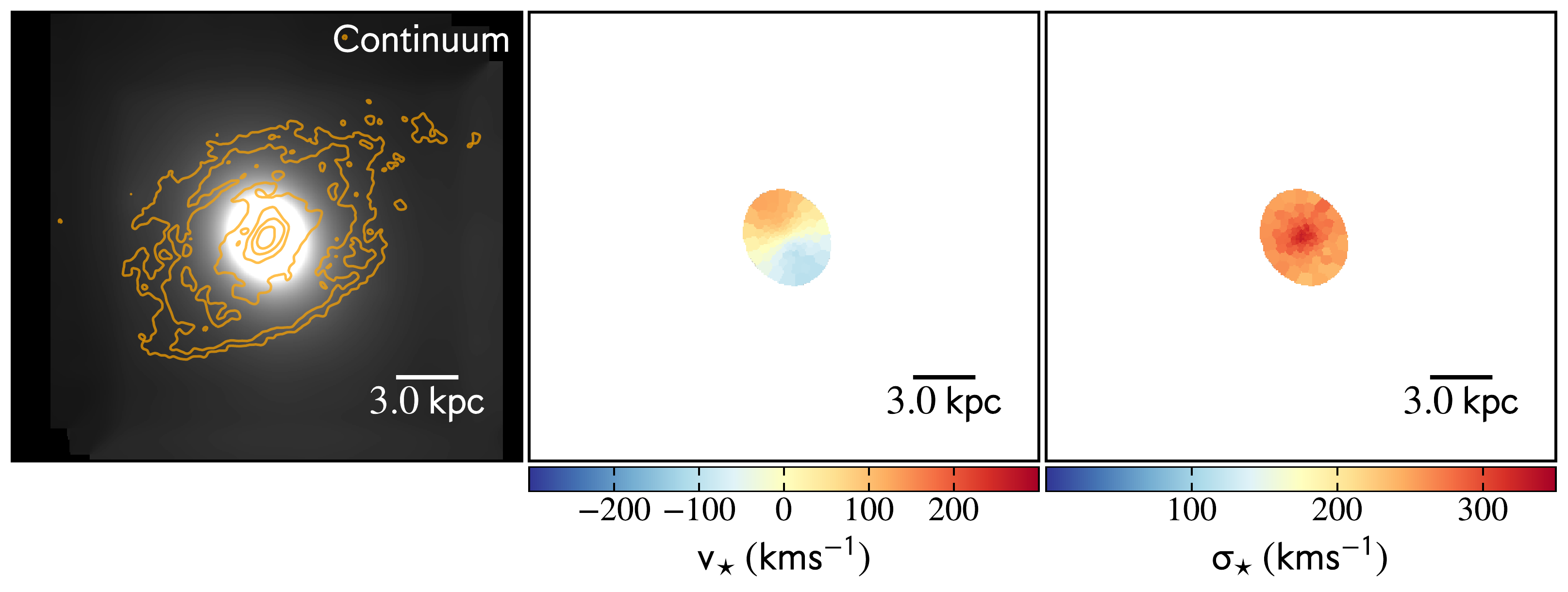}}
        \end{center}
    \vspace{-0.5cm}
        \caption{\noindent Maps of the distribution and kinematics of the ionized warm gas and the stellar components for the BGG sample}{Maps of the spectral fitting to the MUSE data cubes for the sample. The first row, from left to right the panels are: 1) The [NII]$\lambda$6583 flux map in logarithmic scale, overlaid by GMRT 610~MHz contours in green from \citet{kolokythas18}, 2) the line of sight velocity profile of the optical emission lines, 3) the velocity dispersion of the emission lines, overlying [NII]$\lambda$6583 contours in gray. The second row from left to right the panels are: 1) MUSE continuum image, overlaid by [NII]$\lambda$6583 contours in orange, 2) stellar velocity, and 3) stellar velocity dispersion. All sources names are indicated at the top of the first panel of the first row. 
        The 30m beam size from IRAM and APEX observations has been added in white when the sources were non-detected, and in cyan when the molecular gas was detected. {North is up and east is left.}}\label{fig:nii_maps_examples}
\end{figure*}

\begin{figure*}[htbp!]
    \ContinuedFloat
    \captionsetup{list=off,format=cont}
    { \bf  \large Compact Rotating Disks}\\
        \vspace{-0.5cm}
    \begin{center}
        \subfigure{\includegraphics[width=\setwidthsmall\textwidth]{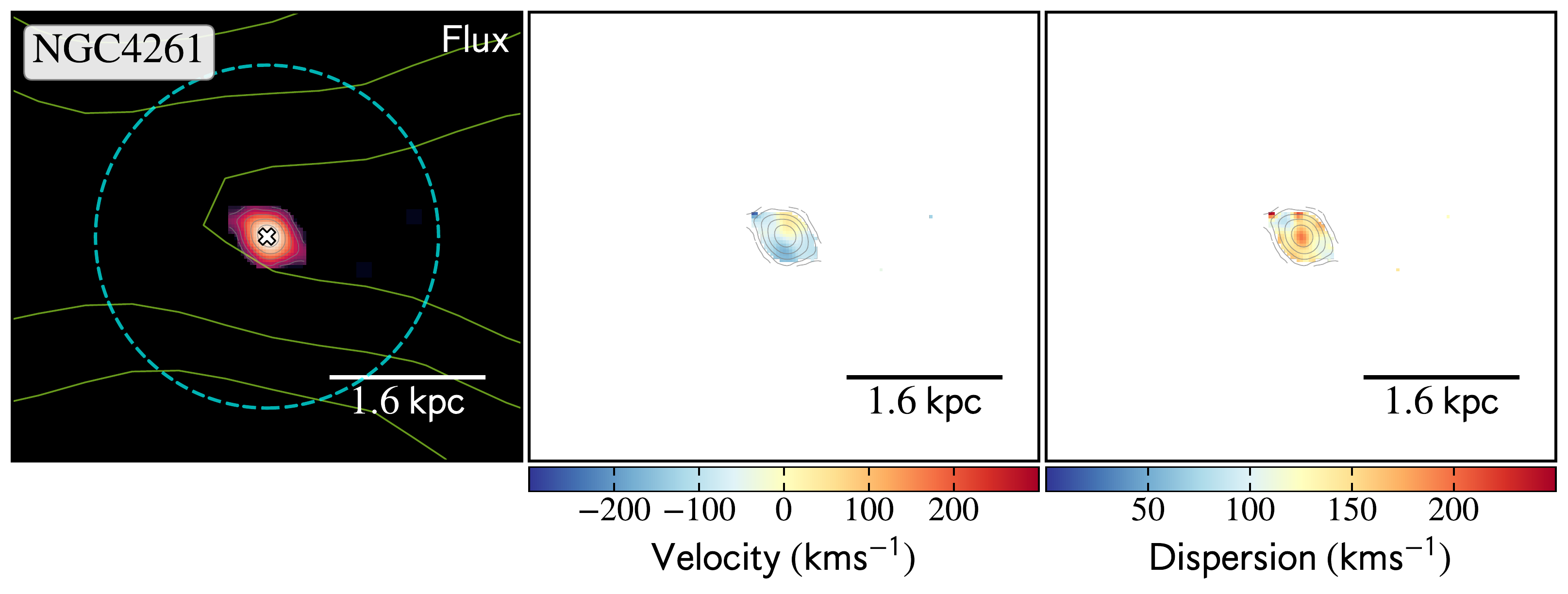}}\\
        \vspace{-0.5cm}
        \subfigure{\includegraphics[width=\setwidthsmall\textwidth]{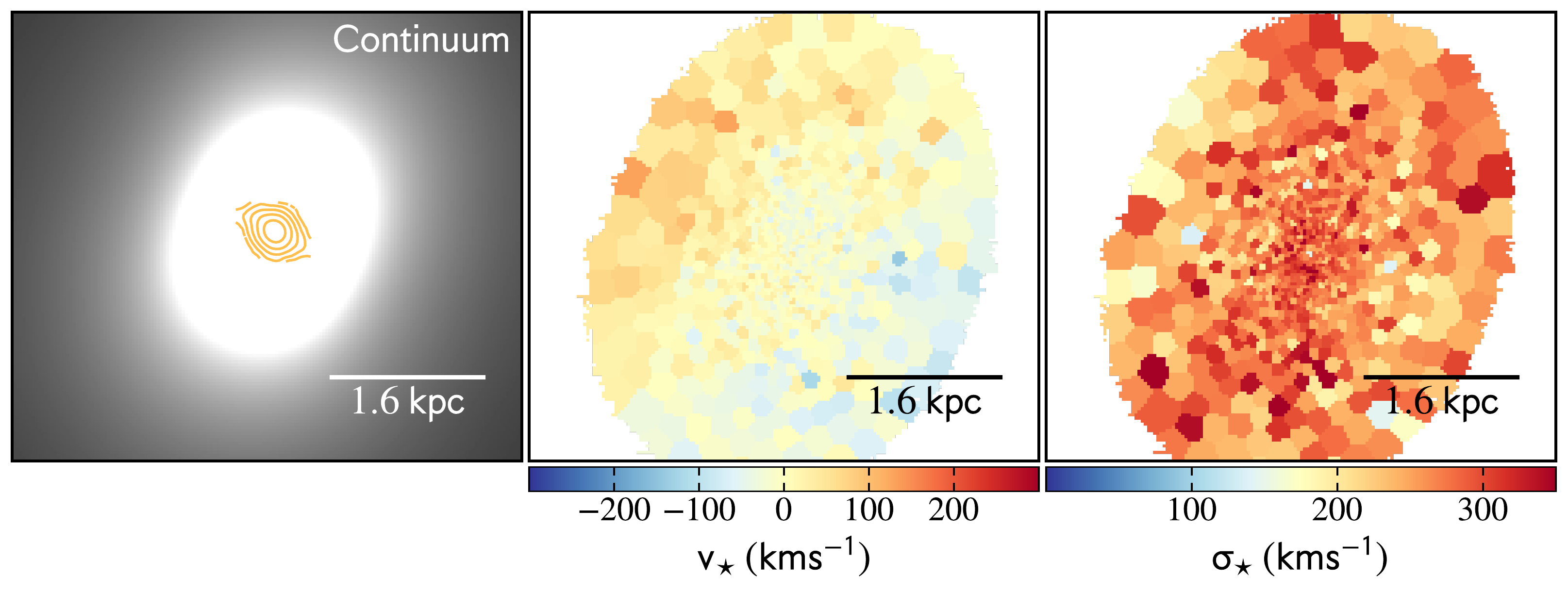}}
        \end{center}
            \vspace{-0.5cm}
    { \bf  \large Extended Filaments }\\
        \begin{center}
                \vspace{-0.5cm}
        \subfigure{\includegraphics[width=\setwidthsmall\textwidth]{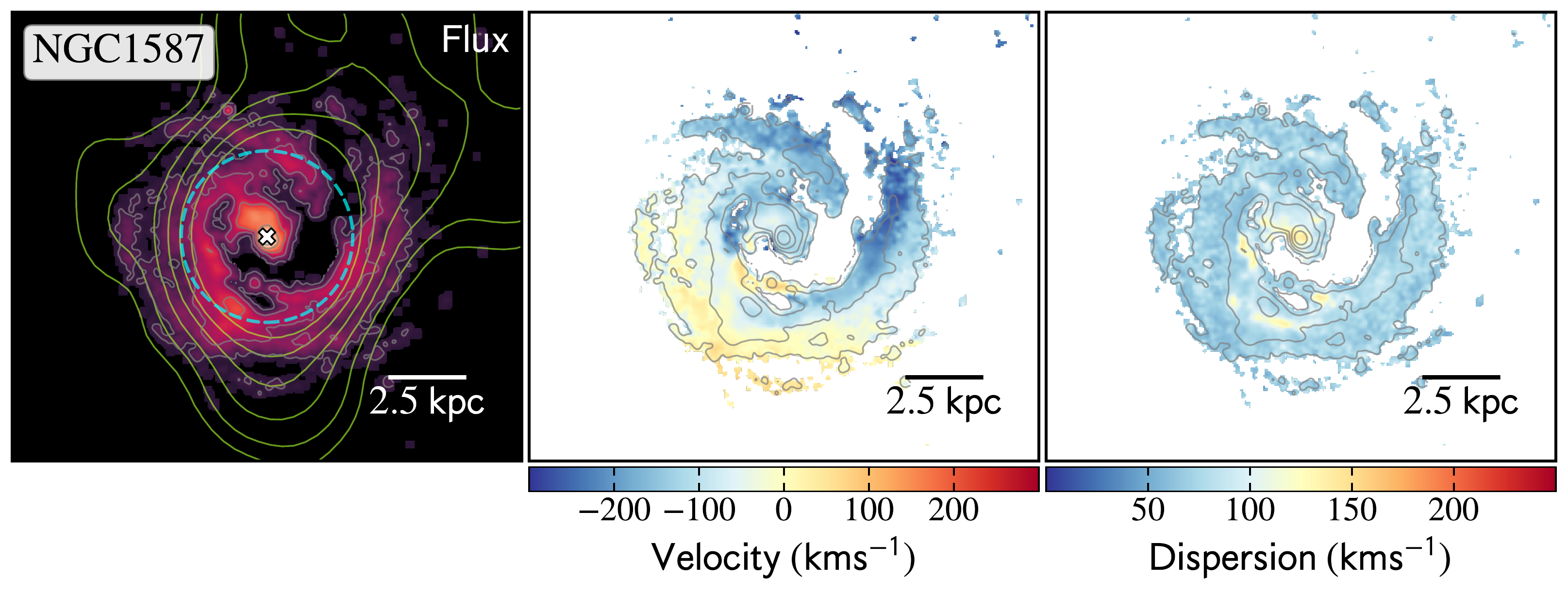}}\\
            \vspace{-0.5cm}
        \subfigure{\includegraphics[width=\setwidthsmall\textwidth]{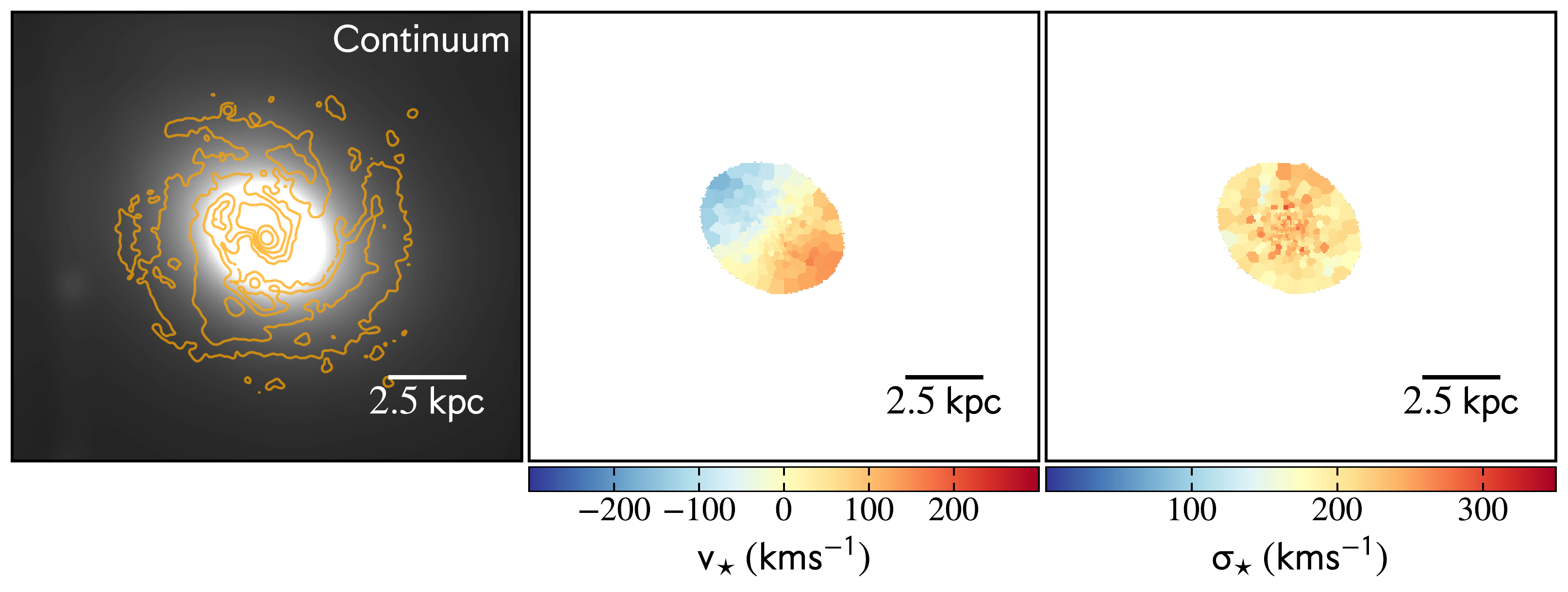}}
        \end{center}
       \caption{Continuation of Fig.~\ref{fig:nii_maps_examples} (see text for details).}
\end{figure*}

\begin{figure*}[htbp!]
    \ContinuedFloat
    \captionsetup{list=off,format=cont}
    { \bf  \large Compact Filaments }\\
            \begin{center}
                \vspace{-0.5cm}
        \subfigure{\includegraphics[width=\setwidthsmall\textwidth]{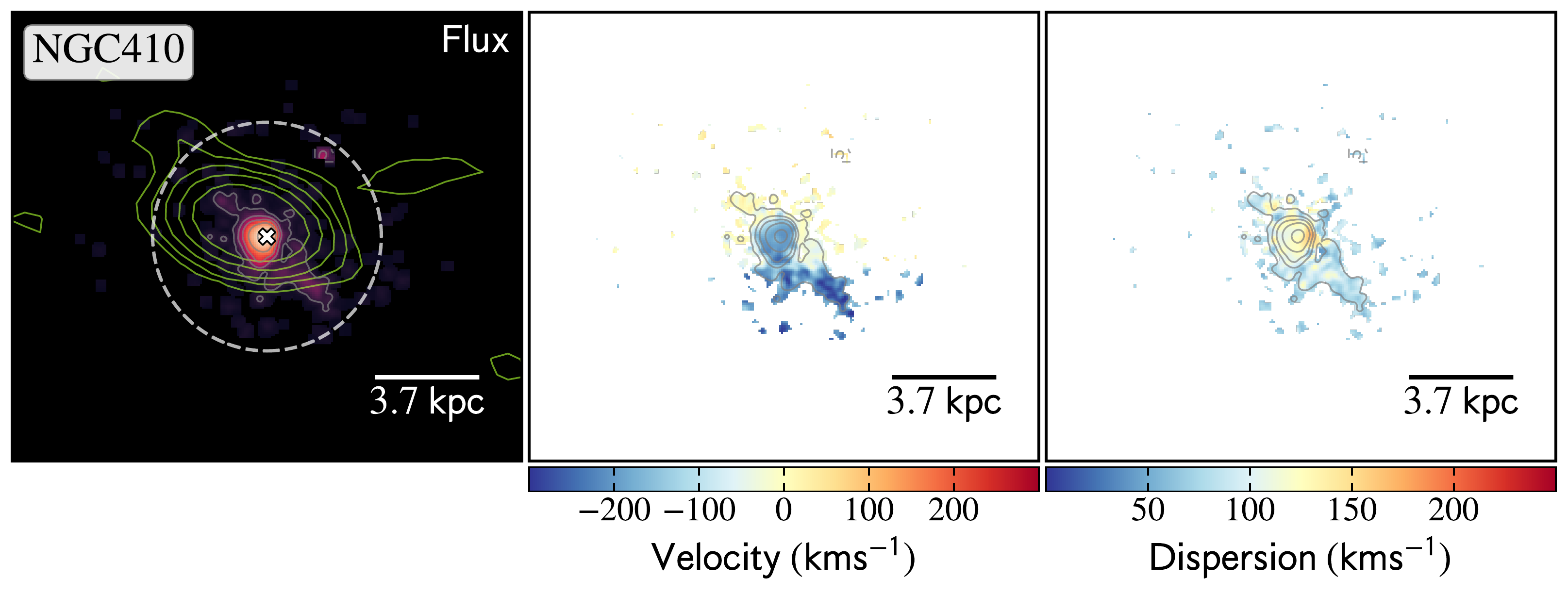}}\\
            \vspace{-0.5cm}
        \subfigure{\includegraphics[width=\setwidthsmall\textwidth]{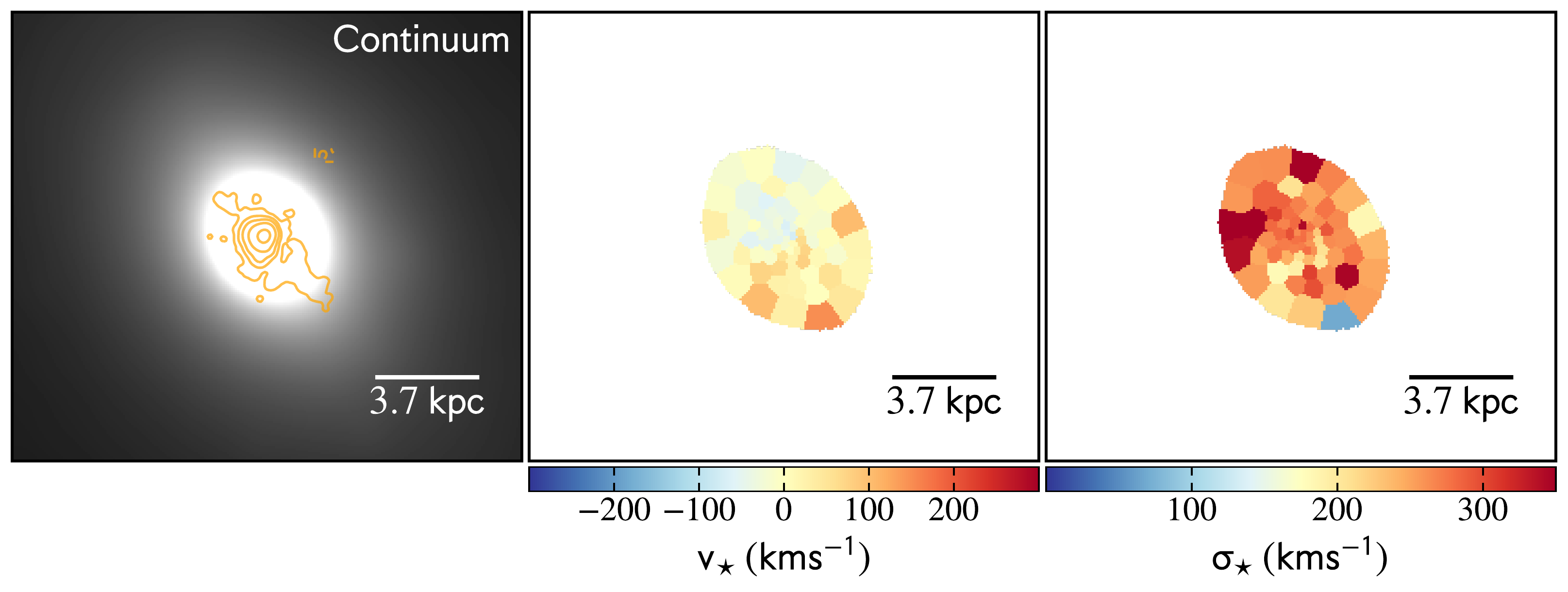}}
        \end{center}
       \caption{Continuation of Fig.~\ref{fig:nii_maps_examples} (see text for details).}
\end{figure*}

In figures~\ref{fig:nii_maps_examples} and \ref{fig:nii_maps_examples_app} we present the distribution of the optical emitting gas traced by [NII]$\lambda$6584 emission line (first panel from the left), line of sight velocity (second panel), and velocity dispersion (third panel) maps on the top row. On the bottom row, we show the stellar continuum (first panel from the left), \texttt{Voronoi}-binned MUSE map of the stellar line of sight velocity (second panel) and velocity dispersion (third panel) for each source in our sample. Only \texttt{Voronoi}-binned spaxels with S/N$>$50 are shown.

The optical continuum images from our MUSE observations unveil an extensive variety of morphologies, a large range of sizes, and shapes for our systems. The galaxies have been classified as E0, E1, E2, E3, E5, S0, and S0a \citep{deVaucouleurs91}. %\footnote{From HYPERLEDA (http://leda.univ-lyon1.fr/)}.
Table~\ref{tab:sample2} lists the galaxy type for each source. The stellar velocity fields reveal that most of the sources in our sample (10/18 sources, e.g., NGC\,940, NGC\,924, NGC\,1453, NGC\,4169, NGC\,978, NGC\,584, NGC\,1587, NGC\,4008, NGC\,7619 and NGC\,6658) show a clear large-scale rotation pattern in their stellar light. In the case of ESO\,507-25, the stellar map reveals a non-regular rotation patter. Another 3 sources exhibit weaker signatures of stellar rotation (NGC\,777, NGC\,677, and NGC\,4261) in their stellar velocity fields. In particular, NGC\,4261 has a prolate morphology with rotation around its major axis \citep[e.g.,][]{davies86}. The detail of the stellar kinematics and angular momentum analysis will be discussed in separate paper (Loubser et al. in prep). 

Contrary to the random motions of the stellar component in BCGs \citep[e.g.,][]{hamer16}, the kinematics of the stellar component lack clear net stellar rotation in only 4 sources of the sample (NGC\,193, NGC\,410, NGC\,1060, and NGC\,5846). Particularly in NGC\,5846, there is a very weak hint of coherent stellar rotation in the inner region of the galaxy (West approaching, East receding), as well as in NGC\,410 (Northwest receding, and Southeast approaching).

For the rotating stellar disk sources, we notice a large range of projected stellar velocities, from $\pm$40~km~s$^{-1}$ up to $\pm$350~km~s$^{-1}$, and the stellar velocity dispersion of 80~km~s$^{-1}$ up to 450~km~s$^{-1}$. Interestingly, we found that the stellar dispersion maps in a few sources with clear stellar rotation (such as NGC\,940, NGC\,924, NGC\,4169 and NGC\,584) exhibit a deeper decrease at large radii. For those systems, the stellar velocity dispersion displays a peak towards the galaxy center, with values of $\sim$180 up to 250~km~s$^{-1}$, while decreasing rapidly at larger radii, reaching velocity dispersions below $<$100~km~s$^{-1}$. 
For the case of NGC\,924 and NGC\,584, spatially resolved {long-slit} spectroscopy found consistent results (steep negative velocity dispersion gradient, see \citealt{loubser18}). Nonetheless, the stellar velocity dispersion map is more homogeneous for most sources, with a shallower radial dispersion decrease. \citet{loubser18} also compare the stellar properties of the high- and low-richness CLoGs sub-samples of BGGs to massive BCGs and found that lower mass systems are dominated by stellar rotation compare to massive systems (BCGs), where dispersion becomes more important. Our results agree with this picture, as more massive BGGs, $>$2$\times$10$^{13}$~M$\odot$, usually either show weak or lack net stellar rotation (with the exception of NGC\,4261).

In summary, the stellar component of the BGGs almost always reveals some patterns of rotation. 
Ten sources ($\sim$55\% of the sample) reveal a strong velocity gradient, while four ($\sim$22\%) other sources have a mild sign of rotation which is very different from local BCGs, where random motions generally dominate the stellar component.

\subsection{Properties of the Ionized Gas}
\subsubsection{Distribution and Velocity of the Gas}
\begin{figure}[htbp!]
    \centering
    \includegraphics[width=0.95\columnwidth]{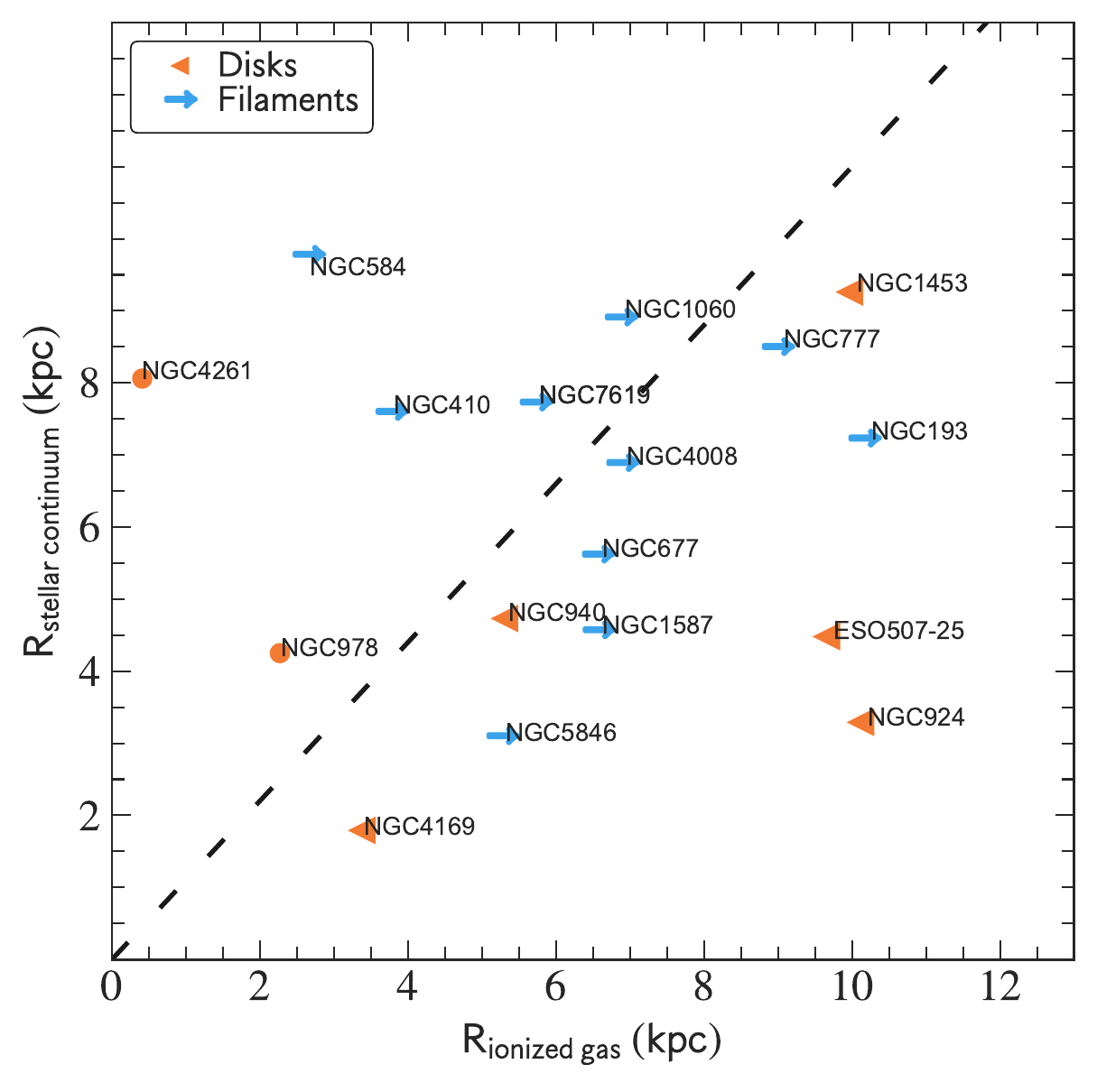}
    \caption{Projected radius measured at a surface brightness of $\rm \sim10 ^{-17}~erg~s^{-2}~cm^{-1}~arcsec^{-2}$ for the stellar continuum versus the ionized gas.
    The dashed black line corresponds to the one-to-one size relation.
    The sources are color-coded as follows: orange left-pointing triangles for extended rotating disks, orange circles for compact rotating disks, and blue right-arrows for systems with filamentary structures. We remark that the extension of the filamentary systems is likely a lower limit due to the projection effects and a sensitivity threshold.}
    \label{fig:galaxy_ha_sizes}
\end{figure}

From the sample of 18 BGGs observed with the MUSE telescope, we have a 3$\sigma$ detection of [NII]$\lambda$6583 emission line in 17/18 sources. In NGC\,6658, optical-emission lines were not detected. The detected optical emission lines for each source are listed in Table~\ref{tab:sample2}. In Table~\ref{tab:MUSE_propserties} we summarize the observed properties of the optical emission lines and the stellar component, such as the morphology of the ionized gas, projected size of the ionized nebula, integrated H$\alpha$ flux, H$\alpha$ luminosity, [NII]$\lambda$6583/H$\alpha$ ratio, semi-major axis and kinematics of the stellar component. Our sample shows a wide range of H$\alpha$ luminosities, 0.04$\pm$0.04--5.01$\pm$0.11$\times$10$^{40}$~erg~s$^{-1}$, and a broad range of ionized gas projected sizes, 1--21~kpc. A detailed description of distribution and kinematics of the ionized gas is presented in Appendix~\ref{app:description_sources} for all sources.
 
With the high-resolution of our MUSE observations ($\sim$2\arcsec, 0.4~kpc), we can adequately compare the relative size of the ionized warm gas distribution to that of the stellar component. The distribution and relative sizes of the optical emission line and stellar components can be used as an indicator to distinguish between an internal and external origin of the gas. For that reason, we compute relative sizes at a given surface brightness to assure consistency in the comparison. We compute the surface brightness profile using circular apertures on the [NII]$\lambda$6583 flux map and the Johnson B filter continuum emission. Figure~\ref{fig:SB_radial_profile} show the surface brightness profiles of the optical emission line and stellar continuum emission (Johnson B filter) for each source. Spaxels with an S/N ratio below 7 in the [NII]$\lambda$6583 flux map were masked to compare the sizes at a given surface brightness value adequately.
Additionally, stars or galaxies presented in the continuum map that do not correspond to the central galaxy were also masked. In Fig.~\ref{fig:galaxy_ha_sizes} we compare the radial extension of the ionized gas and the continuum emission of the galaxy, measured at an apparent surface brightness of $\sim$10$^{-17}$~erg~s$^{-1}$~cm$^{-2}$~arcsec$^{-2}$ (or $\sim$21~mag~arcsec$^{-2}$), which roughly corresponds to the surface brightness of the filaments. {If the surface brightness threshold is set to a higher value, this could lead to a different classification from extended sources to compact.
}

We distinguished four categories for the ionized gas distribution based on the distribution and kinematics of the ionized gas and the comparison between the extension of the gas and the stars. For the sources dominated by rotation, we identify i) {\it compact rotating disks} and ii) {\it extended rotating disks}, while for the disturbed gas distribution, we have iii) {\it extended filamentary} and iv) {\it compact filamentary} sources. Note that the extension of the filamentary sources needs to be taken as a lower limit due to limited sensitivity and projection effects. In figure~\ref{fig:nii_maps_examples} we show an example  for each category. It is important to note that the definitions are not mutually exclusive, as a single object may exhibit a morphology or kinematics of the gas that is consistent with two or more of the categories. We found that 5/18 sources are extended disks, 2/18 corresponds to compact disks, while the great majority 10 of the 18 are filamentary sources. Note that 3/18 of the extended disk-like sources also have elongated structures arising from the main disk, and one source (NGC\,584) shows a smooth velocity field but a highly disturbed gas distribution. In the following, we describe each category. \\

\noindent $\bullet$ \textbf{ Extended rotating disks}: are sources dominated by rotation with relaxed gas distribution, in which the extension of the gas is larger than the stellar component. Five sources of the sample (5/18, NGC\,924, NGC\,940, NGC\,1453, NGC\,4169, ESO\,507-25) belong to this category. The ionized gas distribution in these rotating disks is extended, with projected sizes from 5 to 21~kpc, and projected radii of $\sim$3.2--7.5~kpc, measured from the center of the galaxy to the faintest emission. We note that the peak of the [NII]$\lambda$6583 emission is usually well aligned with the peak of the stellar continuum. We also remark that three systems show a clumpy rotating ring (NGC\,940, ESO\,507-25, and NGC\,924). The gaseous clumps of the rings have projected sizes of $\sim$1--3$\arcsec$ (0.2--1~kpc). Furthermore, we note that several disks display tails and extended structures of ionized gas arising from the main rotating disk (ESO507-25, NGC\,924, NGC\,1453, and NGC\,4169).

\noindent $\bullet$ \textbf{ Compact rotating disks}: are defined as objects in which the distribution and kinematics of the ionized gas is compact and dominated by rotation, and the projected radial extension of the ionized gas is smaller ($\sim$1--3~kpc) than that of the stellar component. Two sources, NGC\,4261 and NGC\,978, in our sample (2/18) belong to this category. 

\noindent $\bullet$ \textbf{ Extended Filaments}: are objects that show elongated structures, unrelaxed gas distributions, the kinematics of the gas is not dominated by rotation, and the optical emission is more extended than the continuum emission. {As turbulence majorly dominates over rotation in these systems (see also Fig.~\ref{fig:taylor_disks}), the CCA model suggests that these systems experience a more extended and filamentary rain of cold/warm gas \citep{Gaspari_2018}}. Six sources (6/18, NGC\,193, NGC\,677, NGC\,777, NGC\,1587, NGC\,4008, NGC\,5846) fall in this category. 

\noindent $\bullet$ \textbf{ Compact Filaments}: Compact gaseous filamentary objects are systems that show a disturbed distribution, but the projected radial extension of the gas is smaller than that of the stellar component. We found that four sources (4/18) (e.g., NGC\,410, NGC\,584, NGC\,1060, NGC\,7619) belong to this category. 

The sources in the filamentary categories display a velocity field dominated by disturbed kinematics, making the warm ionized filaments more challenging to interpret. Those disturbed velocity fields are probably due to a projection of several filaments that are inflowing or outflowing across the line of sight, as it has been shown in the Perseus Cluster through SITELLE observations, and in several cool-core clusters using ALMA and MUSE observations \citep[e.g.,][]{gendron-marsolais18,tremblay18,russell19,olivares19,north21}. The filamentary structures are often clumpy and without coherent velocity structure along the optical emitting nebulae. The best example is seen in NGC\,5846, which hosts a very rich, dense network of filaments extending from the core out to a projected radius of $\sim$40$\arcsec$ (5~kpc). NGC\,584 also presents an exquisite clumpy net of threads extending along $\sim$24$\arcsec$ (3~kpc) from the core of the galaxy. Many sources have an ionized gas peak that coincides with the center of the galaxy, accompanied by one single coherent structure in velocity (NGC\,193, NGC\,1060, NGC\,1587). The projected sizes of these sources go from 4.5 to 14~kpc. Clear examples of very chaotic velocity fields can be seen in NGC\,677, NGC\,4008, and NGC\,5846. In some cases, velocity gradients along the filaments are detected, but the gas is quite disturbed on average. Last, some hint of rotation is detected within the central kilo-parsecs of some BGGs (see for instances, NGC\,193, NGC\,584, NGC\,677). 

\begin{table*}[htb]
\caption{Ionized gas and stellar components properties for each object of the sample.}
\label{tab:MUSE_propserties}
\begin{center}
\scalebox{1.0}{
\begin{tabular}{llrcccc|clr}
    \noalign{\smallskip} \hline \hline \noalign{\smallskip}
    \rm BGG     & \multicolumn{5}{c}{Warm ionized gas component} &&& \multicolumn{2}{c}{Stellar component} \\
    \cline{2-6} \cline{9-10}\\
    \rm         & Morphology & Size & H$\alpha$ flux & Luminosity H$\alpha$ & [NII]/H$\alpha$ & && Kinematics & $a_{\rm stellar}$ \\
    \rm      &             &      (kpc)      & (10$^{-14}$~erg~s$^{-1}$~cm$^{-2}$) & (10$^{40}$ erg~s$^{-1}$) & &  & & & (kpc) \\
    \rm (1) & (2) & (3) & (4) & (5) & (6) & && (7) & (8) \\
    \noalign{\smallskip} \hline \noalign{\smallskip}
    NGC\,193  & E. Filament    & 11.85     & 1.10$\pm$0.04 & 0.53$\pm$0.02 & 1.13 & & & Unordered      & 5.14\\
    NGC\,410  & C. Filament    & 5.78      & 1.08$\pm$0.09 & 0.76$\pm$0.06 & 0.64 & & & Unordered      & 10.76\\
    NGC\,584  & C. Filament    & 7.26      & 2.91$\pm$0.12 & 0.23$\pm$0.01 & 0.88 & & & Rotating disk  & 2.52 \\
    NGC\,677  & E. Filament    & 11.34     & 5.36$\pm$0.24 & 3.50$\pm$0.16 & 0.58 & & & Weak Rotation & 4.61 \\
    NGC\,777  & E. Filament    & 9.91      & 0.19$\pm$0.02 & 0.12$\pm$0.01 & 2.27 & & & Weak Rotation & 6.65\\
    NGC\,924  & E. Disk (+Ring)        & 20.46     & 2.01$\pm$0.10 & 1.01$\pm$0.05 & 0.88 & & & Rotating disk  & 6.18 \\
    NGC\,940  & E. Disk (+Ring)        & 10.41     & 7.41$\pm$0.09 & 4.87$\pm$0.06 & 0.72 & & & Rotating disk  & 4.15\\
    NGC\,978  & C. Disk    & 3.34      & 0.96$\pm$0.04 & 0.54$\pm$0.02 & 0.47 & & & Rotating disk  & 4.64 \\
    NGC\,1060 & C. Filament    & 9.86      & 0.40$\pm$0.05 & 0.27$\pm$0.03 & 1.95 & & & Unordered & 4.62 \\
    NGC\,1453 & E. Disk    & 21.04     & 6.65$\pm$0.85 & 2.50$\pm$0.32 & 1.15 & & & Rotating disk  & 3.51\\
    NGC\,1587 & E. Filament    & 11.61     & 5.07$\pm$0.45 & 1.72$\pm$0.15 & 0.69 & & & Rotating disk  & 3.41\\
    NGC\,4008 & E. Filament    & 6.89      & 0.18$\pm$0.02 & 0.06$\pm$0.01 & 0.45 & & & Rotating disk  & 5.06\\
    NGC\,4169 & E. Disk    & 4.80      & 5.23$\pm$0.09 & 1.86$\pm$0.03 & 0.95 & & & Rotating disk  & 2.76\\
    NGC\,4261 & C. Disk    & 1.16      & 3.23$\pm$0.01 & 0.39$\pm$0.01 & 2.40 & & & Weak Rotation & 3.00\\
    NGC\,5846 & E. Filament    & 9.83      & 14.05$\pm$0.4 & 1.01$\pm$0.03 & 0.55 & & & Unordered      & 1.96\\
    NGC\,6658 & No gas      & --        & -- & -- & -- & & & Rotating disk & 12.52\\
    NGC\,7619 & C. Filament    & 3.28      & 0.29$\pm$0.02 & 0.11$\pm$0.01 & 1.14 & & & Rotating disk  & 7.62\\
    ESO\,507-25 & E. Disk (+Ring)      & 15.04     & 19.3$\pm$0.42 & 5.01$\pm$0.11 & 0.64 & & & Weak Rotation  & 3.81\\
    \noalign{\smallskip} \hline \noalign{\smallskip}
\end{tabular}
}
\end{center}
\tablefoot{
(1) BGG name.\\
(2) Ionized gas classification (see text for details). Here ``C'' refers to compact disk or filaments, while ``E'' to extended disk or filaments.\\
(3) The projected size is the length of the major axis of the ionized gas.\\
(4) Integrated H$\alpha$ flux integrated over all the map. Total fluxes of the measured emission lines are corrected for galactic and local extinction using the O'Donnell extinction curve \citep{odonnell94}, and the dust maps of \citet{schlegel98}.\\
(5) H$\alpha$ luminosity. (6) [NII]/H$\alpha$ ratios.\\
(7) Stellar kinematics.\\ 
(8) Semi-major axis of the stellar component. 
}
\end{table*}

\subsubsection{Velocity Dispersion of the Gas}
In figures~\ref{fig:nii_maps_examples} and \ref{fig:nii_maps_examples_app} (last panel, top row) we present the projected velocity dispersion measured from our MUSE observations\footnote{The velocity dispersion are not corrected by instrumental broadening}. Below, we describe the distribution of the projected velocity dispersion for the filaments and rotating disks categories separately.

\textit{Rotating disks --} The velocity dispersion, $\sigma_{\rm gas}$, for the rotating disks usually peaked at the galaxy center with values of a few 100~km~s$^{-1}$. The $\sigma_{\rm gas}$ of the clumpy rings detected in a few sources is smaller with values of $\sim$30--60~km~s$^{-1}$, implying that the ionized gas is significantly turbulent within those clumps. As predicted by the turbulence cascades from macro to micro scales mainly via Kolmogorov cascades ($ \sigma_{gas} \propto l^{1/3}$), smaller $\sigma_{\rm gas}$ are expected within clumps. Along the disk, the $\sigma_{\rm gas}$ is relatively small, close to 70 -- 120~km~s$^{-1}$. Nonetheless, a few high dispersion velocity structures, $\sigma_{\rm gas}\sim$200~km~s$^{-1}$, sometimes with spiral shapes, are detected in a few sources (e.g., NGC\,924, NGC\,4169 and NGC\,1453). In NGC\,4261, we note that the broader dispersion is reached at the center of the galaxy, with a value of 210~km~s$^{-1}$. The regions at the East and the West of NGC\,4261 have narrower-spectrum gas, with values between 80 and 100~km~s$^{-1}$.

\textit{Filamentary sources --} In this category, the broadest velocity dispersion is, on average, found at the center of the galaxy with values of 150--350~km~s$^{-1}$. Whereas in the extended structures of those sources, the line-width is narrower, with velocity dispersion values of up to 60~km~s$^{-1}$. A broader dispersion spot ($\sim$120~km~s$^{-1}$) is identified in the network of filaments in NGC\,1587 located to the South of the center. Likewise, in NGC\,5846, broader line widths are detected in some regions of the NE network of filaments (``arm''-like structure), likely due to several clumps located in projection along the same line of sight. 

We found that the projected velocity dispersions of the gas are generally in the range of 30--350~km~s$^{-1}$, with lower values within the clumps of the rings and filaments (30--150~km~s$^{-1}$). Whereas, in the center, the gas is more perturbed with velocity dispersion values reaching $\sim$350~km~s$^{-1}$, likely because of an interaction/presence of an AGN, as well as due to an increase of inelastic collision of clouds at the center, the clouds and filaments cancel some angular momentum to later funnel toward the inner SMBH region, as predicted by the CCA feeding simulations \citep{gaspari17}. The superposition of clumps and filaments with a slightly different line of sight velocity could also broaden the optical emission line. {We note that the high dispersion values at the center found in many sources tend to correlate with higher values of [NII]/H$\alpha$ ratio consistent with LINER/AGN emission (see also Lagos et al. in prep).}

\section{Discussion}\label{sec:discussion}

We map the kinematics and distribution of the ionized gas using new MUSE observations of a sample of 18 nearby BGGs selected from the CLoGS sample. We find that most sources have ionized gas kinematics and distribution consistent with (6/18) extended and (4/18) compact filamentary structures when comparing the sizes to the stellar component. Similarly, the rest of the systems show either (5/18) extended and (2/18) compact disks, while one source has no detected optical emission lines using \texttt{pPXF} and assumed stellar libraries. Contrary to the dispersion-dominated nature of the stellar component in BCGs, the stars appear to be mostly dominated by rotation in our sample. In this section, we compare the kinematics of the ionized gas to the stellar component, then we look at the correlation between the ionized gas and the molecular gas, X-ray and radio emission, and last we discuss the potential scenarios that can lead to the formation of the ionized gas in our sample.

\begin{figure}[h!]
    \centering
    \includegraphics[width=0.99\columnwidth]{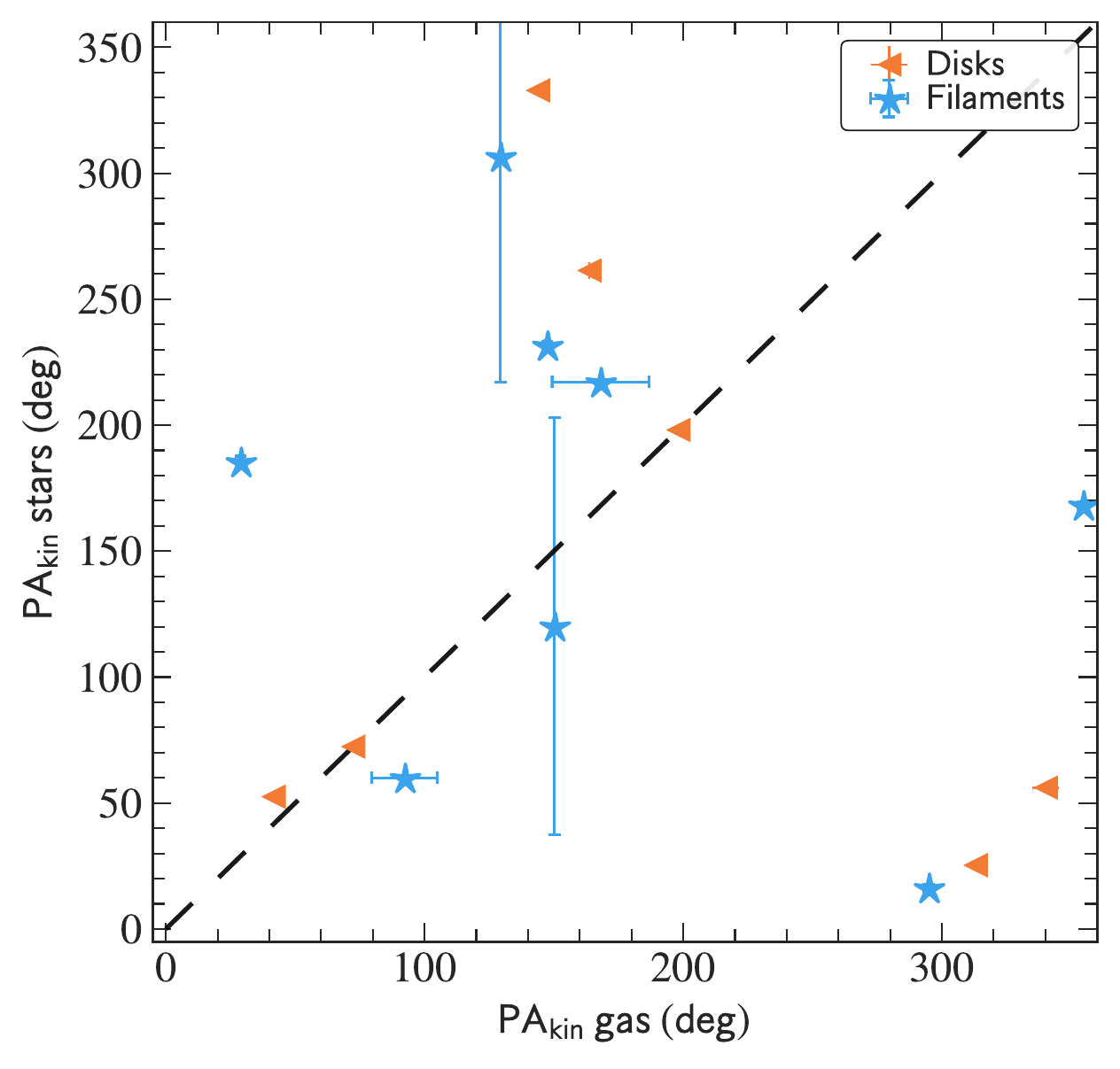}
    \caption{Kinetic PA of the gas versus the stellar component for the rotating disks (orange left-pointing triangles) and some of the filamentary sources (blue stars). The one-to-one relation corresponds to the black dashed line. The gas is kinematically coupled to the stars in three disks. In one disk, the gas and the stars are counter-rotating (open circle).  For the rest of the disks and the filamentary sources, the gas is kinematically decoupled from the stellar component.
    }
    \label{fig:PA_comparison}
\end{figure}

\subsection{Comparison of the Gas and Stellar Kinematics}
\subsubsection{Comparison of the velocity fields}
\label{sec:comparison_kinematics}
\begin{table}[htb]
\caption{Position angle of the ionized gas rotating disks and stellar disks.}
\centering
\begin{tabular}{lccc}
    \noalign{\smallskip} \hline \hline \noalign{\smallskip}
    BGG & PA$_{\rm gas}$ & PA$_{\rm stellar}$ & $\rm \Delta PA$\\
     &  ($\degr$) & ($\degr$) & ($\degr$)\\
    \noalign{\smallskip} \hline \noalign{\smallskip}
    {\it Filaments}    &  &  & \\
    NGC\,193   & 150.2$\pm$0.5  & 120.2$\pm$89.5  & 30.0 \\ % CHECK
    NGC\,410    & 28.9$\pm$ 0.9 & 185.4$\pm$2.3  & 156.5 \\
    NGC\,584    & 92.3$\pm$12.7  & 60.0$\pm$0.6  & 32.3\\
    NGC\,677    & -- & 118.5$\pm$6.3  & -- \\
    NGC\,777    & -- & 319.3$\pm$1.8 & -- \\ 
    NGC\,1060    & 294.9$\pm$0.9 & 16.3$\pm$1.4  & 278.6\\
    NGC\,1587    & 147.4$\pm$0.9 & 231.6$\pm$1.3  & 84.2\\
    NGC\,4008    & 354.6$\pm$0.9 & 168.2$\pm$1.4  & 186.4\\
    NGC\,5846    & 129.3$\pm$0.5  & 306.6$\pm$89.5  & 177.3 \\
    NGC\,7619    & 168.0$\pm$18.7  & 217.1$\pm$0.9  & 49.1 \\
    {\it Disks}    &  &  & \\
    NGC\,924    & 41.6$\pm$0.5  & 52.5$\pm$0.9  & 10.9\\
    NGC\,940    & 198.1$\pm$0.5  & 198.1$\pm$0.5  & 0.0\\
    NGC\,978    & 72.4$\pm$0.5  & 72.4$\pm$0.5  & 0.0\\
    NGC\,1453   & 313.0$\pm$0.5 & 25.3$\pm$1.8  & 287.7\\
    NGC\,4169   & 143.8$\pm$0.5 & 332.9$\pm$0.9 & 189.1\\
    NGC\,4261   & 340.1$\pm$5.0 & 56.1$\pm$1.4  & 284.0\\
    ESO\,507-25 & 163.7$\pm$3.2 & 261.4$\pm$3.2  & 97.7\\
    \noalign{\smallskip} \hline \noalign{\smallskip}
\end{tabular}
\tablefoot{\raggedright The kinetic PA is measured anti-clockwise from the North direction to the {maximum positive velocity}. The errors on the kinetic PA are 3$\sigma$ errors.\\
\raggedright $\rm \Delta PA$: Difference between the stellar and gas kinematic position angles, $\rm \Delta PA = | PA_{\rm stellar} - PA_{\rm gas} |$\\
}
\label{tab:pa_fit}
\end{table}

When comparing the stellar and gas kinematics of the filamentary sources, it is clear that the gas velocity structure is distinct from the ordered motions usually seen in the stellar optical component. The latter suggests that the ionized gas in the filamentary sources is kinematically distinct and decoupled from the BGG stellar components, implying an external origin for the gas. Similar results have been found in galaxy clusters \citep{hamer16}.

The situation is more complex for the rotating gaseous disks, as its stellar counterpart is also rotating. For these sources, we model the velocity fields of the stellar and gas component to determine the kinetic position angles (PAs), using the \texttt{fit$\_$kinematic$\_$pa} routine described in \citet{krajnovic06}. We define the degree of misalignment between the rotation of the stars and that of gas by taking the difference of both position angles, $\rm \Delta PA = | PA_{\rm stellar} - PA_{\rm gas} |$. Following other studies in the literature, (e.g., \citealt{lagos15, bryant19}, we define an object as being misaligned if this difference is higher than $\Delta$PA$>$30$\degr$, and aligned if the $\Delta$PA$<$30$\degr$. We also attempt to fit the kinematic PA for the filamentary sources even when the kinematics of the filamentary, following the same procedure. The PA values for the gas and stellar components of each source can be found in Table~\ref{tab:pa_fit}. Figure~\ref{fig:PA_comparison} we show the kinetic PA angle of the stars and gas for the disks (orange left-pointing triangles) and for some filamentary sources (blue stars) where the fitting of a rotating structure is possible. The kinetic PA of gas for the filamentary sources should, therefore, be taken with care. As previously mentioned, a few systems do not show clear net stellar rotation making it difficult to define the $PA_{\rm stellar}$, leading to large uncertainties.

Three sources with rotating gas components have kinematics aligned with the stars (see NGC\,940,NGC\,978, and NGC\,924). The ionized gas disks might be coupled to their stellar component, indicating the gas is relaxed in the galaxy potential. The latter can be seen in NGC\,940, NGC\,978, where the ionized gas and stellar disk have the same PA, with PA values of about 198.1$\degr$, 72.4$\degr$, respectively. NGC\,924 has a $\rm \Delta PA$ of $\sim$11$\degr$, indicating that this disk may be tilted.

Some disks clearly show velocity structures of the gas that are kinematically decoupled from the stellar components. For example, in ESO\,507-25, the PA of the stars is 261$\degr$, while the PA of gas distribution is 163$\degr$. Similarly, in NGC\,1453, the PA of the ionized gas disk (PA$=$313$\degr$) is very different from the PA of the stars (PA=25$\degr$). In NGC\,4261, where the stellar component shows prolate rotation, we also see that the gaseous disk (PA$=$340$\degr$) is misaligned with the stars (PA$=$56$\degr$).

Note also that in NGC\,4169, the gas (PA$=$143$\degr$) is counter-rotating with respect to the stars (PA$=$332.9$\degr$),{ with a $\rm \Delta PA$ of $\sim$189$\degr$.} In summary, only 3/7 of the rotating gaseous disks are aligned with $\rm \Delta PA \sim$0$\degr$, whereas four disks are dynamically decoupled with $\rm \Delta PA >$20$\degr$. The aforementioned suggests an external origin of the gas for some of the rotating disk-like and most of the filamentary sources.

\subsubsection{Velocity dispersion}
\begin{figure}
    \centering
    \includegraphics[width=0.99\columnwidth]{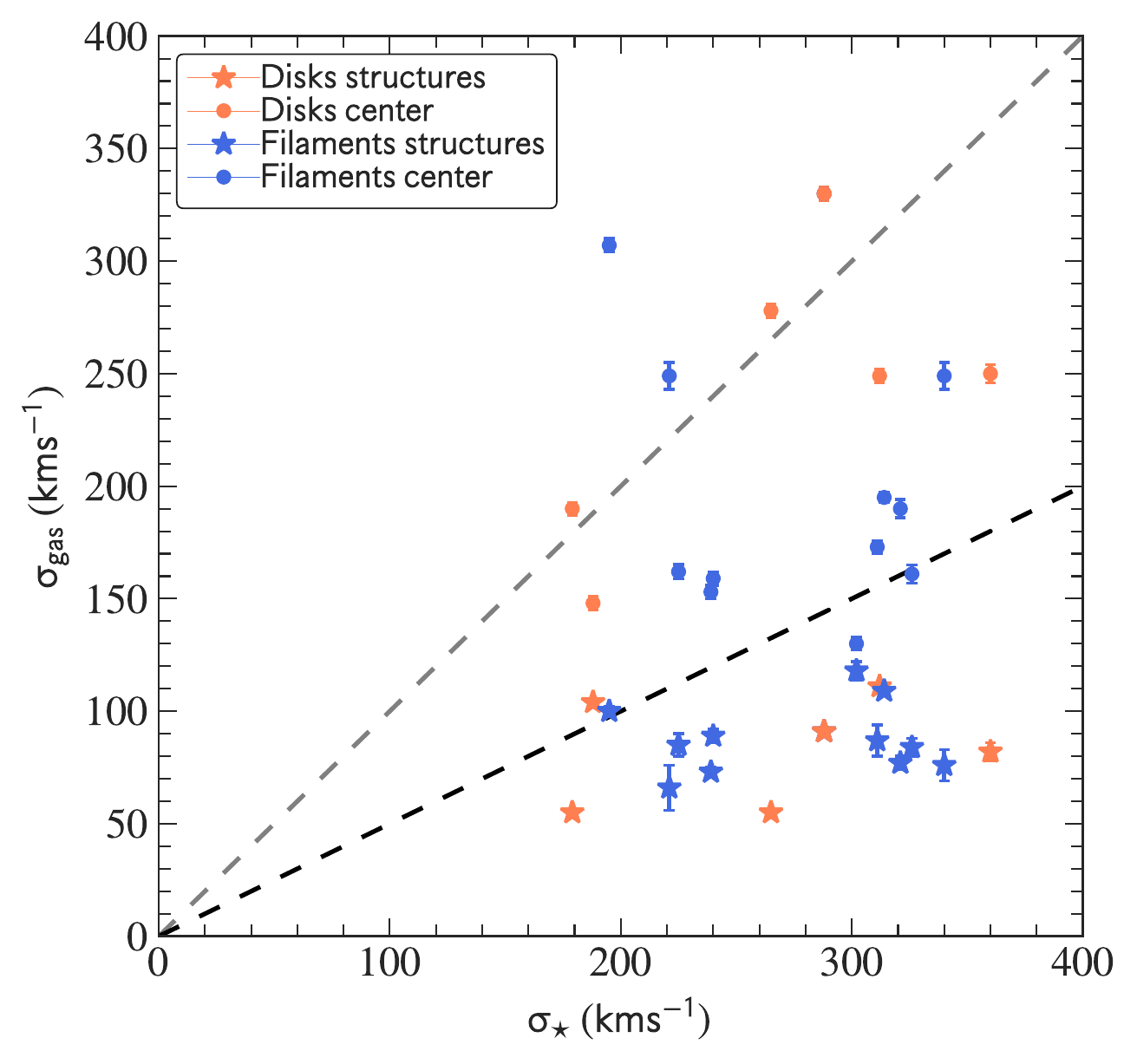}
    \caption{The stellar velocity dispersion of the central group galaxy versus the optical emission line widths from the MUSE sample. The central velocity dispersion of the ionized gas is shown with circles, while the different structures, such as filaments, clumps, rings, and off-central regions, are shown with stars. The gray dashed line represents the one-to-one relation, while the black {dashed} line corresponds to half of the stellar velocity dispersion. Filaments and rotating disks are color-coded with blue and orange color symbols, respectively.}
    \label{fig:visp_gas_stars}
\end{figure}
In Figure~\ref{fig:visp_gas_stars}, we compare the velocity dispersion of the ionized gas ($\sigma_{gas}$) and the central stellar velocity dispersion ($\rm \sigma_{\star}$). We measured the gas velocity dispersion of the ionized gas from the central region (represented with circle symbols) and in different structures of the nebulae (e.g., filaments, clumps, and rings, displayed with star symbols). {A numerical average value of the gas dispersion was taken of each structure for a given galaxy.} We noticed that the dispersion of the structures ($\sigma_{gas}$, star symbols) are at least two times smaller than the central stellar velocity dispersion ($\rm \sigma_{\star}$). Whereas the dispersion of the structures at the centers ($\sigma_{gas}$, circle symbols) is, on average, smaller and sometimes equal to the central stellar dispersion. {However, sometimes for disk center (orange circles), the points appear to be sitting on the equal relation (gray dashed line), and the $\sigma_{gas}$ for the filament center (blue circles) a few times also exceed the central stellar dispersion.} The latter indicates that the gas is likely not dynamically relaxed in the gravitational potential of the galaxy, {but may also indicate modulation of the gas sigma around the relaxed value, likely by intermittent AGN outburst.}

{In the framework of hot-halo cooling, two related models -- warm AGN-driven outflows \citep{qiu20} and turbulence-driven raining \citep{Gaspari_2018} -- show a somewhat different scatter for the correlation between hot-warm gas velocity dispersions. In the former, the velocity dispersions of both phases are only intermittently correlated, with the hot gas typically displaying values in the 100--140~km~s$^{-1}$ range, except during peak activity.
In the latter, while still being variable via the AGN feedback loop, the cold and warm gas share a similar dispersion over most of time, thus showing a tighter correlation scatter. Turbulent-driven precipitation simulations by \citet{prasad18} also find similarly varying hot-phase velocity dispersions tied to the feedback loop (50--200~km~s$^{-1}$). 
Hitomi observations gave us already a sneak peek at the hot gas line-of-sight velocity dispersion in Perseus cluster core of $164\pm10$~km~s$^{-1}$ \citep{hitomi16}, which results to be fully consistent with that found by SITELLE H$\alpha$ detection (see Figure 1 in \citealt{Gaspari_2018}), thus favoring turbulent raining, at least in Perseus cluster.
Analogous dispersion values have been found at the center of our filamentary systems (see Fig.~\ref{fig:visp_gas_stars}, blue circles). However, future high-resolution X-ray IFU observations such as Lynx and Athena X-ray satellites are required to build the hot-gas sample and to compare with our MUSE observations of cool gas. Indeed, besides Perseus core, only upper limits are available on the hot-gas velocity dispersion; e.g., even \textit{XMM-Newton} observations of the NGC\,5044 BGG (using the Reflection Grating Spectrometer) provide wide upper limits of 320--720~km~s$^{-1}$ via the FeXVII line broadening.}

\subsection{AGN X-ray bubbles, Radio Emission versus Ionized Gas Comparison}
\label{sec:radio_gas}

\begin{figure}[htbp!]
    \centering
    \includegraphics[width=0.48\textwidth]{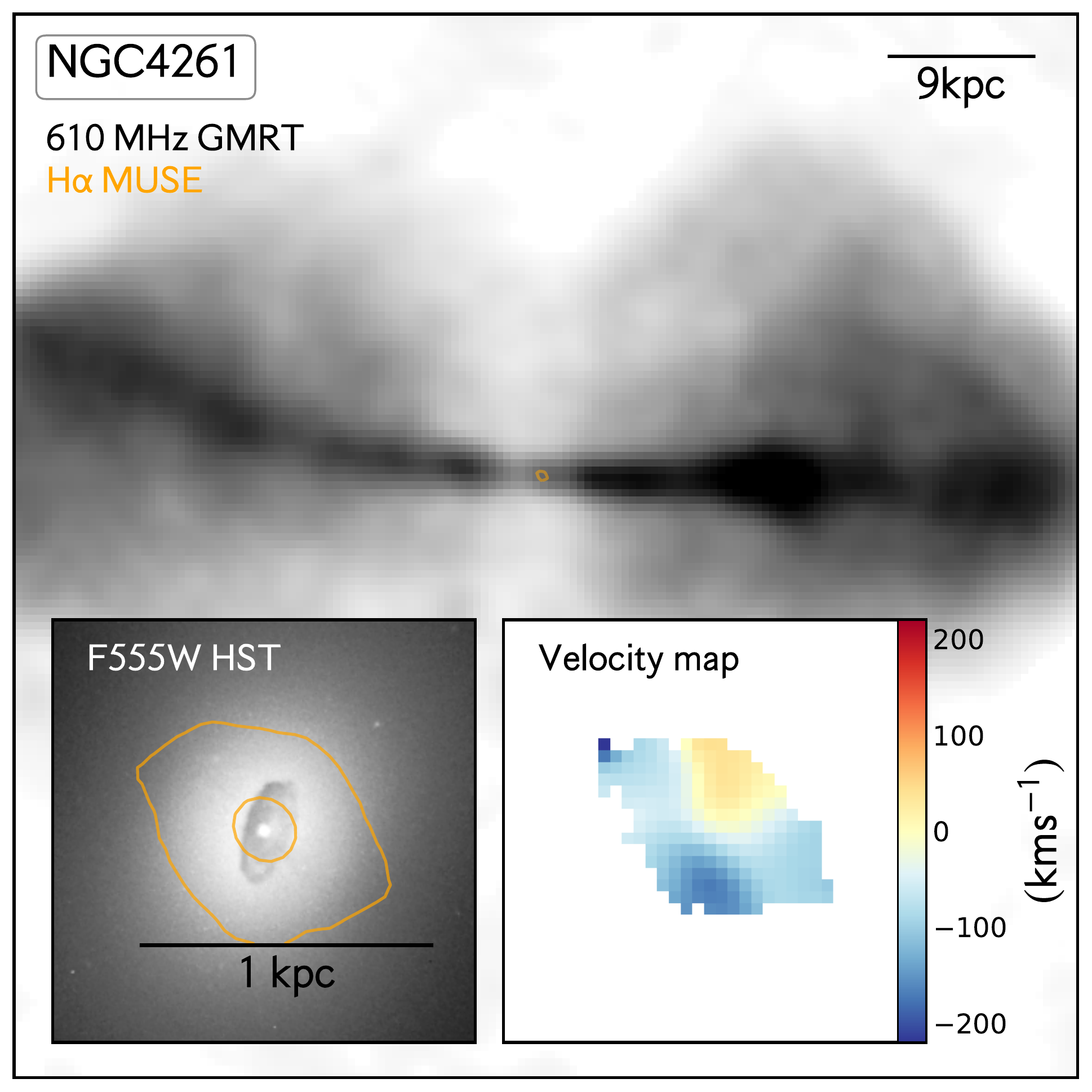}
    \caption{610 MHz GMRT image of NGC\,4261 \citep{kolokythas15}, overlaid with optical emission-line gas contours in orange. Bottom-left panel:
    Hubble Space Telescope (HST) F555W image of NGC\,4261 overlaid with H$\alpha$ emission in orange. Bottom-right panel: velocity map of the ionized gas.}
    \label{fig:ngc4261_hst}
\end{figure}

\def\setwidthsmall{0.3}
\begin{figure*}[htbp!]
        \vspace{-0.2cm}
        \centering
        \subfigure{\includegraphics[width=\setwidthsmall\textwidth]{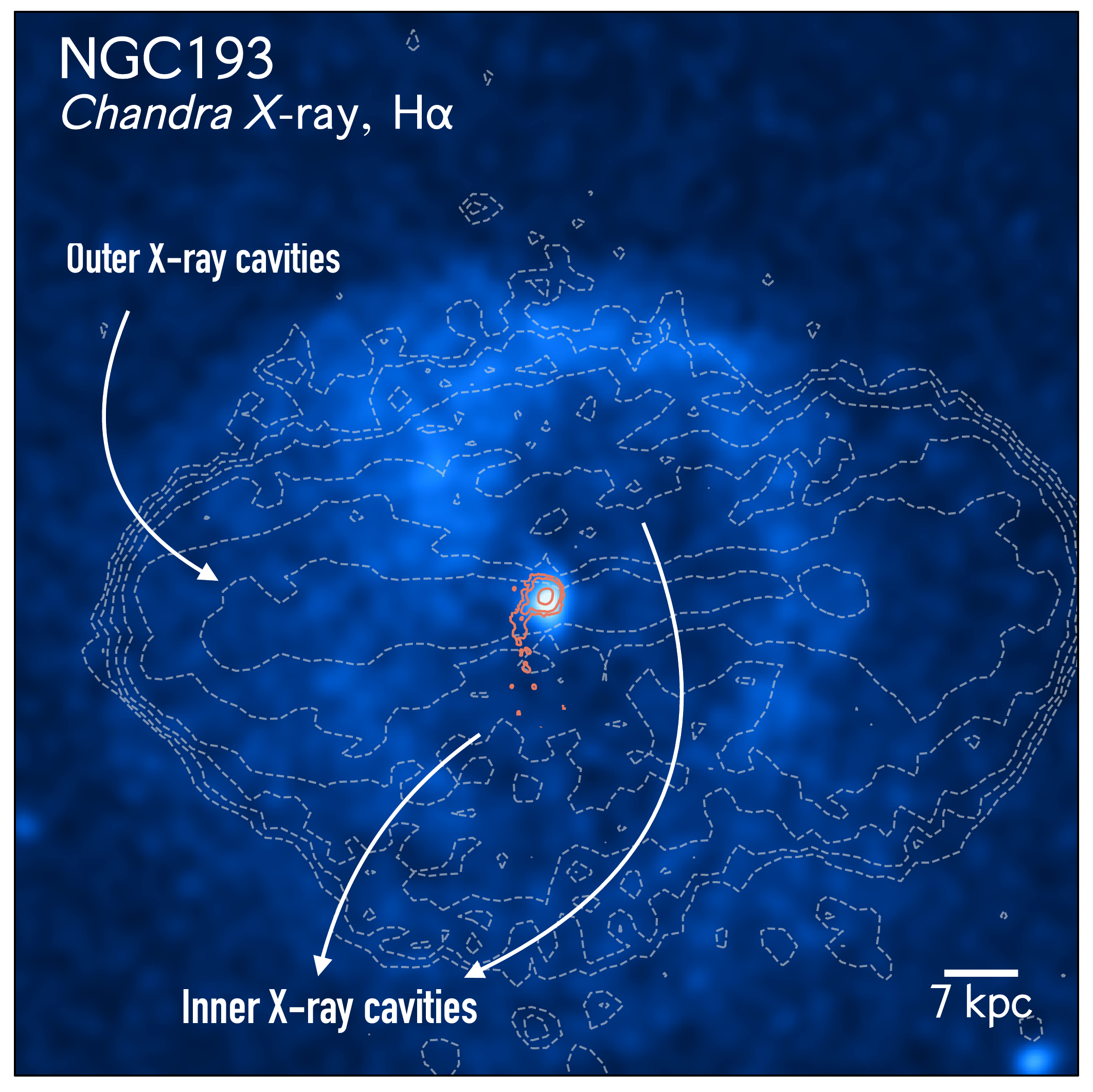}}
        \subfigure{\includegraphics[width=\setwidthsmall\textwidth]{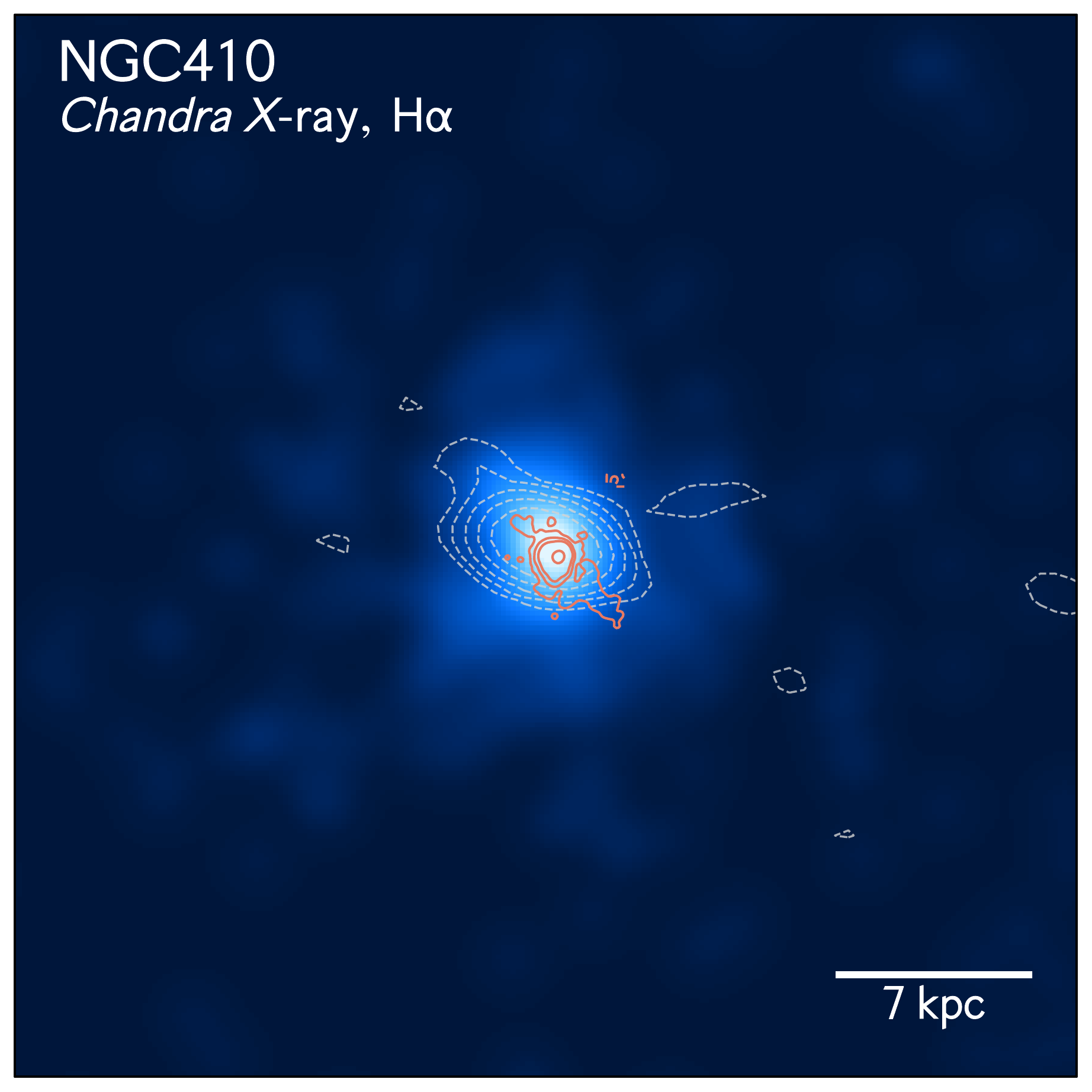}}
        \subfigure{\includegraphics[width=\setwidthsmall\textwidth]{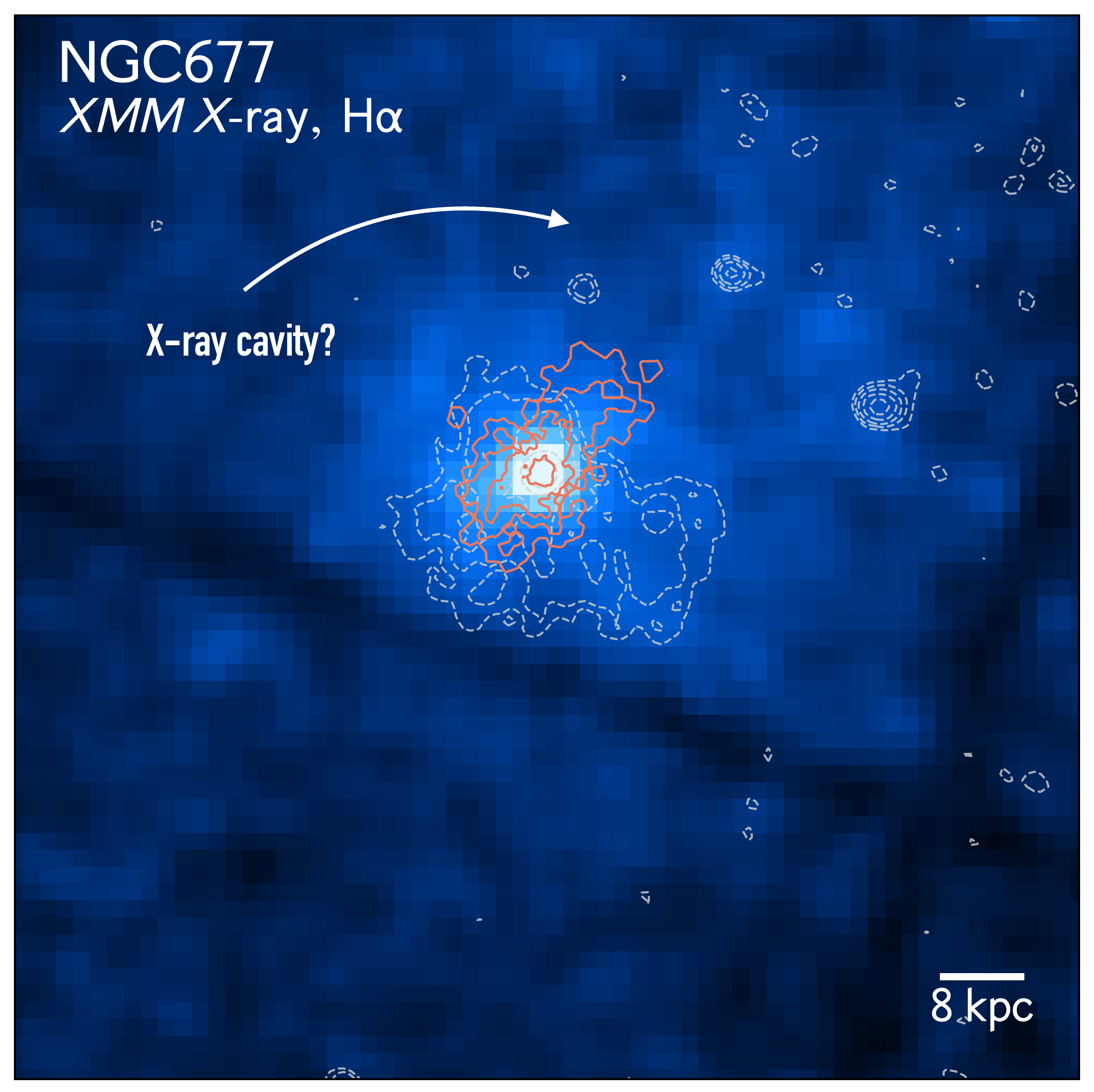}}\\
        \vspace{-0.4cm}
        \subfigure{\includegraphics[width=\setwidthsmall\textwidth]{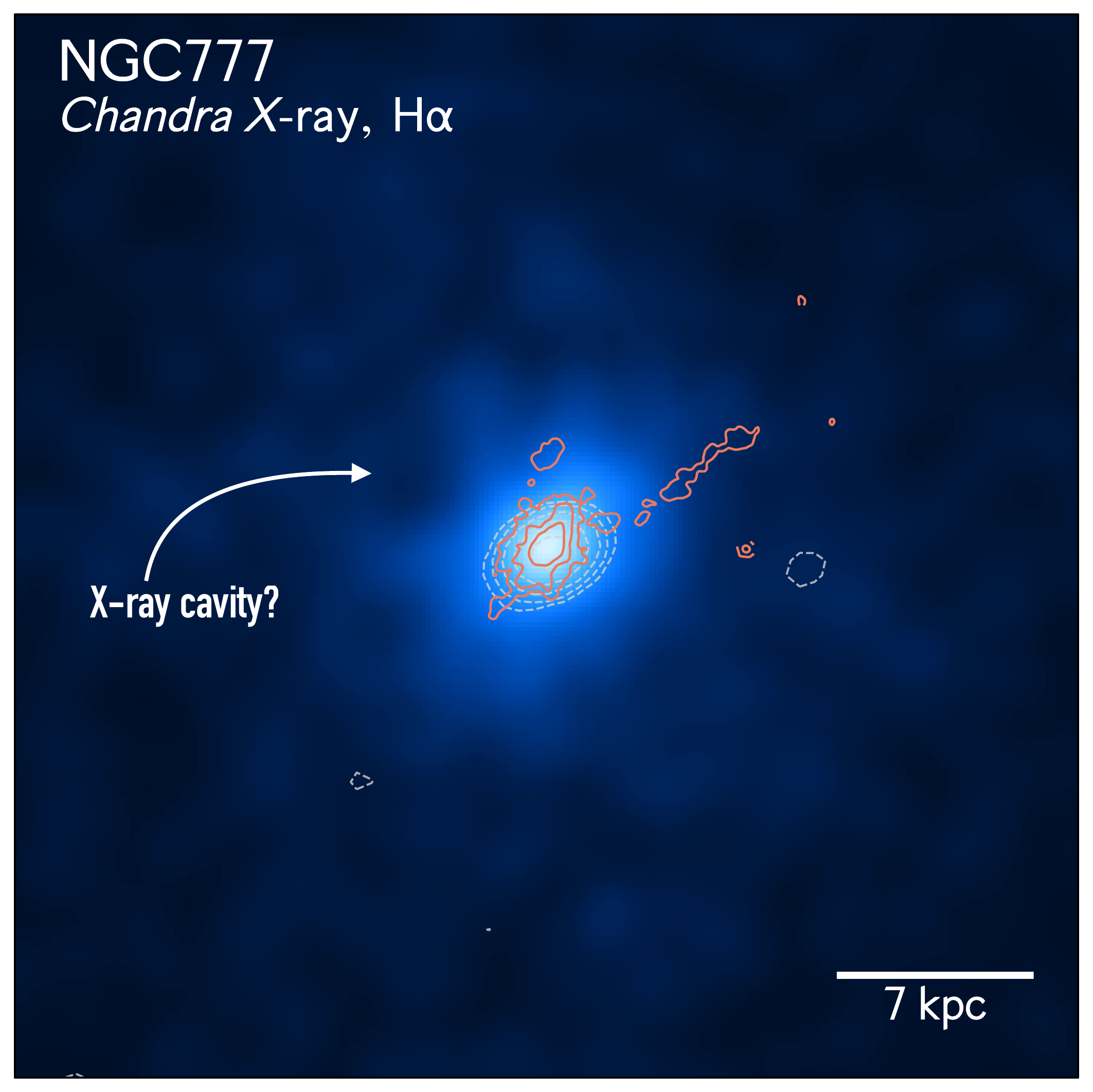}}
        \subfigure{\includegraphics[width=\setwidthsmall\textwidth]{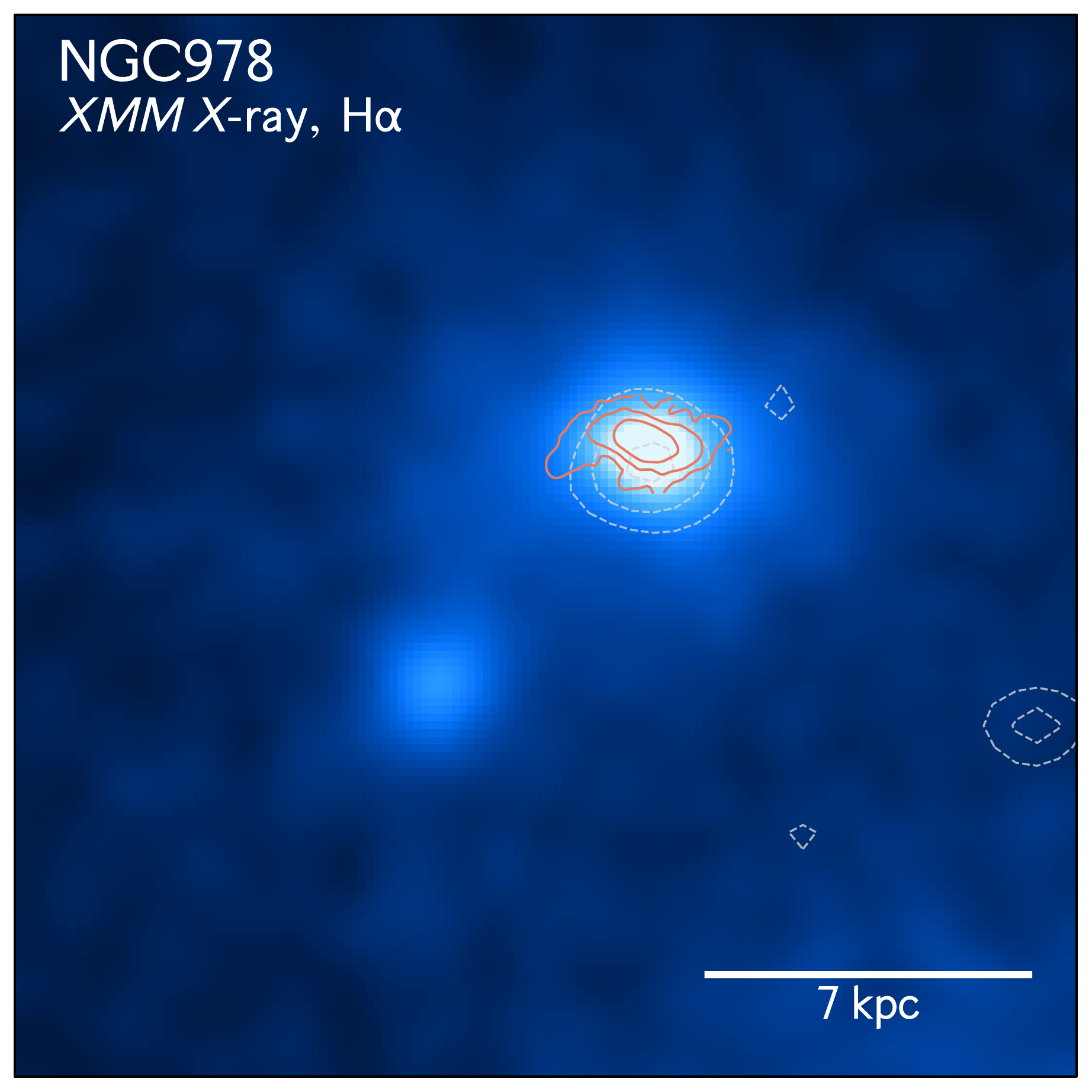}}
        \subfigure{\includegraphics[width=\setwidthsmall\textwidth]{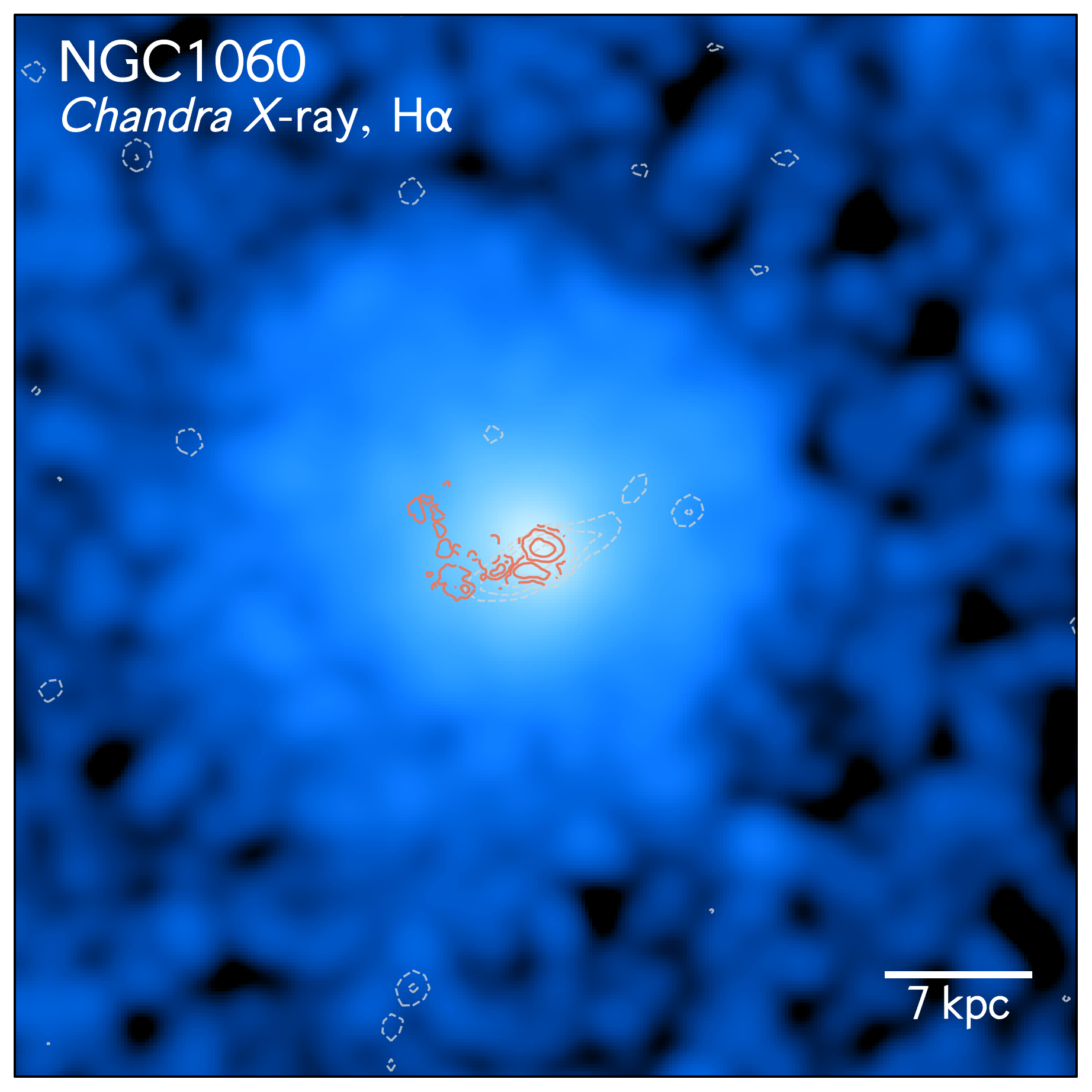}}\\
        \vspace{-0.4cm}
        \subfigure{\includegraphics[width=\setwidthsmall\textwidth]{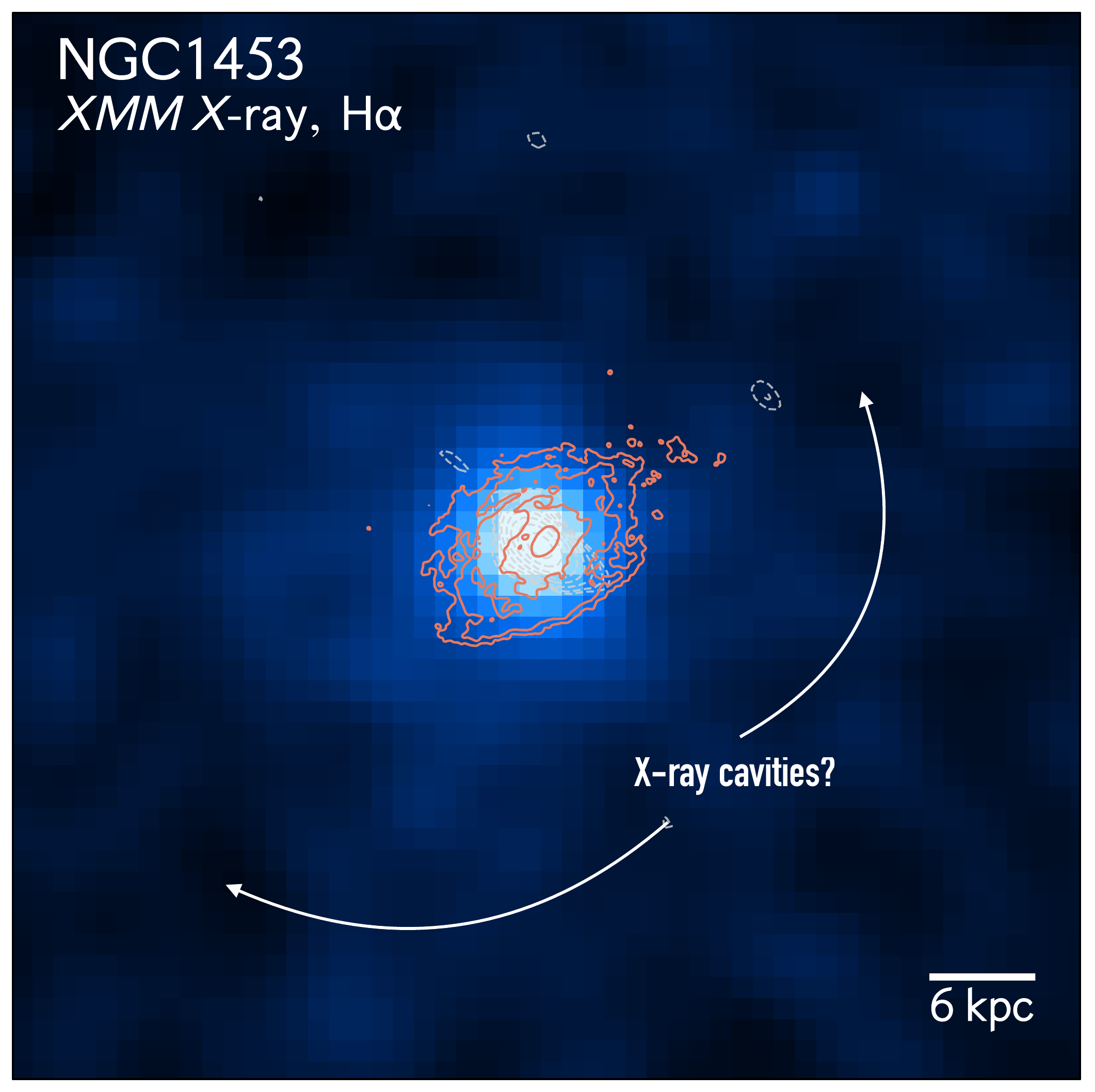}}
        \subfigure{\includegraphics[width=\setwidthsmall\textwidth]{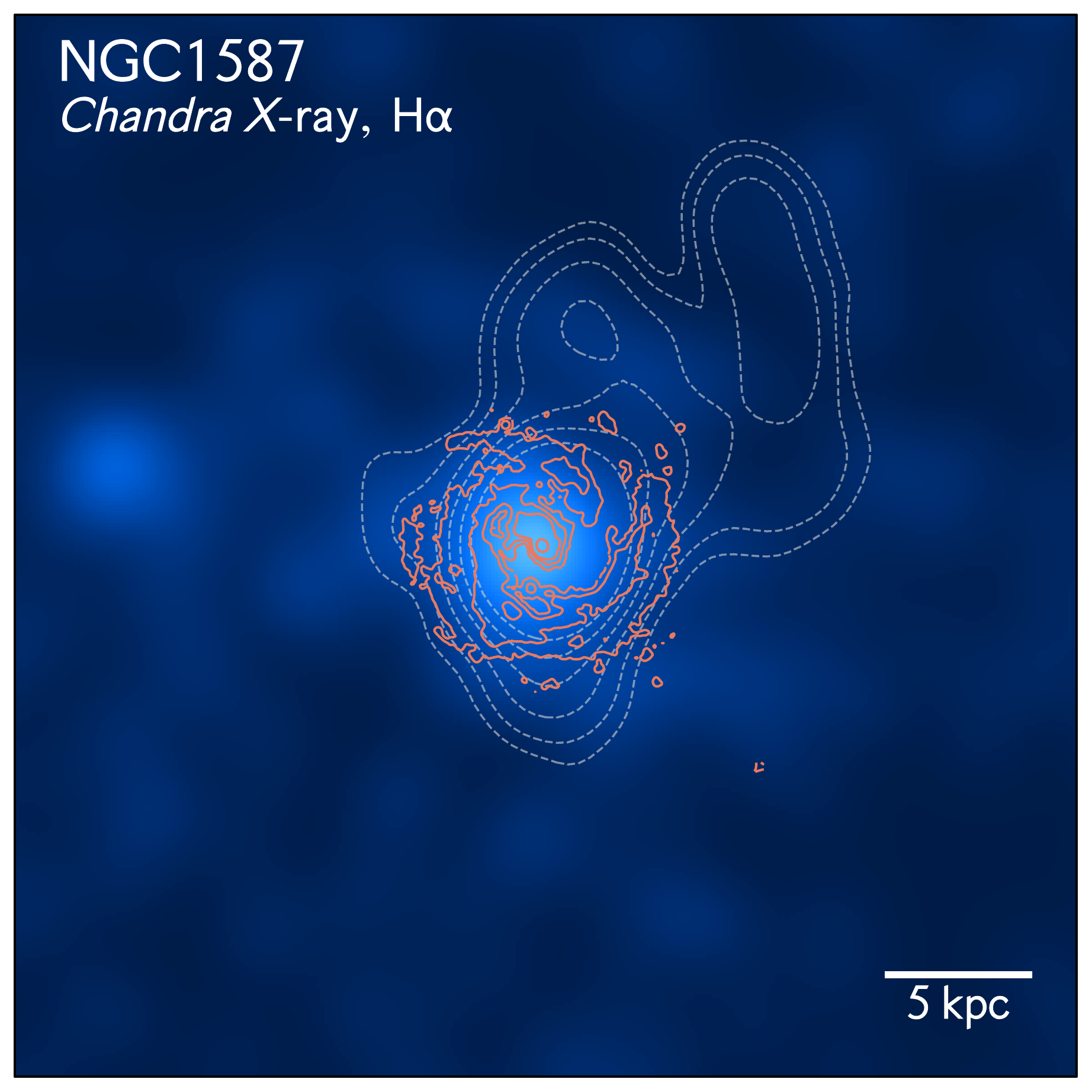}}
        \subfigure{\includegraphics[width=\setwidthsmall\textwidth]{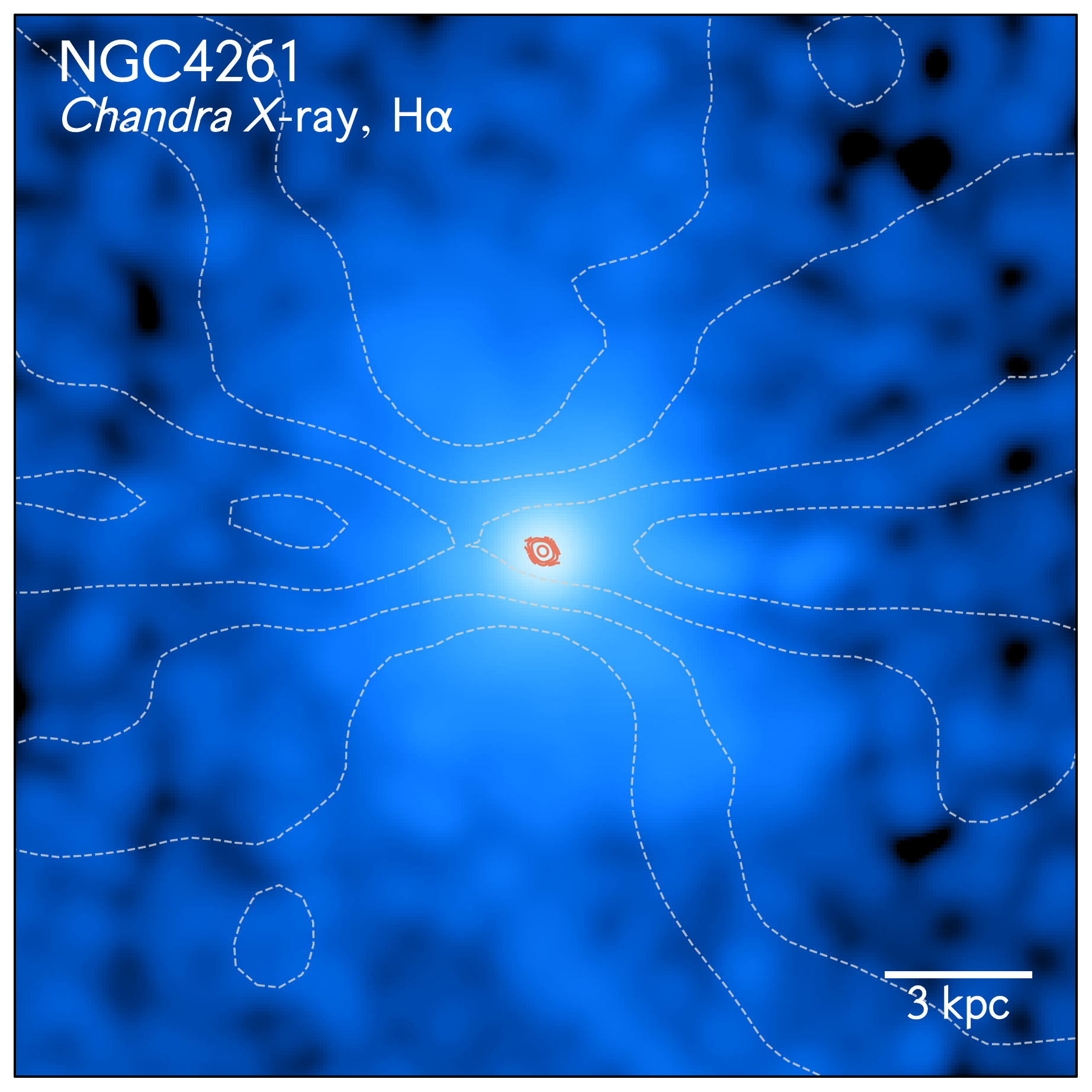}}\\
        \vspace{-0.4cm}
        \subfigure{\includegraphics[width=\setwidthsmall\textwidth]{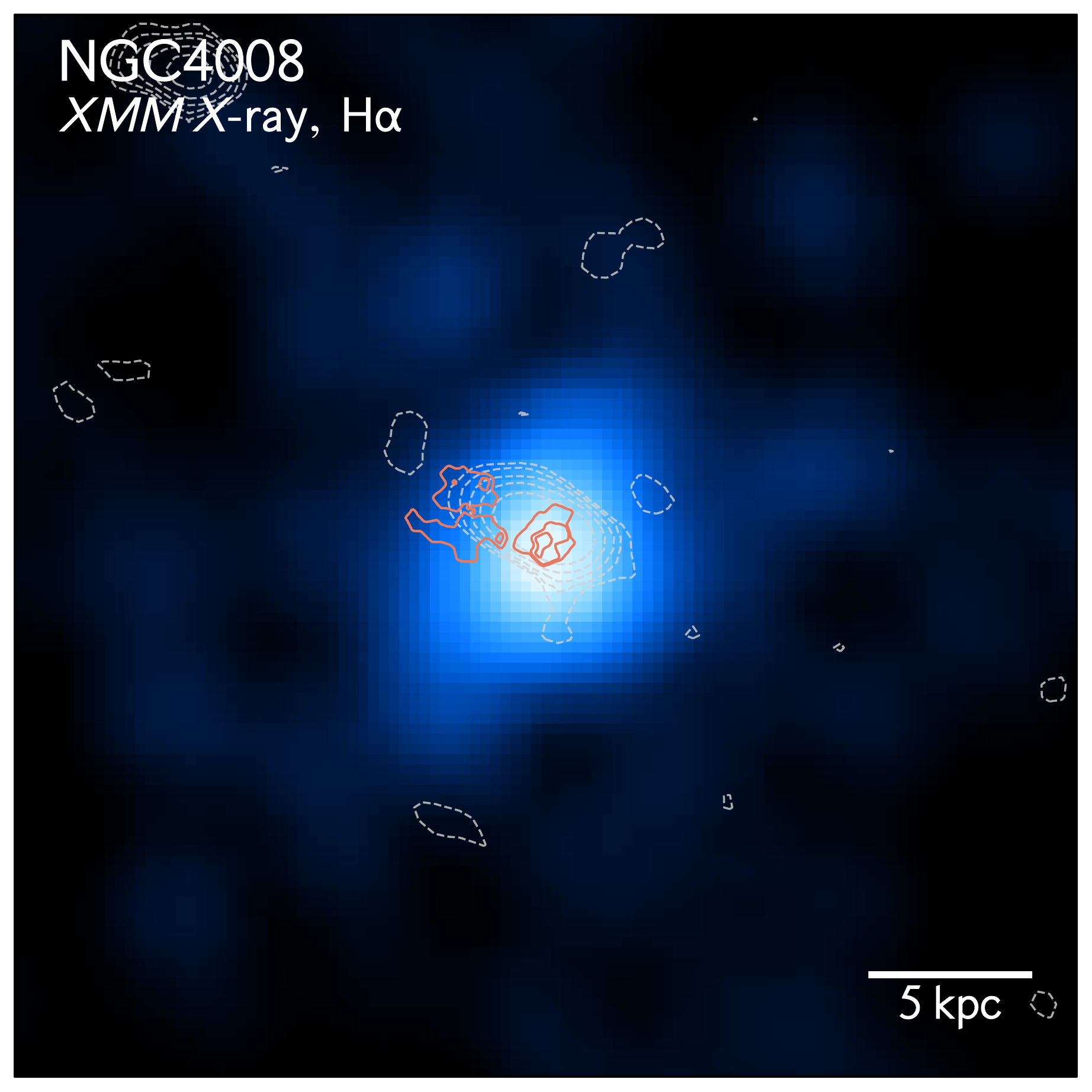}}
        \subfigure{\includegraphics[width=\setwidthsmall\textwidth]{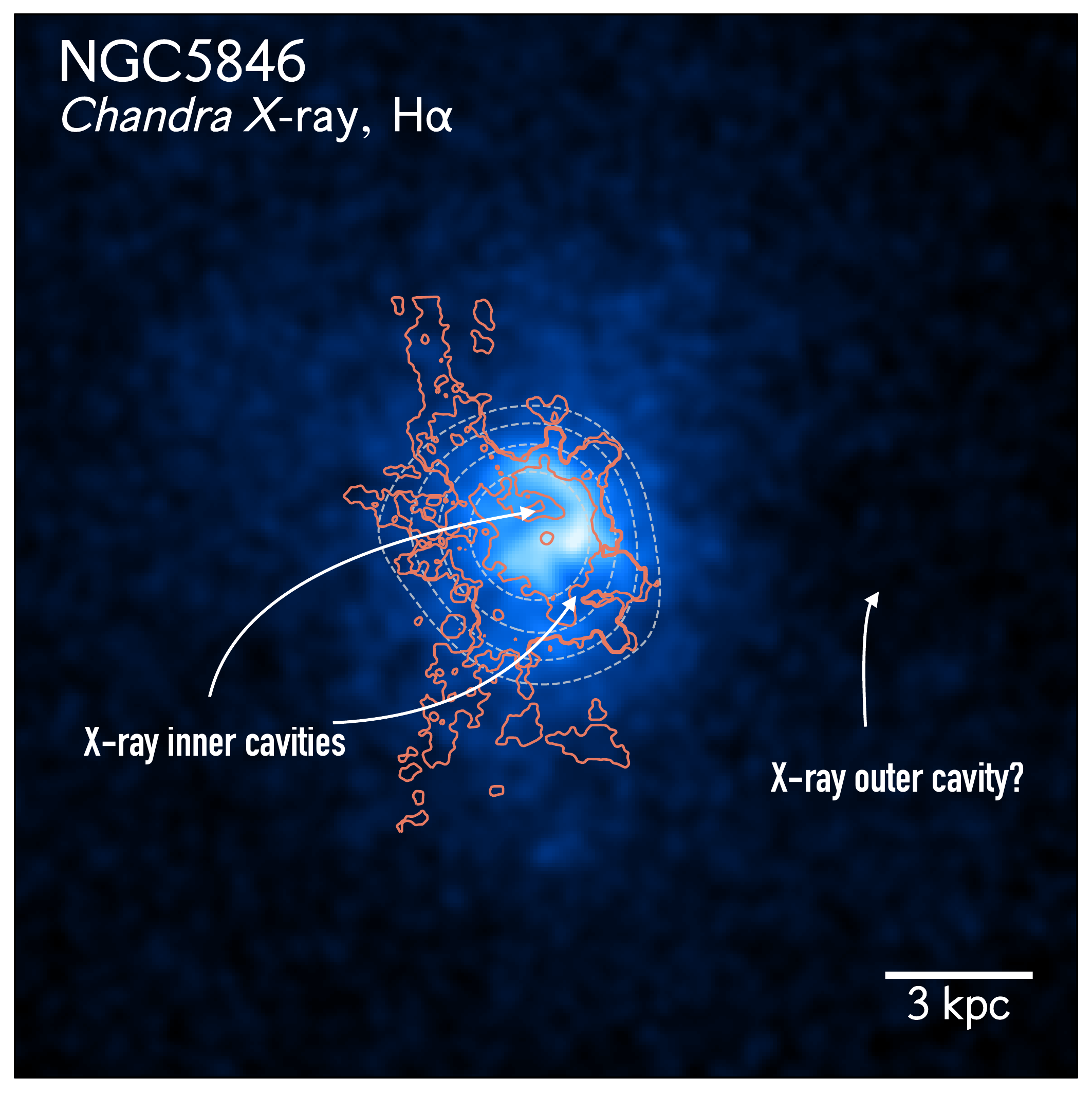}}
        \subfigure{\includegraphics[width=\setwidthsmall\textwidth]{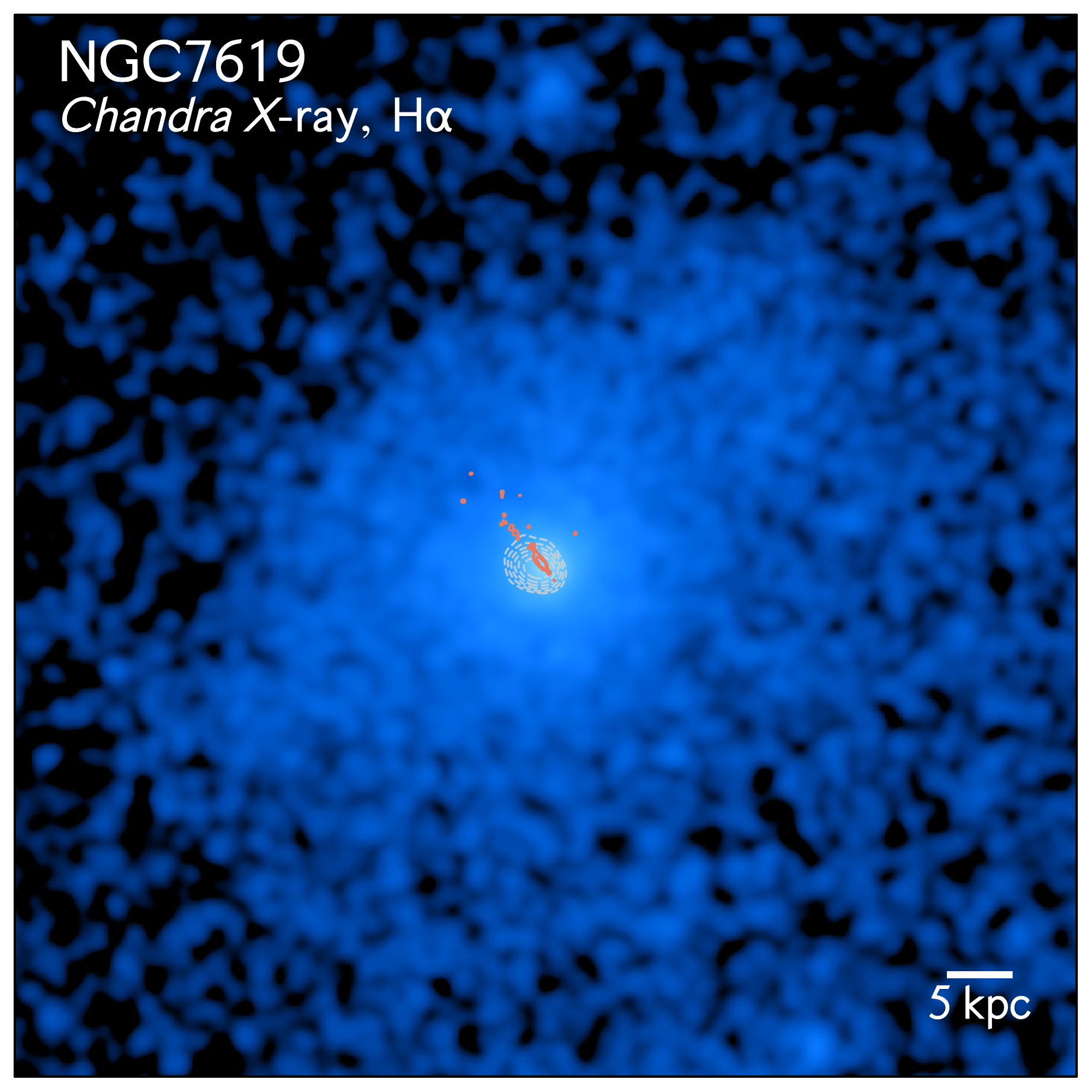}}
        \caption{
        X-ray \textit{Chandra} or \textit{XMM-Newton} images \citep{osullivan17} for the sources that have a detected IGrM. H$\alpha$ contours are overlaid in orange from our MUSE observations. Contours from 610 MHz GMRT radio emission are shown with dashed grey lines. Buoyant X-ray cavities are marked with white arrows. The X-ray bubbles for NGC\,5846 and NGC\,777 have been identified by \citet{panagoulia14b}, whereas for NGC\,193 by \citet{bogdan14}. Possible X-ray cavities are also shown for NGC\,677 and NGC\,1453. {North is up, east is left.} %X-ray features due to group-merger and -interaction are labeled for NGC\,1060 and NGC\,7619.
        }
        \label{fig:chandra}
\end{figure*}

This section examines the spatial distribution of the ionized gas in connection with the radio emission and IGrM features. Radio jets in central clusters and groups galaxies are deemed to heat the IGrM/ICM compensating the radiative losses maintaining the long-term balance (see \citealt{mcnamara06, fabian12}), plus potentially induce cold gas condensation via turbulence and uplifting of the low entropy gas \citep[e.g.][]{Revaz_2008,pope10,mcnamara16,Gaspari_2018}.
The 235 and 610 MHz GMRT radio data used was drawn from \citet{kolokythas18}, while for NGC\,4261 was drawn from the study of \citet{kolokythas15}, NGC\,1587 and NGC\,193 from \citet{giacintucci11}. \citet{kolokythas18} reported large- ($>$20~kpc) and small-scales ($<$20~kpc) radio jets in 4 of our systems, diffuse radio emission in 3 sources, whereas ten systems have a point-like radio emission ($\leq$11~kpc) (see table~\ref{tab:prop_multiwave} for details of each source). In the left-top panel of Figures~\ref{fig:nii_maps_examples} and \ref{fig:nii_maps_examples_app}, we show the ionized gas maps overlaid with contours from the GMRT 610~MHz radio emission in green. Whilst Figure~\ref{fig:chandra} displays the X-ray emission from \textit{Chandra} or \textit{XMM-Newton} observations for the sources that have a detected X-ray halo, overlaid with contours from the ionized gas (red contours) and 610~MHz radio emission (dotted gray contours).

In NGC\,193, the SW set of filaments are located in projection behind the rim of one of the inner Southern X-ray cavities excavated by the large-scale (80~kpc, \cite{giacintucci11}) radio jets (see Fig.~\ref{fig:chandra}). Noteworthy, any trace of ionized gas is found neither at the Northern inner bubble or the outer bubbles. According to \citet{bogdan14}, the inner cavities correspond to a weaker old ($\sim$ 70~Myr) AGN outburst, while the outer cavities to a strong younger ($\sim$ 10--20~Myr) outburst. Based on the kinematics of the ionized gas and the spatial correspondence with the inner younger cavity, it is plausible that the SW filaments formed in a previous AGN outburst, and the cooled gas is now inflowing towards the galaxy center. 

NGC\,677 hosts a highly diffuse radio emission, but it does not seem to be aligned with the optical-line emitting gas. On the other hand, X-ray \textit{XMM-Newton} observations show X-ray surface brightness depression, we interpreted as a possible cavity at the NW of the group's center (see Fig.~\ref{fig:chandra}). Note that the data quality is not the best to make a reliable identification. The NW set of optical filaments found in NGC\,677 (red contours) are located in projection behind this putative X-ray cavity. 

The shallow X-ray \textit{Chandra} observations of NGC\,777 reveal the presence of a potential X-ray cavity on the hot atmosphere detected by \citet{panagoulia14b}, located in projection near the NE H$\alpha$ filament (see Fig.~\ref{fig:chandra}). The 4.85~GHz VLA and 610~MHz GMRT observations show no indication of jets \citep{kolokythas18}. However, \citet{ho01} found a slightly resolved radio emission at 5 GHz, which appears to be aligned with the direction of the SE-NW ionized filaments.

NGC\,1060 hosts a small-scale jet, although the low spatial resolution of \textit{XMM-Newton} observations are insufficient to resolve any X-ray cavities at the scale of the jets \citep{kolokythas18}. Nevertheless, the distribution of the SE filament seems to be spatially linked to one of the radio lobes, raising the possibility that AGN outburst could have triggered the formation of the cold gas and produced the smooth velocity gradient in the SE clumpy filament (see Fig.~\ref{fig:nii_maps_examples_app}). 

From visual inspection, we found the \textit{XMM-Newton} X-ray image of NGC\,1453 shows hints of two depressions along the NW-SE axis indicating the presence of possible X-ray cavities (see Fig.~\ref{fig:chandra}). {The two depressions are chosen out of the others for being the biggest; however, deeper X-ray observations are needed to confirm the presence of any cavity.} These two depressions are aligned with the major axis of the ionized gas disk. Therefore, if these are cavities, the redshifted NW and SE blueshifted structures coming out of the disk could be related to these possible X-ray depressions. Note that NGC\,1453 hosts a radio point emission with no evidence of jets \citep{kolokythas18}.

The 610~MHz radio observations reveal that NGC\,1587 hosts a central radio source surrounded by a diffuse emission extending out to the NW, whose nature is unclear \citep{giacintucci11,kolokythas18}. Consistently, the 1.4~GHz VLA observations show a small-scale radio emission (1~kpc, 4$\arcsec$) to the NW, hinting an interaction with the ionized filaments (see Fig.~\ref{fig:nii_maps_examples}). 

\begin{figure*}[htb!]
    \includegraphics[width=0.5\textwidth]{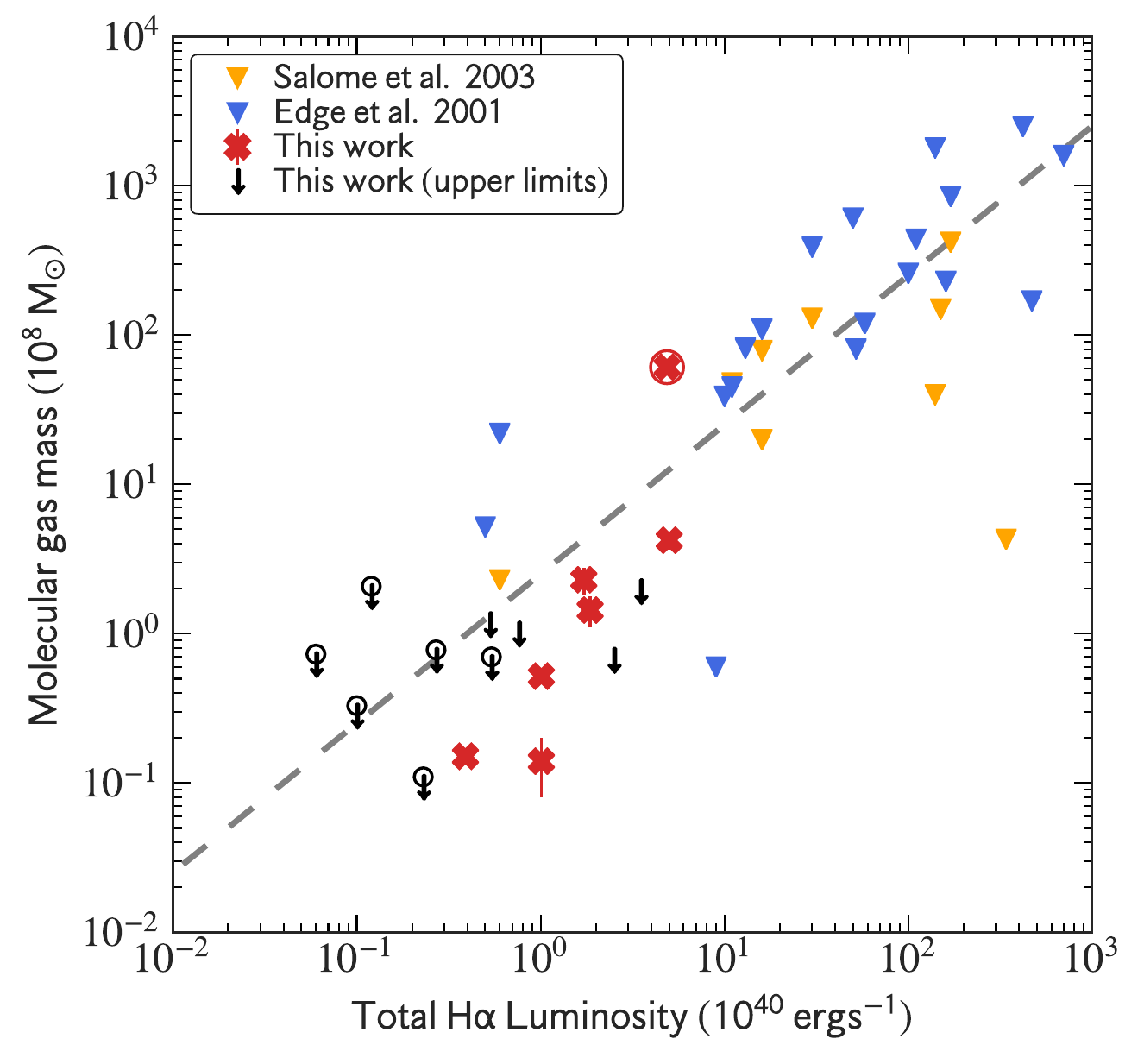}
    \includegraphics[width=0.5\textwidth]{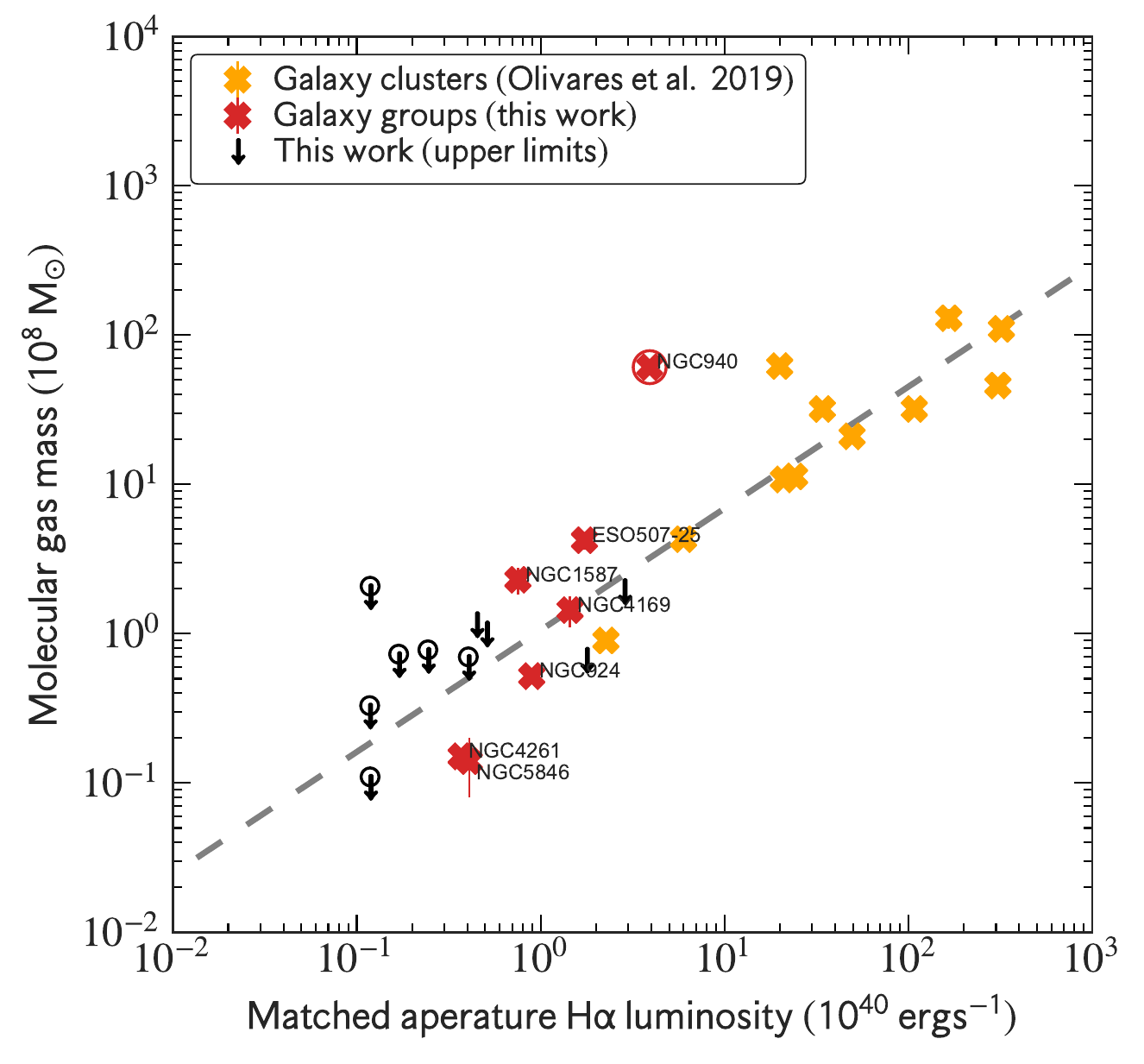}
    \caption{Left panel: Molecular mass versus H$\alpha$ luminosity. The molecular masses were taken from either IRAM 30m or APEX telescope observations \citep{osullivan15,osullivan18b}. %The molecular gas was detected in the CC and NCC groups.
    We compare this relation with higher mass systems (BCGs). Observations from \citet{edge01} and \citet{salome03} are plotted with inverted blue and yellow triangles, respectively. Upper limits from previous works are not included in this plot. A clear correlation between H$\alpha$ luminosity and molecular gas masses is seen, in agreement with the previous detection from BCGs. This correlation is consistent with the top-down multiphase condensation and CCA rain. The gray dashed line corresponds to a linear fit considering only BCGs. Right Panel: Molecular mass versus H$\alpha$ luminosity extracted from the same aperture as the CO measurements. The dashed gray line corresponds to the linear fit using measurements from BCGs. The CO molecular gas masses for BCGs are taken from \cite{olivares19} using ALMA observations, whereas H$\alpha$ luminosities are measured using MUSE observations. {BGGs showing H$\alpha$ in absorption at the galaxy center are marked with circles}.}
    \label{fig:H2_Ha_corr}
\end{figure*}

In NGC\,4261, the large-scale radio jets are being ejected by the SMBH nearly perpendicular to the plane of the ionized gas disk (see Fig.~\ref{fig:ngc4261_hst}) excavating cavities on the IGrM \citep{osullivan11}, similar to the BCG in Hydra-A \citep{hamer14,rose19}. Long-lived disks revealed by simulations are shown to play a crucial role in driving strong jets perpendicular to the plane of the disk \citep{beckmann19, li15}. The little amount of raw material and compact gas distribution is potentially a consequence of a disbalance between the jet power and cooling, where the former is about two orders of magnitude greater than the radiative losses \citet{kolokythas15,kolokythas18}.

The ionized gas in NGC\,5846 is distributed along the rims of the inner X-ray cavities that have been detected within the inner 2~kpc of the IGrM filled out with radio emission (\citet{dong10}, see also Fig.~\ref{fig:chandra}). The small-scale radio jets might entrain the North set of filaments, distributed in an ''arm-shape`` (see Fig.~\ref{fig:nii_maps_examples_app}), while the South network of filaments is spatially coincident with the radio lobe. The extended NE filaments are also preferentially located along the radio lobes. There is also evidence of a larger outer cavity located west of the central galaxy \citep{machacek11}, but no cold gas is found close to this cavity.

In NGC\,7619, the 610~MHz GMRT display an unresolved radio emission, and no X-ray bubbles have been identified for this interacting system \citep{dong10}. 

Note that spatial coincidence between the radio jets/emission and the warm ionized gas does not necessarily indicate a direct uplift of cold gas. The radio jets could also produce strong turbulence and compression on the hot gas, triggering non-linear thermal instabilities. We also recall that the putative X-ray cavities mentioned in this paper are only potential as deeper X-ray observations are needed to confirm those. NGC\,6658, the source without optical emission lines, also lacks radio emission at any radio frequencies observed with the GMRT telescope. The rest of the sources have an unresolved point-like radio distribution. As well as a diffuse (ESO\,507-25) radio emission with no clear association with the ionized gas. We recall that feeding can generate compact radio synchrotron emission close to the inner gravitational region, even without active feedback in action.

\subsection{Is the Ionized Gas linked to the Molecular Gas?}
The sources of our sample have been observed with the single-dish IRAM 30m or APEX telescope \citep{osullivan15, osullivan18b}, but CO(1-0) or CO(2-1) emission lines were only detected in six of the galaxies (see Table~\ref{tab:prop_multiwave} for the details of each source). The sources with detected CO emission lines have molecular gas masses within a range from 0.14$\times$ $10^{8}$ up to 61$\times$ $10^{8}$~M$_{\odot}$. Half of the sources (NGC\,940, NGC\,924, and ESO\,507-25; \citealt{osullivan18b}) show double-horn profiles in agreement with the ionized gas distribution. In particular, two sources with filamentary structures, NGC\,5846 and NGC\,1587, have CO emission line detections. Molecular emission from filaments is much fainter and harder to detect, making it more challenging to detect molecular clumps in extended structures. Plus, there is likely a bias that favors the detection of molecular gas when it has settled into a massive rotating disk. Numerical simulations of CCA predict that in groups, the condensed gas mass should be about 2~dex lower than in clusters \citep{gaspari17}, making it more challenging to detect it, particularly in small clouds. Therefore, cooling in the less luminous X-ray galaxy groups should provide less condensation of cold gas, as they have less hot gas available. Another possibility has been pointed out by \citet{liang16}, where the gas could have been expelled by either SN or AGN-driven winds up to 3--4~R$_{\rm 200}$ at early times, and it has not been entirely re-accrete yet. Deeper observations are needed to detect the fainter molecular gas (if present). This is a current limitation that will be further explored with future ALMA and NOEMA observations. 

In Figure~\ref{fig:H2_Ha_corr} we compare the molecular gas mass with the H$\alpha$ luminosity. The left panel shows the overall correlation between molecular gas mass against the total H$\alpha$ luminosity for our MUSE sample. We have also included BCGs taken from the literature \citep{edge01,salome03,pulido18,olivares19}. In the right panel of Figure~\ref{fig:H2_Ha_corr}, the H$\alpha$ luminosity has been estimated from the same aperture as the CO measurements using ALMA and IRAM observations for the BCGs and BGGs, respectively. The CO aperture of each source is shown in the second panel of Figures~\ref{fig:nii_maps_examples} and ~\ref{fig:nii_maps_examples_app} with a white a cyan circle for upper limits and detections, respectively. As shown in Fig.~\ref{fig:H2_Ha_corr}, the scatter is reduced when both quantities are measured within the same aperture, preserving the tight correlation between these two gas phases, including less massive systems.

Previous studies of cool-core clusters using ALMA and MUSE observations in synergy have found that the molecular and warm ionized gas are co-spatial and co-moving, consistent with the hypothesis that H$\alpha$ emission comes from warm ionized envelopes of cooled molecular clouds \citep[e.g.,][]{tremblay18, olivares19}. The aforementioned suggests a common origin for these two gas phases likely via the top-down multiphase condensation cascade, and thus the cold and clumpy gas may have formed from the hot ICM or IGrM condensation through the CCA mechanism \citep{gaspari17}. The molecular-gas-mass-to-H$\alpha$-luminosity correlation which follows our sample supports this scenario (see Fig.~\ref{fig:H2_Ha_corr}). However, further kinematic and spatial information of the cold molecular gas is required to unveil whether these two phases are tightly connected as expected from this scenario \citep{Gaspari_2018}. In particular, the turbulence driven in the hot halo by recurrent AGN feedback \citep{Wittor20} is expected to percolate through the phases, with the ensemble warm filaments acting as the best kinematical tracers for the CCA condensation rain (see also Sec.~\ref{sec:origin}).

Our preliminary \textit{NOrthern Extended Millimeter Array} (NOEMA) observations of NGC\,940 reveal a cold molecular gas disk (Olivares et al. in prep), which appears to be co-spatial and co-moving with the optical emission lines in a rotating disk structure. Likewise, ALMA observations of NGC\,4261 show a cold molecular rotating disk \citep{boizelle21}, consistent with MUSE observations. Besides that, ALMA observations of NGC\,5846 show sub-kpcs molecular CO clouds of 10$^{5}$~M$_{\odot}$ associated with some of the H$\alpha$ filaments and dust structures unveiled by the HST dust extinction maps \citep{temi18}, comparable to the gas distribution observed in cool-core clusters. Qualitatively similar results have been reported in several local BGGs by \citet{werner14}, using [CII]$\lambda$157 emission line (a tracer of 100~K cold gas) and H$\alpha$+[NII] optical emission line, showing co-spatiality between the two temperature gas phases. Hence, molecular and ionized gas are either associated inside a disk or in clumps within the filaments. A careful study of the morphology of ionized gas excitation can help shed light on the origin (or evolutionary stage) of these different gaseous structures (HII regions vs shock or energetic particles heating).

\subsection{Origin of the Gas}\label{sec:origin}

The question of how galaxies acquired their gas is still a matter of debate. Several scenarios have been discussed for the origin of cold gas in elliptical galaxies over the past years. Those scenarios are the acquisition of cold gas from the cooling of gas ejected by the stellar population (stellar-mass loss), through mergers or interactions with gas-rich galaxies, or cooling from the hot atmosphere (cooling-flows). A mixture of these different processes is also, and likely possible. \citet{davis11} suggested that in elliptical galaxies, the gas produced by the stellar-mass loss should form a kiloparsec-scale rotating disk aligned with the stellar component, while the gas brought into the system through mergers is likely to be misaligned or to create multiple tails, rings, or disks. Numerical simulations of cool-core clusters and groups also found kiloparsec rotating disks of raw material at the center of the galaxy. The formation of the cold rotating disk may thus also occur via the condensation of the hot gas via thermal instabilities \citep[e.g.,][]{gaspari15, prasad18, Gaspari_2018, beckmann19} or merger events.

\subsubsection{Stellar-mass loss} 

%We explore the possibility that the gas could have originated through stellar mass loss. 
In most cases, there is a kinematics misalignment between the gas and the stars in discrepancy with the stellar-mass loss scenario, where a kinematic alignment between the gas and the bulk of the stellar population is expected due to angular momentum conservation \citep{davis11}. Only in three rotating-disks (NGC\,940, NGC\,978, and NGC\,924), the gas is kinematically aligned with the stellar component by at least 10$\deg$. Furthermore, the distribution of the gaseous disks differs from the stellar population, with some sources even displaying gaseous structures arising (e.g., NGC\,978) from the main disk and outer rings (e.g., NGC\,924), making it difficult to reconcile with this scenario. In particular, only in the lenticular galaxy NGC\,940, the distribution of both components appears to be alike.

\begin{figure}[htb!]
\centering
        \includegraphics[width=0.5\textwidth]{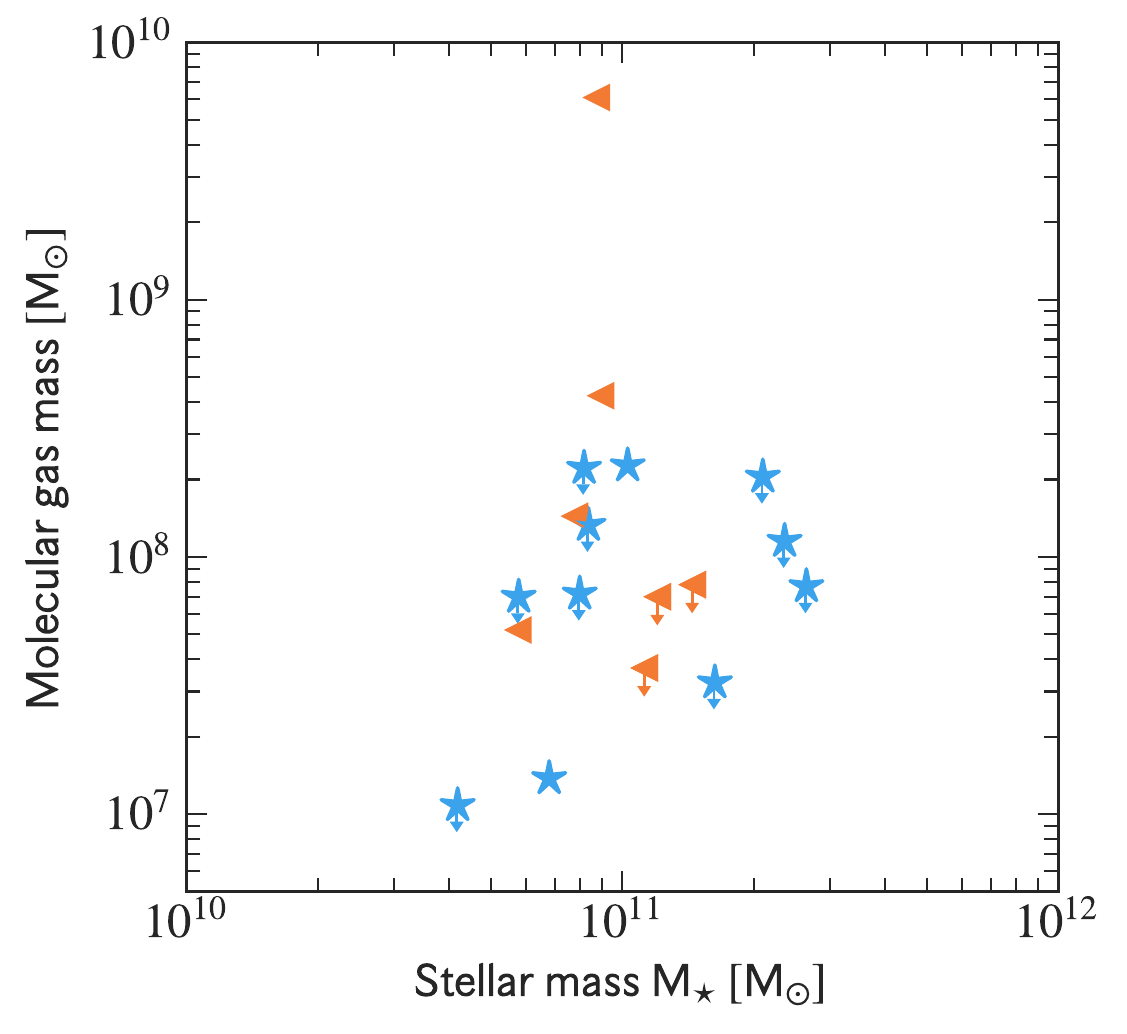}
        \caption{Molecular gas mass versus stellar mass. Filaments sources are represented with blue stars, whereas gaseous rotating disks are shown with orange left-pointing triangles. Upper limits are shown with an arrow. }\label{fig:Mcold_vs_Mstars}
\end{figure}

Most of the material produced by stellar mass loss is expected to be thermalized by shocks and will join the hot galaxy halo \citep{parriott-bregman08, bregman-parriott09}, though a fraction might contribute to the raw material settling onto a rotating disk. Most internal processes that can return material to the IGrM are related to the mass of the stars present in the galaxy, {and its star formation history. \citet{canning13} suggested that evolved stellar population is expected to provide about $\sim$1~M$_{\odot}$~yr$^{-1}$ of gas per 10$^{11}$~M$_{\odot}$, which corresponds to roughly the stellar mass of the sources in our sample. 
Considering that the cooling time is $\sim$10$^{8}$~yr, the mass rate is not constant, and only a fraction of expelled gas condenses into molecular gas (without producing star formation), the galaxies would take at least 1 Gyr to build such cold gas mass budget.}
It is also important to know whether the reservoir of gas correlates with stellar mass \citep{Young_2011, davis19}. As shown in Fig.~\ref{fig:Mcold_vs_Mstars}, we do not find any correlation between the amount of cold gas mass and stellar mass, indicating that the stellar mass loss is likely not the dominant source of cold gas and that the gas might originate through external processes. 

\subsubsection{Gas-rich mergers and interaction}
The distinct kinematic and spatial distribution between the stellar and the gas component may suggest another dominant mechanism for acquiring cold gas in our sample. In particular, we found that four gaseous rotating disk systems have misaligned stars and gas kinematics with $\Delta$PA$\sim$90 \degr (NGC\,1453, NGC\,4261, NGC\,4169, and ESO\,507-25), while three sources display extended rotating rings (NGC\,940, NGC\,924, and ESO\,507-25), in addition to the filamentary source, NGC\,584, whose kinematics show sign of rotation. The oxygen abundances distribution of the circumnuclear star-forming regions detected in the rotating rings show a chemically homogeneous ISM (Lagos et al. 2021, in prep.), with metallicity near Solar for these systems. This could indicate short time scales for the accretion process.

It is also worth mentioning that most of the extended rotating disks and NGC\,584 lack a detected IGrM and have a point-like radio source (NGC\,940, NGC\,924, and ESO\,507-25, NGC\,4169). The latter may be indicative of cold gas formed through gas-rich mergers or galaxy interactions, which could potentially form central disks and rings in the final stages of mergers \citep{Eliche-Moral10}, as also suggested by studies on elliptical galaxies \citep{young08, crocker11, davis11}.

Another possible mechanism comes from interactions with satellites. Cosmological ROMULUS simulations show that satellite galaxies are not required to merge to deliver their gas to the central BGG. Instead, the gas from the satellites gets stripped (Babul et al. in prep) forming gaseous streams, {proving another source of filaments formation}, that could eventually end up raining down onto the central galaxy. If the satellites are sufficiently low mass, they can continue orbiting until they lose most of their gas. From a kinematic point of view, galaxy satellites also generate wakes and turbulence, producing disturbed kinematics. Still, some extended rotating structures and rings of a few ten kpcs lengths are also formed, as the gaseous streams naturally contain significant angular momentum allowing for the gas to settle in an extended rotating structure. 

In terms of timescales, the gaseous disks have surprisingly short depletion timescales ($\tau_{\rm dep}=\rm M_{\rm gas}$/SFR), implying a very rapid replenishment of their gas reservoirs on timescales of $\sim$1$\times$10$^{8}$~yr. Moreover, based on FIR measurements the specific star formation rate ($\rm sSFR = SFR/M_{star}$) is about 0.020~Gyr$^{-1}$ and 0.0008~Gyr$^{-1}$ for the extended and compact disks, respectively (see for more details \citealt{osullivan15,osullivan18b}). Note that SFR derived from UV measurements are typical of the order of 10$^{-2}$~M$\odot$year$^{-1}$ \citep{kolokythas18}, with a negligible contribution to the total SFR. The latter suggests that most galaxies have formed their stars in earlier star formation episodes \citep{osullivan15}. The replenishment of the cold gas must be in very short timescales of $\sim$1$\times$10$^{8}$~yr, making it difficult to reconcile with a merger scenario. On the other hand, these timescales are of the same order of magnitude as the cooling time seen in these objects, $\lesssim$0.1~Gyr. 

\begin{figure}[htb!]
    \centering
    \subfigure{\includegraphics[width=0.25\textwidth]{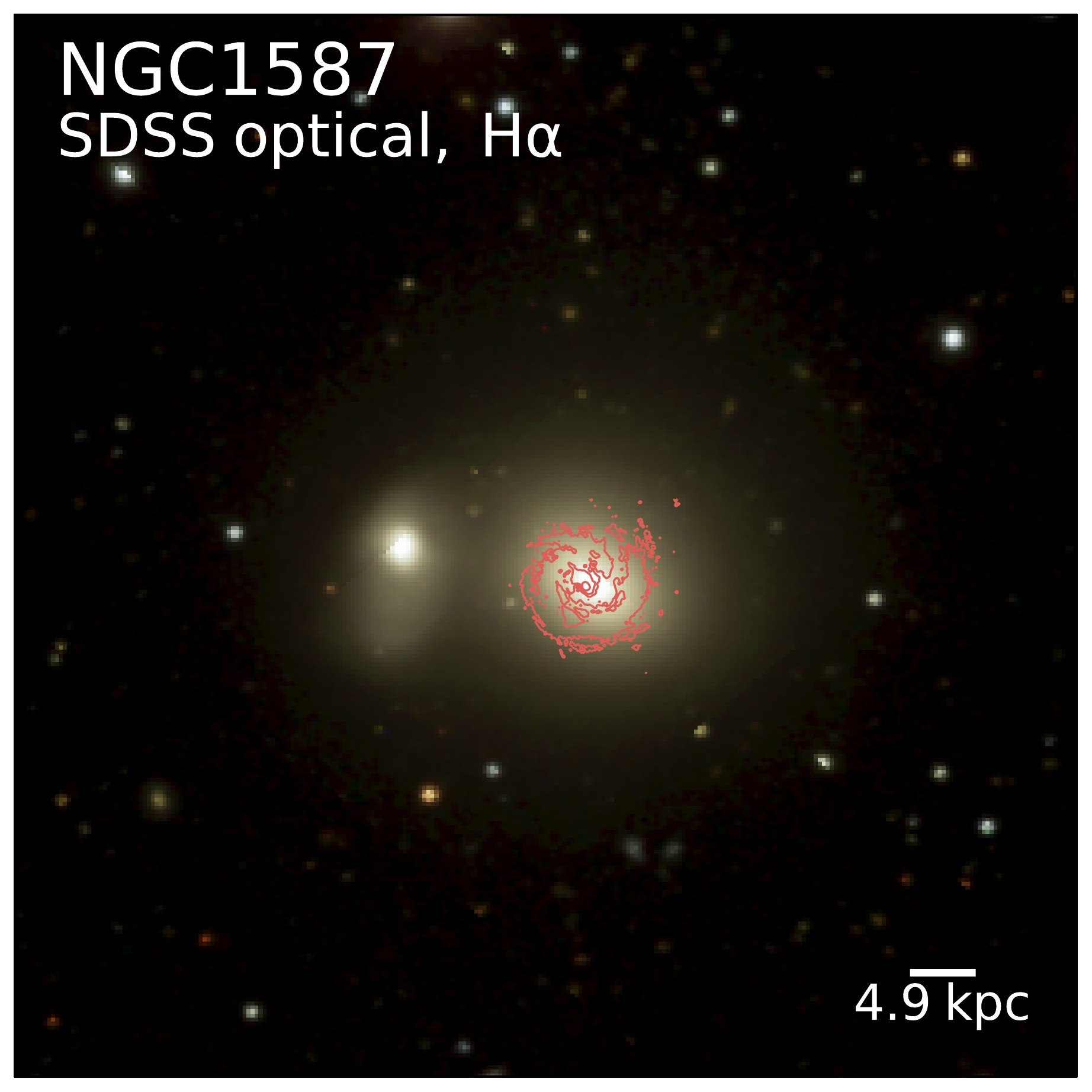}}
    \hspace{-0.4cm}
    \subfigure{\includegraphics[width=0.25\textwidth]{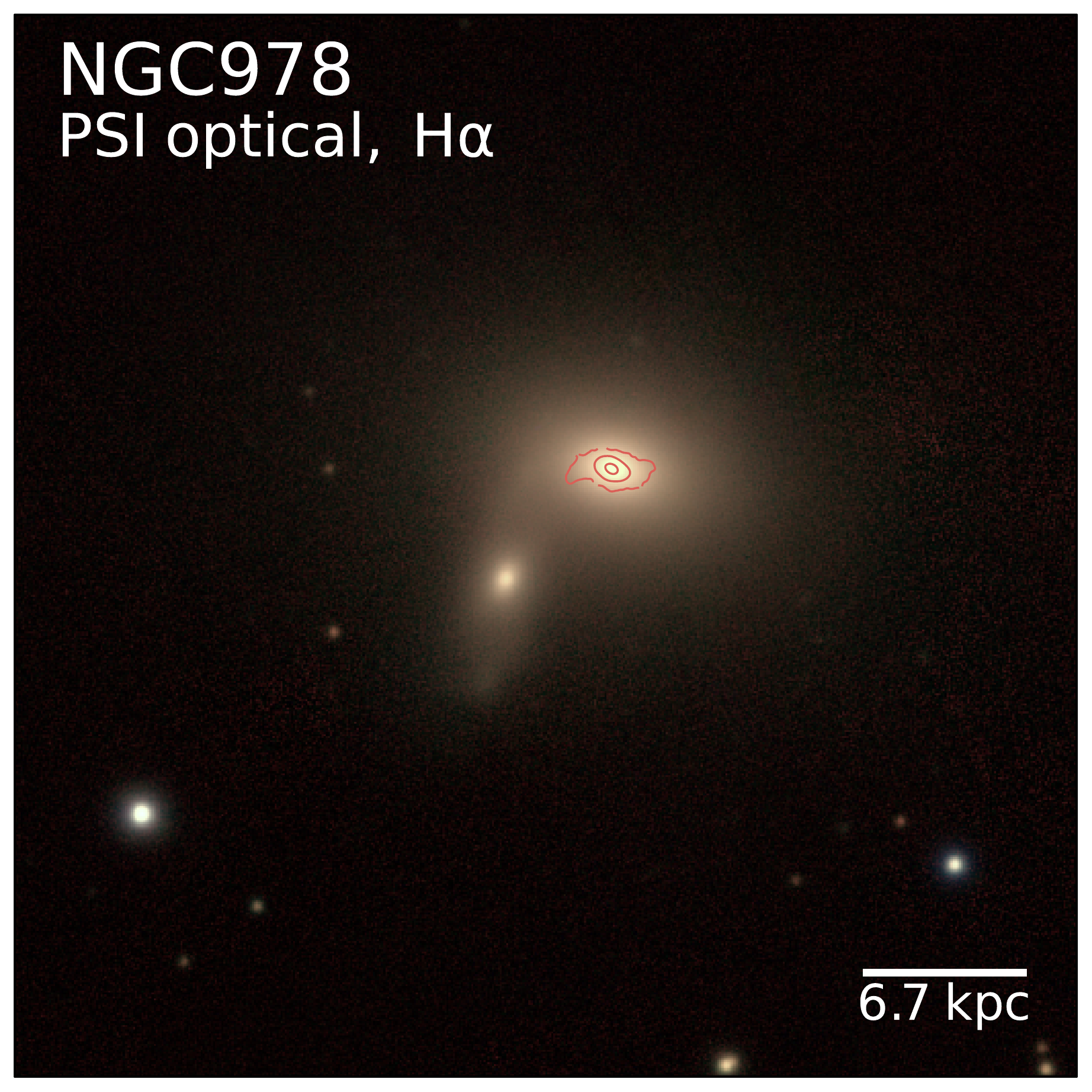}}
    \caption{Optical images of the interacting galaxies of our sample. Left panel: SDSS9 optical image of NGC\,1587. Right panel: SDSS9 optical image of NGC\,978. The distribution of the optical emission-line gas is shown with red contours.}\label{fig:optical}
\end{figure}

We do not find any evidence of long tidal tails and streams of stars seen in the optical images, reminiscent of ongoing galactic collision of two galaxies \citep[e.g.,][]{saviane08}, but note that the FOV of MUSE is limited to the central galaxy. Only NGC\,1587 and NGC\,978 reveal a hint of interaction with its neighboring elliptical, NGC\,1588 and MCG+05-07-017, respectively. The set of optical filaments in NGC\,1587 appears spiral inflowing into the central galaxy, indicating signs of tidal stripping (see Fig~\ref{fig:optical}, left panel). This single example is perhaps indicative of the merger role in gas accretion in these systems. On the contrary, in NGC\,978, a rotating disk system, no sign of cold gas is found between the two galaxies (see Fig~\ref{fig:optical}, right panel). Instead, the optical-emission line gas shows some emission arising of the disk to the SE-NW direction, although this emission is only extending $\sim$1.5~kpc (4$\arcsec$) in projection. The filamentary gas in our sample usually has disturbed velocity structures within their short extension, $<$10~kpc, in contrast to what might be expected from the kinematics of tidal tails in interacting galaxies \citep[e.g.,][]{bournaud04}. Some shallow velocity gradients are found along the filaments. However, these filament morphologies are more consistent with the scenario of a condensing hot halo \citep[e.g.,][]{beckmann19} rather than being from a ``wet'' merger event. We also do not see evidence of shells structures in the optical images, features expected in post-merger galaxies \citep{prieur90}. We note that even if they are present, these structures are very faint and thus difficult to detect.

\subsubsection{Accretion from a hot IGrM}
\def\setwithtotal{0.5}

\begin{figure}[htb!]
        \centering
        \includegraphics[width=\setwithtotal\textwidth]{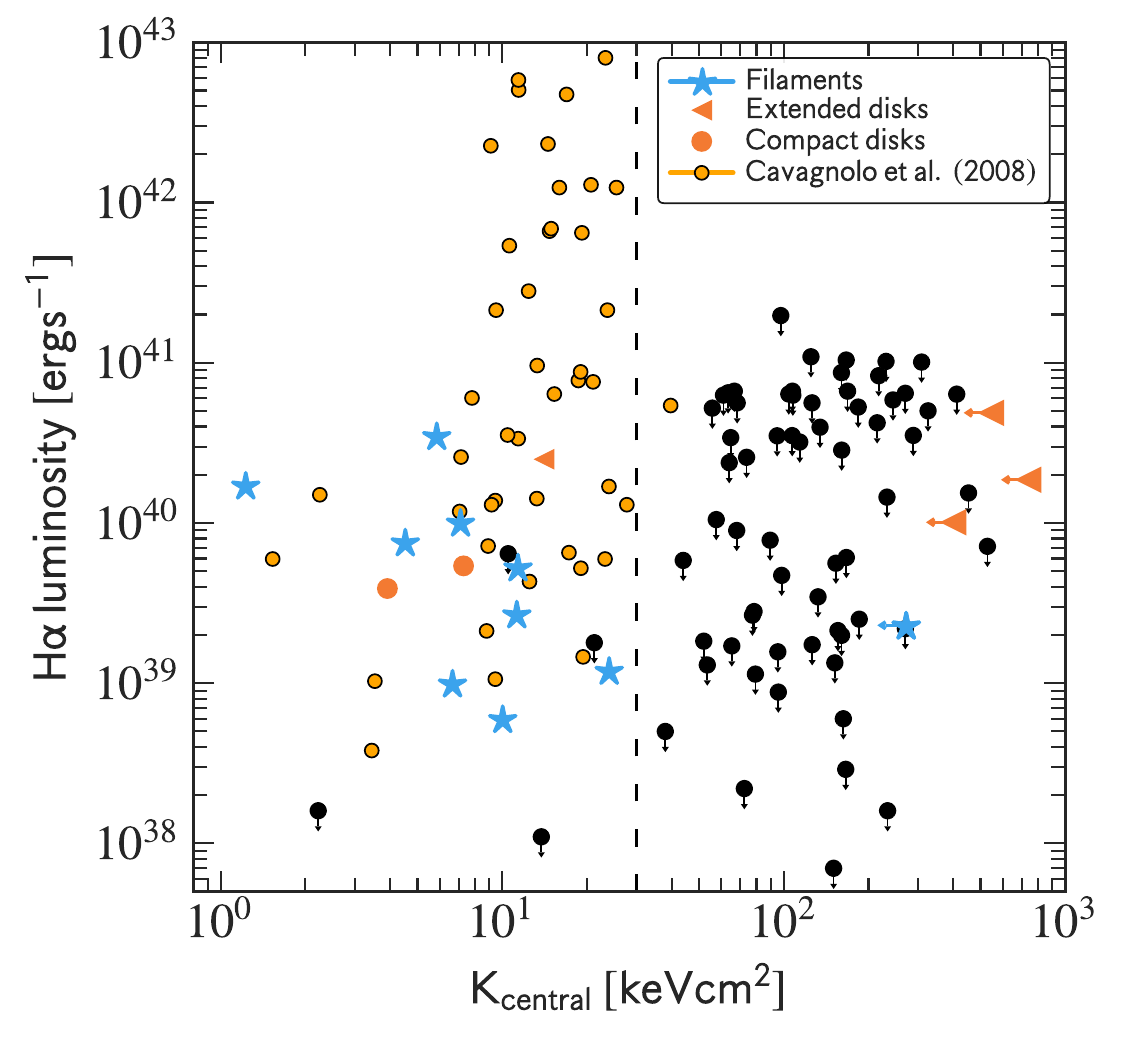}
        \caption{H$\alpha$ luminosity versus central entropy values measured at 1~kpc for our sample. Filamentary sources are shown with blue stars, compact disks with orange circles, and extended disks with orange left-pointing triangles. As a matter of comparison, we have included the sample from \citet{cavagnolo08} as a reference for BCGs with yellow circles, H$\alpha$ upper limits are shown with black circles. The entropy values correspond to core entropy, K$_{0}$, for the Cavagnolo sample. The dashed gray line is the entropy value of 30~keV~cm$^{2}$. Note that this entropy threshold should be a wide range of values as K can vary enormously in groups in comparison to clusters that are more self-similar \citep{sun09a,oppenheimer21}.
        %Right panel: Central cooling time at 10~kpc versus H$\alpha$ luminosity for our sample.
        }
        \label{fig:Entropy_tcool_vs_LHa}
\end{figure}

\begin{figure*}[htb!]
\centering
        \includegraphics[width=1.0\textwidth]{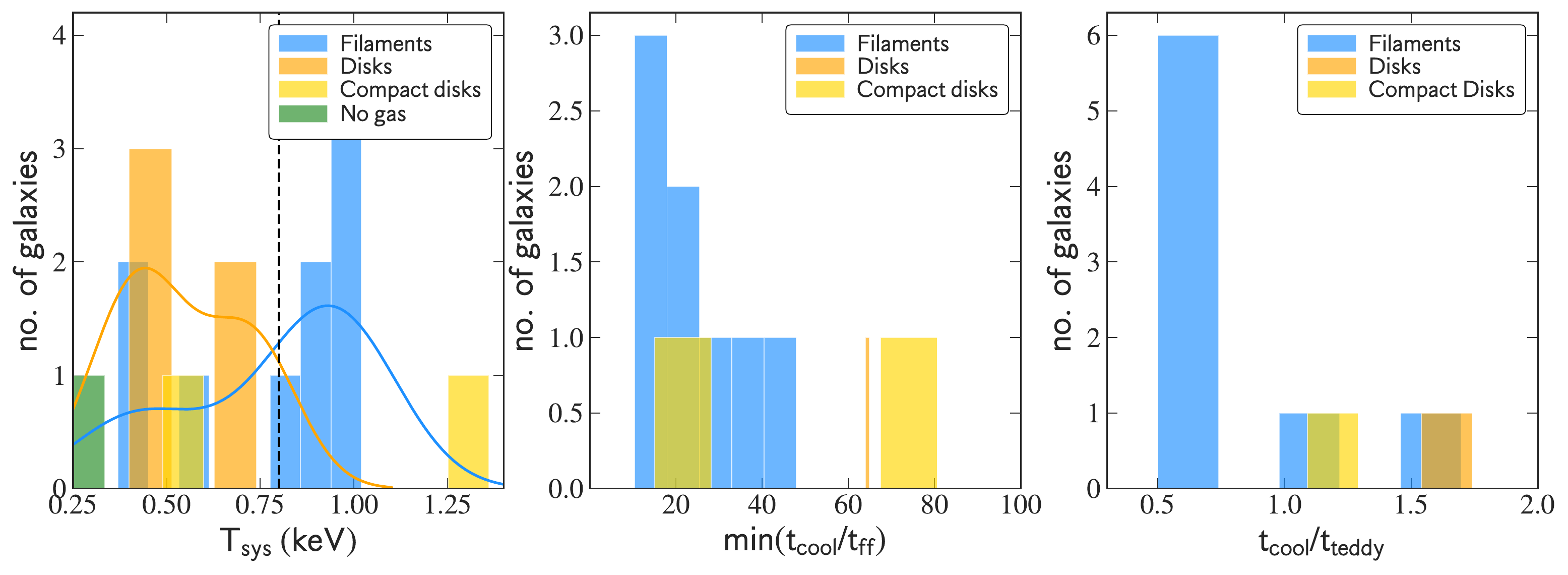}
        \caption{Left panel: Histogram of the X-ray temperature. To visualize the distribution of observations, we plot the kernel density estimate (KDE) that expresses the dependence of the disks (orange curves) or filaments (blue curves) on the system temperature, $T_{\rm sys}$. Vertical dashed line corresponds to temperature system of 0.8~keV that separate hotter and cooler groups \citep{osullivan17}. Middle panel: Histogram of the min(t$_{\rm cool}$/t$_{\rm ff}$).
        Right Panel: Histogram of the C-ratio , t$_{\rm cool}$/t$_{\rm eddy}$, at 10~kpc. The bars are color-coded by source classification as disk-like systems with orange bars, compact disk-like systems with yellow bars, filaments with blue bars, and non-gas detection with a green bar.
        \label{fig:distribution}}
\end{figure*}

In terms of kinematics, misaligned angular momentum between stars and gas may not necessarily imply a merger origin. The same can occur in an IGrM cooling scenario, including counter-rotation, by an angular momentum misalignment among the cooling IGrM, stellar disks, and dark matter halos \citep{lagos15}. Counter-rotation is also possible for cooling gas as it is not known how the angular momentum may develop as it sinks to the center of the potential.

Observational studies found the presence of nebular and cold molecular gas, in galaxy clusters, groups, and even elliptical galaxies, in systems with shallower X-ray entropy profile and lower inner entropy values, K$_{10}$, below $<$ 35~keV~cm$^{-2}$ \citep[e.g.,][]{rafferty08, cavagnolo08, lakhchaura18, babyk19}. Numerical simulations and theoretical studies have interpreted the latter, as a consequence of IGrM condensation, which develops a multiphase medium via {thermal instabilities triggered by the influence of the AGN feedback}, when the cooling time is short enough, or when the ratio of the cooling time over a dynamical time-scale such as the free-fall time, $t_{\rm cool}/t_{\rm ff}$, is below 10  \citep[e.g.,][]{sharma12,mccourt12}.

{Subsequent simulations and theoretical studies show that condensation may develop when $t_{\rm cool}/t_{\rm ff}\sim$5--20, as an outcome of precipitation-regulated feedback \citep[e.g.,][]{gaspari12,voit17,prasad18,prasad20,wang19}, which corresponds to a central entropy threshold of 30--40~keV~cm$^{-2}$. 
It is important to note that, unlike in early thermal instability simulations, this does not appear to be a line-in-the-sand criterion, but rather a wide band with large scatter.}

In a similar perspective, the C-ratio, defined as the ratio of the gyration time scale of turbulent eddies, $t_{\rm eddy}$, and the cooling time, $t_{\rm cool}$, is considered to be crucial for the development of non-linear thermal instabilities and the onset of gas condensation via turbulence. Cold gas precipitation occurs when the C-ratio is $\sim$1 \citep{Gaspari_2018}. The eddy turnover timescales, defined as $t_{\rm eddy}= 2 \pi \frac{r^{2/3}L^{1/3}}{\sigma_{v,L}}$, is inversely proportional to the gas velocity dispersion, ${\sigma_{v,L}}$, measured at the injection scale, $L$. As described in \citet{Gaspari_2018}, we use the length of the filaments as a proxy for the injection scale, and the velocity dispersion, ${\sigma_{v,L}}$, from the MUSE observations. Recently, \citet{wang19} found strong evidence of multiphase gas in one of their galaxy group simulations, which has comparatively shallower $t_{\rm cool}/t_{\rm ff}$ and entropy profiles than single-phase systems. \citet{osullivan17} found that CC and NCC groups have very similar entropy and $t_{\rm cool}/t_{\rm ff}$ profiles. {An alternative is the formation of cold gas via stimulated-feedback \citep{Revaz_2008,pope10,li14,mcnamara16}, or similarly, AGN-driven outflows \citep{qiu19,qiu20}, in which the cold gas condenses out of the uplifted low-entropy gas by the radio bubbles (see  Section~\ref{sec:stimulated_feedback}).}

While most of clusters have an X-ray emitting ICM, in our optically-selected groups, only twelve sources (12/18, $\sim$65\%) have a detected extended ($>$10~kpc) hot gas halo that allows us to investigate the condensation criteria. The rest of the groups have an extension smaller than the (XMM) PSF (3/18, point-like), and two lack X-ray detection \citep{osullivan17}. 
Note that X-ray observations were constrained to detect very faint X-ray halos with X-ray luminosities of a few 10$^{40}$~erg~s$^{-1}$. Noteworthy, all point-like and non-detection corresponds to extended disk-like sources and NGC\,584, a source categorized as filamentary but dominated by rotation. The detection of an IGrM for those sources will be explored with future X-ray facilities (e.g., Athena). Detailed X-ray properties of each source are listed in Table~\ref{tab:prop_multiwave}.

{In Fig.~\ref{fig:Entropy_tcool_vs_LHa}, we plot the central entropy values at 1~kpc as a function of H$\alpha$ luminosity. We found that twelve sources (12/17, $\sim$70\%) with cold gas have low central entropy values below $<$30~keV~cm$^{-2}$. As shown in Fig.~\ref{fig:Entropy_tcool_vs_LHa}, all of them correspond to filamentary sources (blue stars) and compact disks-like systems (orange circles). On the contrary, if we estimate the entropy at 10~kpc, only five sources are consistent with the entropy threshold, while another six sources have central entropy values between 30 and 40~keV~cm$^{-2}$. Bear in mind that the fixed radius of 10~kpc to measure the inner entropy values may be larger than the cooling region in some cases, as the radial projection sizes of the ionized gas are usually smaller.}

{Taking as a strict onset of TI the criterion of min($t_{\rm cool}/t_{\rm ff}$)~$\leq$~10, only one source may have condensed their cold gas through thermal instabilities (see Fig.~\ref{fig:distribution}, middle panel).} {Note that most of the the t$_{\rm cool}$/t$_{\rm ff}$, never goes below 10, as found in BCGs \citep[e.g.,][]{hogan17b,olivares19}. This shows that the TI-ratio has a large scatter and is likely not the primary criterion of condensation.} One needs to note that those predictions are based mostly on idealized simulations, where other important cosmological processes such as satellites accreting onto groups, orbiting and being stripped, and possibly merging, may be dominant in galaxy group environments (see, for example \citealt{jung21}). 

On the other hand, the ratio of the cooling time to the eddy turn-over timescales, $t_{\rm cool}/t_{\rm eddy}$, are within the range of 0.5--1.6 at 10~kpc in seven sources (see Fig.~\ref{fig:distribution}, last panel), consistent with the range of values predicted by numerical simulations (C=0.6--1.8, \citet{Gaspari_2018}). Again, all the sources belong to the filamentary category, including one compact disk (NGC\,4261). Only in NGC\,1453 with an extended-rotating disk, the $t_{\rm cool}/t_{\rm eddy}$ is close to unity (0.6 at 10~kpc). However, its min($t_{\rm cool}/t_{\rm ff}$) ratio is high, $\sim$64. If the emission line gas in this system has formed through cooling, this indicates that the C-ratio criterion is likely a more reliable threshold to dissect the condensation than TI-ratio.

From the criteria presented above {($t_{\rm cool}/t_{\rm eddy}$ and $\rm K_{0}$}), we found that the filamentary sources have an IGrM that appears to be cooling according to either {i) $t_{\rm cool}/t_{\rm eddy}$, or ii) the low central entropy values, $\rm K_0$}, except for NGC\,584, where the IGrM was not detected by the \textit{XMM-Newton} observations \citep{osullivan17}. In the compact rotating disk systems, NGC\,978 and NGC\,4261, {both low central entropy values, $K_0$ and the cooling time to the eddy turnover time ratios, $t_{\rm cool}/t_{\rm eddy}$,} suggest that the warm ionized gas condensed through the cooling of the IGrM. Similar results have been found for X-ray and optically bright selected elliptical galaxies presented in \citet{babyk19} and \citet{lakhchaura18}, who investigate the connection between the thermodynamical properties of hot X-ray emitting gas, using \textit{Chandra} observations, the H$\alpha$+[NII] emitting phase, and cold molecular gas, respectively.

\subsubsection{Could extended disks also be condensing from rotation-dominated hot atmospheres?}
\begin{figure}[h]
        \centering
        \includegraphics[width=\setwithtotal\textwidth]{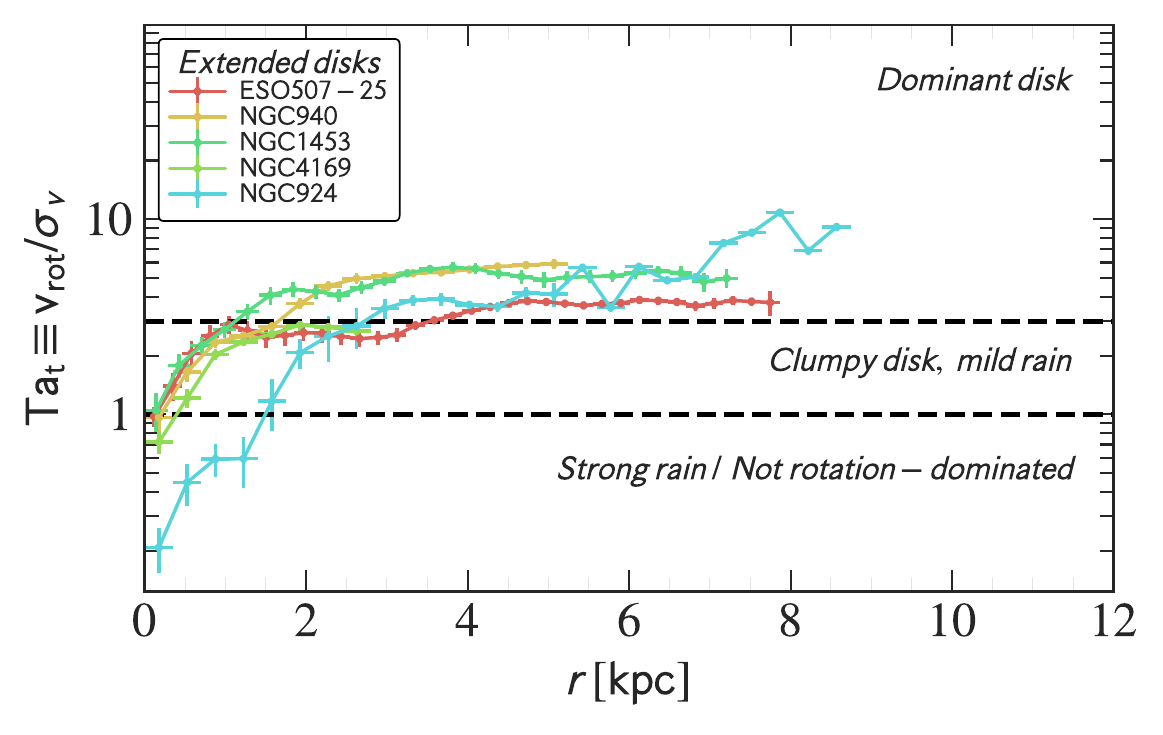}\\
        \includegraphics[width=\setwithtotal\textwidth]{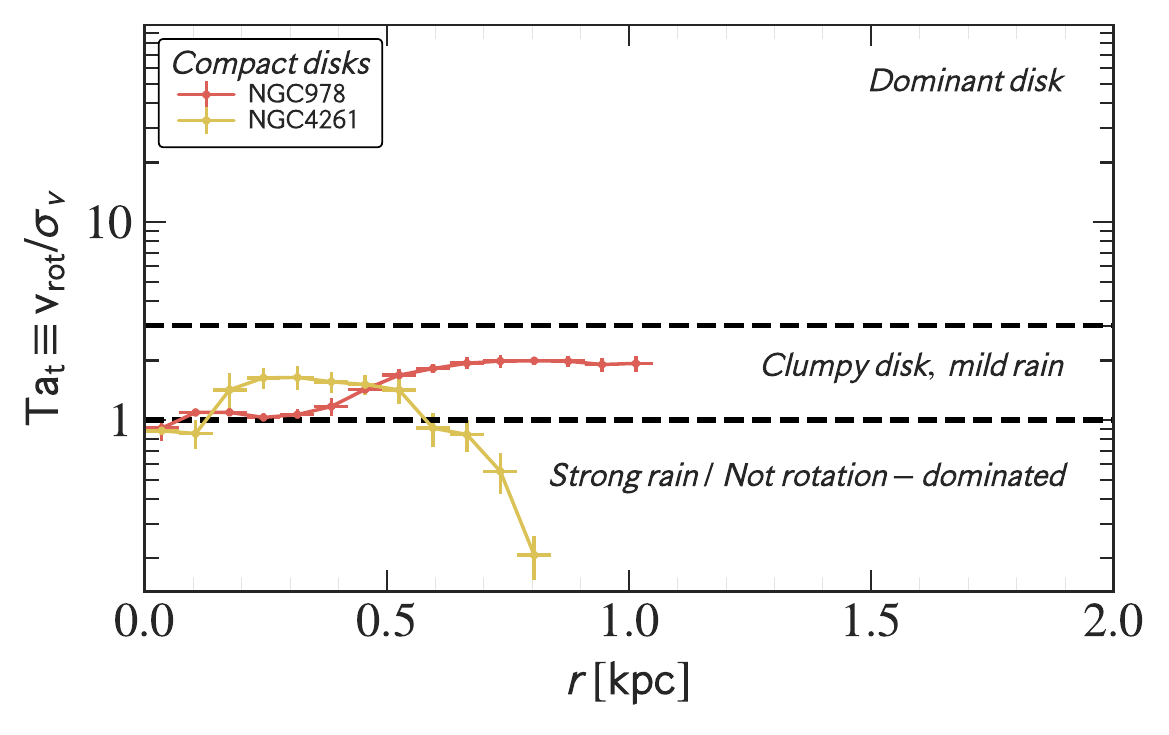}\\
        \includegraphics[width=\setwithtotal\textwidth]{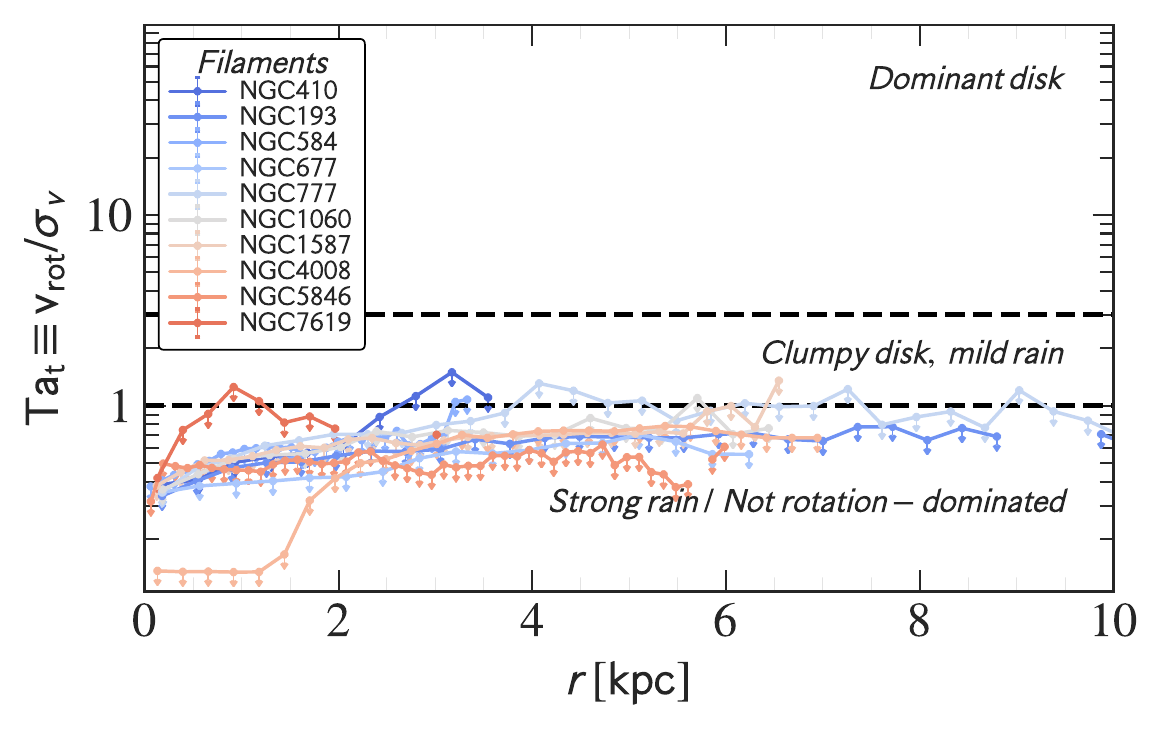}
        \caption{Top panel: The turbulent taylor number $\rm Ta_{\rm t} \equiv v_{\rm rot}/\sigma_{v}$ of the extended rotating disks for the optical-emitting gas. Middle panel: The turbulent taylor number for the compact rotating disks.
        Bottom panel: Upper limits for the turbulent taylor number for the filamentary sources.
        The black dashed lines denote the transition between strong rain at $\rm T_{\rm a_{\rm t}}<1$ (turbulence dominated), mild rain (clumpy disk) at $1<\rm Ta_{\rm t} <3$ (turbulence and rotation), and dominant disk $\rm Ta_{\rm t}>3$ (rotation dominated).}
        \label{fig:taylor_disks}
\end{figure}

Several arguments favor hot gas cooling for the origin of the filamentary structures. However, discussing the origin of the extended rotating disk systems in the context of hot gas condensation is more complicated since these systems possess the highest cold molecular gas masses of the sample but usually lack IGrM detections. It might be thus possible that the dominant origin of the gas for most of these systems is either from wet mergers or gas spilled by cosmological satellites, as discussed above. The following section discusses arguments supporting a possible contribution from hot gas condensation for the extended disks. 

Only NGC\,1453 has a detected IGrM, which appears to be cooling according to the criteria discussed above. It is noteworthy that, as shown in Fig.~\ref{fig:distribution}, the extended rotating disk systems tend to have lower global X-ray temperature, consequently fainter X-ray halos, compared to the filamentary systems, although radio jets are not found in any of the observed radio frequencies. In smaller systems than clusters, other physical processes are also important in controlling the thermal state of the X-ray halos \cite{mathews03}. SN heating can be similar to the radiative cooling losses in some elliptical galaxies \citep{voit20}, and it also has been shown that SN explosions can facilitate precipitation, as well as driving turbulence in elliptical galaxies \citep{limiao20, limiao20b}.

The lack of a hot halo detection in some of the extended rotating disks can be due to the angular momentum effect of the galaxy on the existence of a hot atmosphere. In this sense, angular momentum reduces the hot gas density, lowering the X-ray surface brightness and temperature in rotating galaxies to non-rotating galaxies, as shown by numerical simulations of group halos and individual ellipticals \citep[e.g.,][]{negri14a, negri14b, gaspari15}. On the other hand, theoretical studies show that hot gas content and baryon fraction is more susceptible to feedback processes than clusters \citep[e.g.][]{liang16}. A more likely explanation for the lack of hot gas in some systems arises from gas removal driven by SN winds (or also AGN-driven), which may be able to expel the gas up to 3--4~R$_{200}$ at early times as shown by simulations \citep{liang16}. At the same, it is worth to note that several of these systems are expected to host halos emitting mainly in the very soft X-ray band, with current instruments having difficulty in retrieving strong constraints even with a deep exposure.% (which is also often lacking).

\citet{werner14} found similar peculiarities in their two disk-like systems of its BGG sample. The two sources are the X-ray faintest and, controversially, the strongest [CII] and H$\alpha$+[NII] emitter galaxies of their sample. However, the bolometric X-ray luminosity of these two sources is about ten times higher than of our extended gaseous disk-like sources. The authors also argue that hot gas may be more prone to condensation in rotating systems as rotational support may prevent gas from infalling, and thermal instabilities may develop more easily (see also \citealt{juranova19,juranova20} for S0s rotating galaxies). Another option that could explain this dichotomy of high molecular mass and lack of IGrM medium in the extended disks is that these systems could be in a heating-dominated phase, where the IGrM gas has been recently heated up by either AGN-outbursts or SN heating, challenging the detection of the X-ray emitting gas. However, the low SFR ($<$0.5~M$\odot$~yr$^{-1}$), as well the lack of radio jets on these systems makes it less plausible. Note that even in the case that these galaxies do not have a hot halo detected, they could have it in the past. Thus the massive sink of raw material could have originated through previous cooling episodes.

\citet{gaspari15} investigate the influence of the large-scale rotating hot atmosphere in the AGN feedback and feeding. The simulations predict the formation of multiphase filaments when the turbulent Taylor number, $\rm {Ta_{\rm t}} \equiv v_{\rm rot}/\sigma_{\rm LOS}$, is below $<1$ (turbulence-dominated), here $v_{\rm rot}$ and $\sigma_{\rm LOS}$ are the rotational velocity and velocity dispersion of the gas, respectively. As rotation is introduced in the hot atmosphere, the condensation shifts from turbulence-driven to rotationally-driven, leading to the formation of a central disk. As the rotation dominates, the accretion changes from clumpy disks with mild condensation when $\rm Ta_{\rm t}>1$ to a coherent rotating disk. The rotation suppresses the accretion onto the SMBH due to the angular momentum barrier, leading to a longer condensation onto the equatorial direction, when $\rm Ta_{\rm t}>3$. The feedback is also very weak due to the angular momentum barrier, agreeing with the lack of radio jets in these systems. In the absence of major CCA rain, the extension of the rotating disks will be consistent with the duration of the condensation. In other words, the larger the disk, the longer the condensation cascade \citep{gaspari17}. 

We compute the Taylor number profiles, assuming that the gaseous disks form out of the hot gas condensation. We obtained the rotational velocity, $v_{\rm rot}$, by building a disk model using Barolo$^{3D}$ \citep{DiTeodoro15} that describes the kinematics of the ionized gas. While the velocity dispersion, $\sigma_{\rm LOS}$, was measure as the average value in radial bins of 1$\arcsec$. The radial profiles of $\rm Ta_{\rm t}$ of the extended and rotating disks shown in Fig.~\ref{fig:taylor_disks} (top panel) are consistent with the predictions from the simulations \citep{gaspari15}, as $\rm Ta_{\rm t}>$1--3 over most of their extensions. At the inner region of some extended disks, $\rm Ta_{\rm t}$ is below 1 (Fig.~\ref{fig:taylor_disks}), indicating that the disks are influenced by turbulence and could be raining. The galaxy dynamics could also increase the velocity dispersion of the gas at the center, decreasing the $\rm Ta_{\rm t}$ value. The $\rm Ta_{\rm t}$ radial profile of NGC\,978, the compact disk, is $\sim$1--2 along most of the nebula (Fig.~\ref{fig:taylor_disks}, middle panel), indicating a clumpy disk with mild condensation. In NGC\,4261, the $\rm Ta_{\rm t}$ is $<$1 along the extension of the blueshifted NE and SW filaments ($>$0.5~kpc) seen in Fig.~\ref{fig:ngc4261_hst}. As for the filamentary structures identified in some of the extended rotating disks, the turbulent Taylor number, $\rm Ta_{\rm t}$, is typical $>$ 3 as they follow the kinematics of the disk.

Lastly, in the case of the filamentary sources, the turbulent Taylor number, $\rm Ta_{\rm t}$, is $<$1 since the kinematics is not dominated by rotation, then the rotational velocity is minimal. By assuming a conservative upper limit for the rotational velocity of $50$~km~s$^{-1}$ for the filamentary sources, we derived an upper limit for Taylor number of $<$1 along most of the optical emission-line gas (Fig.~\ref{fig:taylor_disks}, bottom panel).

\subsubsection{Predictions from CCA simulations}

\begin{figure}[h]
        \centering
        \includegraphics[width=\setwithtotal\textwidth]{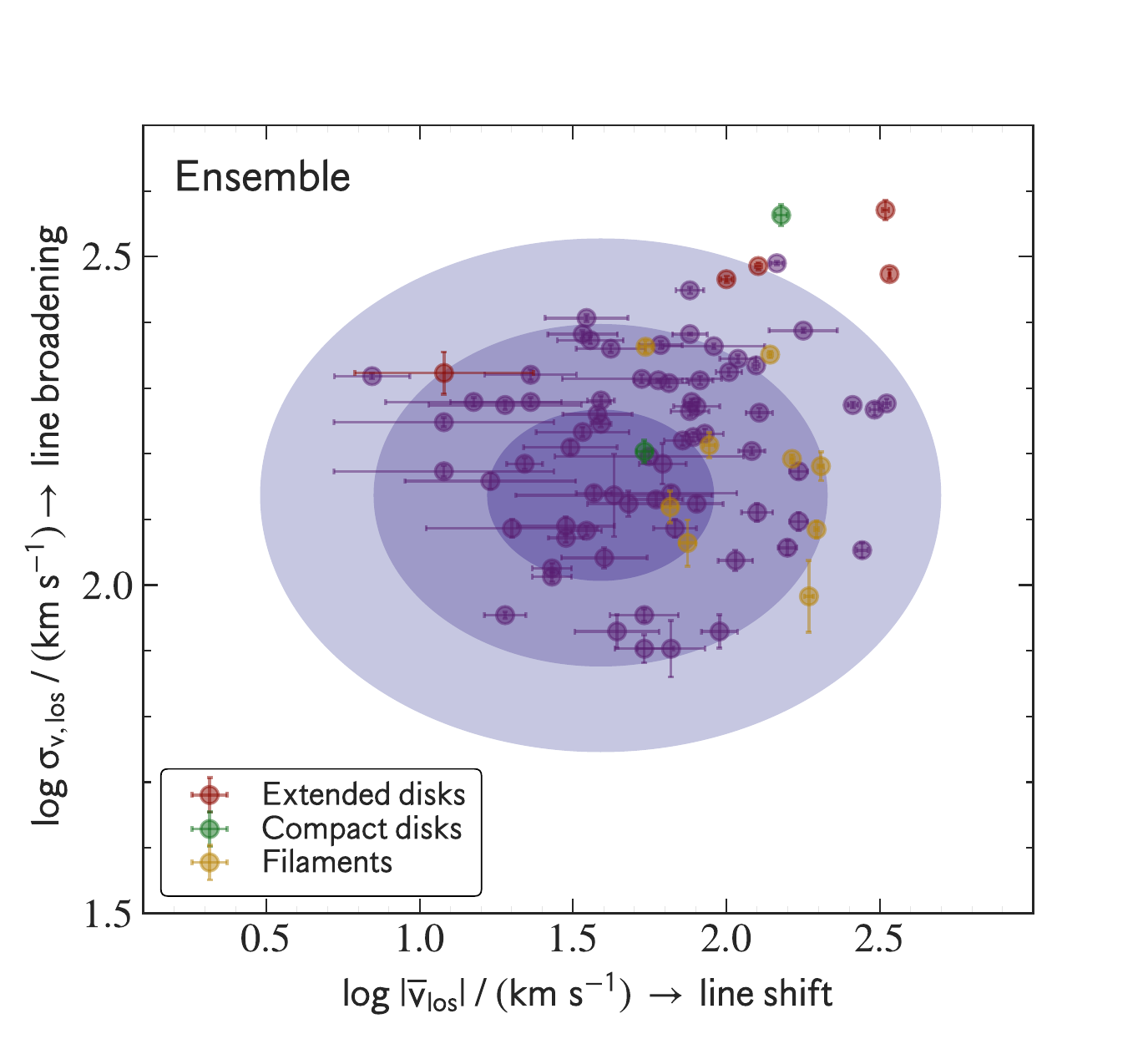}
        \caption{Line shift, $|v_{\rm los}|$, versus velocity dispersion, $\sigma_{\rm los}$, of the ionized gas. Comparison of observations with the CCA model \citep{Gaspari_2018}. The purple shaded region show ensemble (single spectrum) predictions from the CCA simulations at 1$\sigma$, 2$\sigma$, and 3$\sigma$ confidence intervals. Data points from our sample are shown with yellow, red, and green circles for filamentary, extended, and compact disks, respectively. Whereas data from the literature are displayed with purple circles \citep{Gaspari_2018}. This diagnostic is also known as kinematical plot, in short ``k-plot''.\label{fig:CCA_comparion}}
\end{figure}

Simulations of CCA predict the cold gas to be in a certain location of the velocity shift versus velocity dispersion diagram \citep{Gaspari_2018}. In Fig.~\ref{fig:CCA_comparion}, we compare observation from the ensemble region for our sample with prediction from the simulations. The ensemble gas, which comprises the kinematics of the cold gas at large scales from the AGN, gives us information on the macro-evolution of the global halo condensation. To compute the ensemble points displayed in Fig.~\ref{fig:CCA_comparion}, we followed the same procedure described in Sec~4 of \citet{Gaspari_2018} performed for the VIMOS observations of 73 BCGs taking the \citet{hamer16} sample. We found that the ensemble gas observations are consistent with that of H$\alpha$+[NII] ensemble detection in other BCGs and BGGs (purple points) from \citet{hamer14} sample, as well with the prediction from the CCA simulations (purple shaded regions). As for the rotating disks, the ensemble gas detection is slightly deviated to a higher velocity shift compared to the predictions, as expected in rotating-driven systems (see also \citealt{maccagni21}). {The shift is mainly produced by the large velocity gradient observed in the those disks.}

This finding suggests a tight correlation between the turbulent hot atmosphere and colder gas phases \citep{Gaspari_2018}. The data points for our BGG sample predict values for the hot gas turbulence. These predictions can be tested in the near future with the upcoming X-IFU satellites (such as XRIMS and Athena).

\subsubsection{Stimulated feedback and AGN-driven outflows}\label{sec:stimulated_feedback}
Based on the observed spatial distribution of the molecular gas flows around the radio bubbles or X-ray cavities in several galaxy clusters, {several authors proposed the uplifting mechanism or so-called stimulated feedback model (see  \citealt{Revaz_2008,pope10,li14,mcnamara16,qiu20,qiu21})}. In this model, the cold molecular clouds condense from the uplifted X-ray low-entropy gas in the wake of the radio bubbles inflated by the central AGN. As a result, the uplifted X-ray low-entropy gas will become thermally unstable, and it will develop a multiphase gas at an altitude where the ratio of the cooling time over the infall time, t$_{\rm cool}$/t$_{\rm infall}$, is below or close to unity.

The interaction between the AGN bubbles and the ionized gas has been observed in a few of our sources with filamentary structures (previously mentioned in Section~\ref{sec:radio_gas}). The most remarkable cases are seen in NGC\,5846 and NGC\,193, and potentially in NGC\,677, where the H$\alpha$ network of filaments trails the wake of the X-ray cavities. Comparable features have been observed previously in other galaxy groups, such as NGC\,5044, and NGC\,4636 \citep{david14,werner14,temi18}. Note that higher-sensitivity X-ray observations are needed to detect potential X-ray cavities on the fainter X-ray galaxy groups. Alternatively, other bands (such as radio SZ) can be used to further probe cavities \citep[e.g.,][]{yang19}. It is also possible that the cold gas condensation may not need radio jets (or cavities), but when present, there is clear evidence that it enhances the precipitation. Additionally, spatial distribution of the ionized gas correlates with the one from the radio emission in some sources as discussed in Section~\ref{sec:radio_gas}, indicating that the AGN feedback may play a role in the gas condensation.

%\subsubsection{Link to the AGN power?}
\begin{figure}
    \centering
        \includegraphics[width=0.52\textwidth]{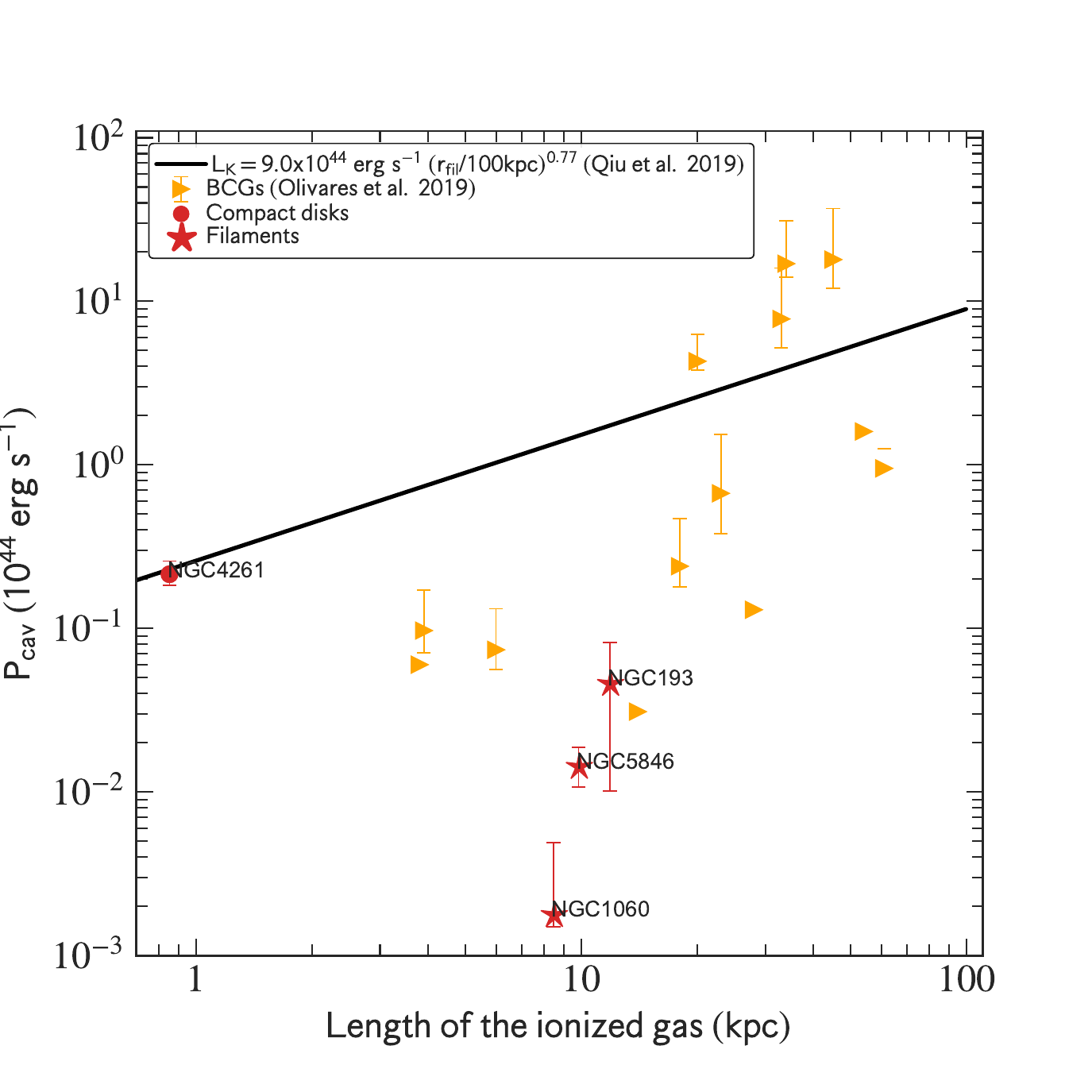}
    \caption{{AGN power versus spatial extent of the ionized gas. The black solid line corresponds to the relation predicted by \citet{qiu19} simulations. BGGs with $P_{\rm cav}$ estimated using their X-ray cavities properties from \citet{kolokythas18} are shown with red symbols.
    %while $P_{\rm cav}$ estimated from  $P_{\rm 235MHz}$ are shown with grey symbols.
    Filaments sources and compact disks are displayed with stars and circles, respectively. Yellow triangles mark the BCGs sample from \citet{olivares19}. Rarely the points follow the correlation, thus ruling out a dominant role of AGN-outflow driven condensation. 
    }}
   % \label{fig:vmax_sigma_classification}
    \label{fig:pcav_filament_extent}
\end{figure}

{In the context of radiatively cooling AGN-driven outflows \citep{qiu20}, simulations by \citet{qiu19} find a positive correlation between the AGN feedback power and the spatial filaments extent (see Fig~\ref{fig:pcav_filament_extent}, solid black line). We plot this relation for our sample using cavity power, $P_{\rm cav}$, as a proxy for the AGN feedback power from \citet{kolokythas18} for the four BGGs with confirmed X-ray cavities (red stars). %For the rest of the systems (gray symbols), we estimate their cavity powers using the $\rm P_{\rm cav}$ versus radio power at 235 MHz, $P_{\rm 235MHz}$, relation presented in \citet{osullivan11b}, with radio power at 235 MHz estimated from \citet{kolokythas18}. 
For a comparison, we have also included 12 BCGs from \citet{olivares19} (yellow inverted triangles). As shown in Fig~\ref{fig:pcav_filament_extent}, the BGGs rarely sit along the relation except for NGC\,4261, a very compact system with a mighty AGN. The cavity powers of the BGGs are a few orders of magnitude below the predicted values by the simulations. The BCGs tend to follow a bit more the relation, albeit the scatter is still substantial. We recall that the correlation was found in simulations of galaxy clusters, in which the AGN feedback is thought to be well coupled to the ICM, as well they do not include the effect of possible mergers and stirring by orbital galaxies (see Section~5.3.4 of \citealt{sarazin88}, and references therein). In summary, the lack of a tight and significant correlation shows that AGN-outflow condensation likely plays a sub-dominant role.}

\subsection{Evolutionary transition?}
\def\setwithtotal{1.0}
\begin{figure*}[htb!]
        \includegraphics[width=\setwithtotal\textwidth]{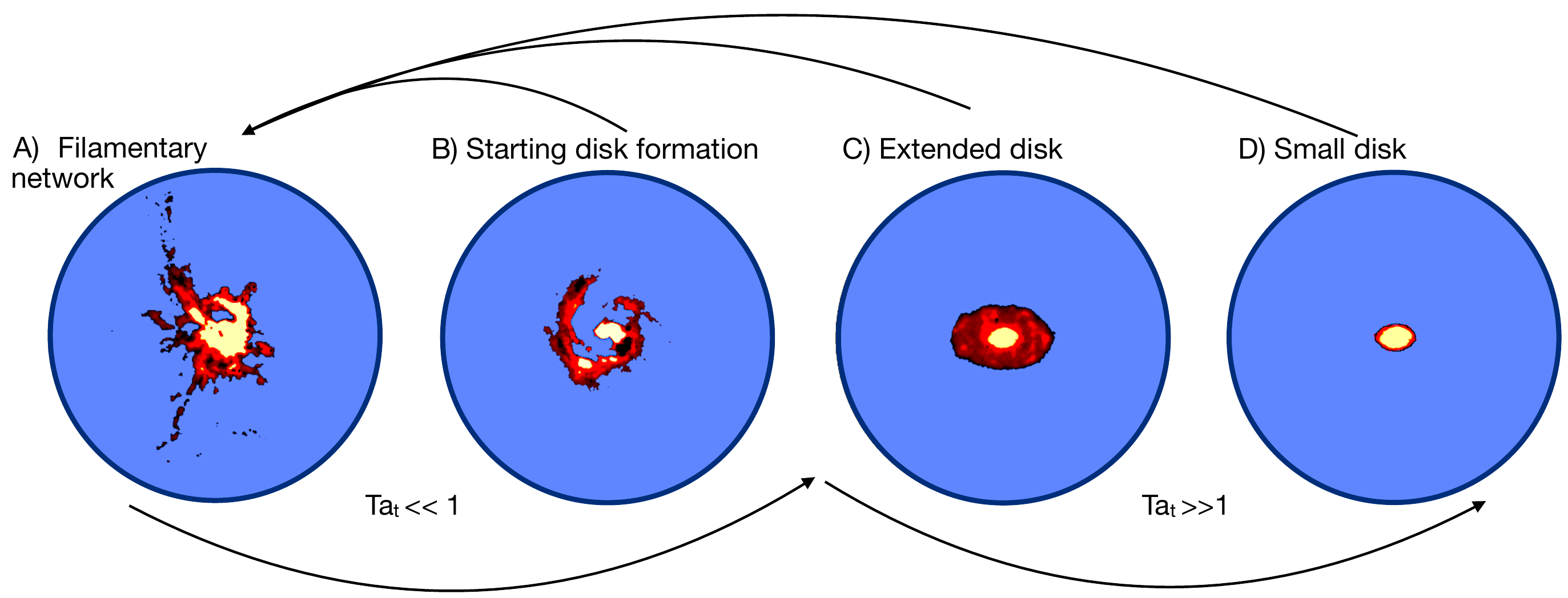}
        \caption{Possible evolutionary transition of the gas in BGGs in our sample (see text for details).}
        \label{fig:evolution}
\end{figure*}

In figure~\ref{fig:evolution}, we propose a possible evolutionary transition to explain the variety of ionized gas distributions in our sample. This evolutionary transition for the cold gas is consistent with an extensive suite of high-resolution hydrodynamical simulations \citep{gaspari16_IAU,gaspari15}. In this scenario, the hot gas cools via CCA or stimulated feedback, forming (i) multiphase filaments via thermal instabilities. At this stage, the turbulence drives the dynamics of the gas if the turbulent Taylor number T$_{\rm a_{\rm t}}<$1 \citep{gaspari15}. The filamentary phase is likely to be the longest, based on the large fraction of systems ($\sim$56\%) found in our sample that stands in this phase. As the gas precipitates (ii), the filaments experience chaotic collisions at the inner region of the galaxy, boosting the accretion rate.
As a consequence, the turbulence drops, the turbulent Taylor number becomes T$_{\rm a_{\rm t}}>1$, and the thermal instabilities become weaker, resulting in the formation of a (iii) gaseous disk. In this systems the feeding is not efficient due to the angular momentum barrier, thus the feedback is also very weak, allowing for longer condensation sink from the hot halo onto the expanding equatorial. Star formation is more likely to be detected at this stage. %The cold disk can grow a few tens of kpc if the angular momentum dominates the condensation.
This extended disk phase is potentially shorter ($\sim$28\%) than the filamentary stage. The star formation may provide some feedback and limit the cooling. More importantly, the angular momentum lowers the X-ray surface brightness, making it more difficult to detect the X-ray halo in this phase, as well as preventing the cold gas from being rapidly accreted. %During this phase, the angular momentum prevents the gas from being rapidly accreted.
The residual is a quiescent, (iv) compact disk, poor in molecular gas and star formation. The compact disk phase is likely very short (11\%). Later during this stage, the cold gas will be accreted onto the SMBH, potentially triggering the jets (as seen in NGC\,4261), and may restart hot gas cooling (see also \citealt{wang19}). As previously mentioned, there are maybe some exceptions where the origin of the gas is not clear, such as NGC\,584, where an X-ray halo has not been detected.

{In principle, the evolutionary transition might be explained the other way around, starting with disks or compact structures to extended filamentary structures (see upper arrows in Fig~\ref{fig:evolution}): as the cold gas accumulates at the center of the BGG and feeds the central engine, an AGN outburst is triggered which expels some of the ionized gas to further distances, forming extended filamentary structures as suggested by AGN-driven outflow models \citep{Revaz_2008,pope10,li14,mcnamara16, beckmann19,qiu20,qiu21}. However, the direct uplift of cold gas is extremely difficult to achieve given the large mass ratio between the cold gas and the very light jet (akin to moving a brick with wind).} 

The systems presented in this work are likely at different stages of the AGN feedback cycle. It has been shown that the duty cycle can be pretty irregular and even shorter in groups compared to clusters, on the order of a few 10s Myr \citep[e.g.,][]{gaspari11, sharma12, prasad15, wang19}. This could be reflected in the complex and diverse range of optical morphologies and properties presented in this work.

\section{Conclusion}
\label{sec:conclusion}
This paper presents the morphology and kinematics of the ionized gas and stars using MUSE observations of 18 BGGs selected from the high-richness CLoGS sub-sample. Characterizing the properties of the ionized gas in galaxy group cores is an important aspect to develop a complete picture of the nature of the accretion of the cold gas in groups and clusters. We summarize our main findings below:

\begin{enumerate}
    \item The optical emitting gas shows filamentary structures in most (10/18) of the sources of our sample, with a variety of shapes and projected sizes, $\sim$3--12 kpc, similar to what we found in the central cluster galaxies. The projected sizes of the filaments in groups are shorter than those found in central cluster galaxies, consistent with the different sizes of the condensation regions of the hot halos \citep{mcdonald11a}. Another significant difference with cluster galaxies is the presence of several (7/18) disks and rings dominated structures found in our sample, of which two disks are very compact, $\sim$1--3~kpc, whereas the rest of the disks can reach projected sizes of up to $\sim$21~kpc. The extended disks also reveal clumpy rings and extended structures. The velocity structure of the ionized gas in the sources with filaments is often chaotic, but some shallow velocity gradients are noticed along some filaments. The velocity dispersion maps of the ionized gas show a peak towards the center of the galaxy. The velocity dispersion peak is generally consistent with the peak of the gas emission, with values of 100--300~km~s$^{-1}$, although it can be below 100~km~s$^{-1}$ on the filaments and rings, similar to what we find in clusters of galaxies. Both chaotic kinematics and increased central velocity dispersion are consistent with the macro-scale CCA rain and inner related inelastic collisions.
    
    \item The stellar kinematics of the BGGs in our sample seem to be rotation-dominated compared to BCGs where the stars are consistent with random motions \citep{hamer16}. At least 14 of the 18 systems ($\sim$78\%) show some sign of rotation within their stellar light. Only in four sources ($\sim$22\%), is the stellar component dominated by random motions. The velocity field of the stellar light varies for each system and ranges from $\pm$40 and up to 350~km~s$^{-1}$.
    
    \item In most cases, the ionized gas is kinematically decoupled from the stellar component. In three of the rotation-dominated systems (3/7), the ionized gas is kinematically coupled (including one disk that may be tilted by at least $\rm \Delta PA \sim$10$\degr$); whereas four systems (4/7) are dynamically decoupled. In one of these disks, the gas and the stellar component are counter-rotating. 
    This all indicates that in most cases the stellar population and the ionized gas are not associated, supporting an external origin of the gas.
    
    \item We find that the H$\alpha$ luminosity correlates strongly with the cold molecular gas. Moreover, the scatter on this relation is highly reduced by 0.5~dex when these two physical quantities are measured on the same aperture in both galaxy clusters and groups. As shown for galaxy cluster cores, this tight correlation suggests a common origin for these two gas phases. 
    
    \item According to the {t$_{\rm cool}$/t$_{\rm eddy}$ ratios, and central entropies} values of the X-ray atmosphere, at least 11 sources in the sample are possibly cooling gas from their hot atmospheres via thermal instabilities. They all correspond to filaments, compact disk-dominated sources, and one extended rotating disk (NGC\,1453). The only exception is NGC\,584 (a filamentary source), where X-ray observations do not detect a thermal emission, likely because of sensitivity limits. However, a merger origin can not be ruled out. Besides, we find the $C$-ratio ($t_{\rm cool}/t_{\rm eddy}$) criterion is a more reliable threshold to unveil the condensation rain. 

    \item NGC\,1587 shows signs of tidal interaction with its elliptical neighbor. Accordingly, we suspect that at least a fraction, if not all, of the colder gas detected in this source, may have been acquired through a galaxy interaction. Still, based on the condensation criteria, the IGrM is also compatible with gas cooling.
    
    \item The larger fraction of rotating disks in central group galaxies than in clusters may hint towards a non-negligible contribution of mergers or gas stripped from cosmological satellites, that are more prone to happen in low-mass halos (extended gas disks are found in systems where the IGrM is not detected by the current X-ray observations and they lack radio-jets). We thus discuss the possibility that the presence of extended disks could trace a less intense cooling phase of the feedback, likely allowing for a longer condensation sink from the hot halos. The latter is consistent with the turbulent Taylor number, $\rm T_{a_{t}} >1$, which indicates such condensation may occur on non-radial orbits, gradually forming a disk.    
    
    \item As supported by high-resolution hydrodynamical simulations (see \citealt{gaspari16}, for a review), we propose a possible evolutionary sequence for the multiphase gas: the filaments first condense out of the hot gas via CCA, followed by a transitional episode, where the gaseous threads will start to accrete onto the center to form a massive extended kiloparsec rotating star-forming disk in the galaxy. The raw material of the disk can then be consumed either by star formation activity (which can also provide feedback) or by fueling the center of the AGN, leading to a compact and quiescent rotating disk.
\end{enumerate}

% Acknowledgements
\begin{acknowledgements}
This work was supported by the ANR grant LYRICS (ANR-16-CE31-0011). Based on observations collected at the European Organisation for Astronomical Research in the Southern Hemisphere under ESO programme 097.A-0366(A), and/or data products created thereof. M.G. acknowledges partial support by NASA Chandra GO9-20114X and HST GO-15890.020/023-A. E.O'S. acknowledges support from NASA through XMM-Newton award 80NSSC19K1056. PL (contract DL57/2016/CP1364/CT0010) is supported by national funds through Funda\c{c}\~ao para a Ci\^encia e Tecnologia (FCT) and the Centro de Astrof\'isica da Universidade do Porto (CAUP).
\end{acknowledgements}

\bibliographystyle{aa} % style aa.bst
\bibliography{master_volivares_BGGs} % your references Yourfile.bib

\appendix
\section{Distribution and kinematics of the ionized gas}\label{app:description_sources}

%\FloatBarrier

\subsection{Compact and Extended Rotating Disks}
\def\setwithtotal{0.8}
\def\setwidthsmall{0.8}
\begin{figure*}[!b]
 %   {\bf Extended Rotating Disks}\\
    \centering
            \subfigure{\includegraphics[width=\setwidthsmall\textwidth]{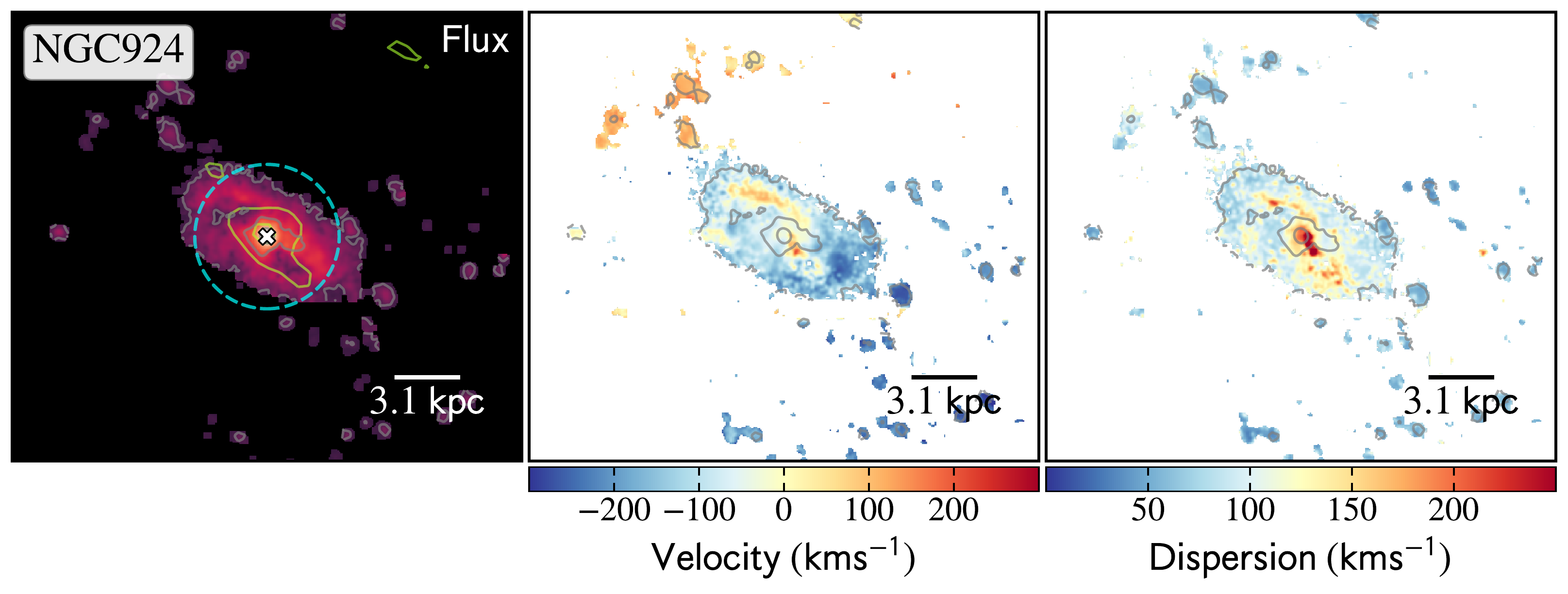}}\\
        \vspace{-0.5cm}
        \subfigure{\includegraphics[width=\setwidthsmall\textwidth]{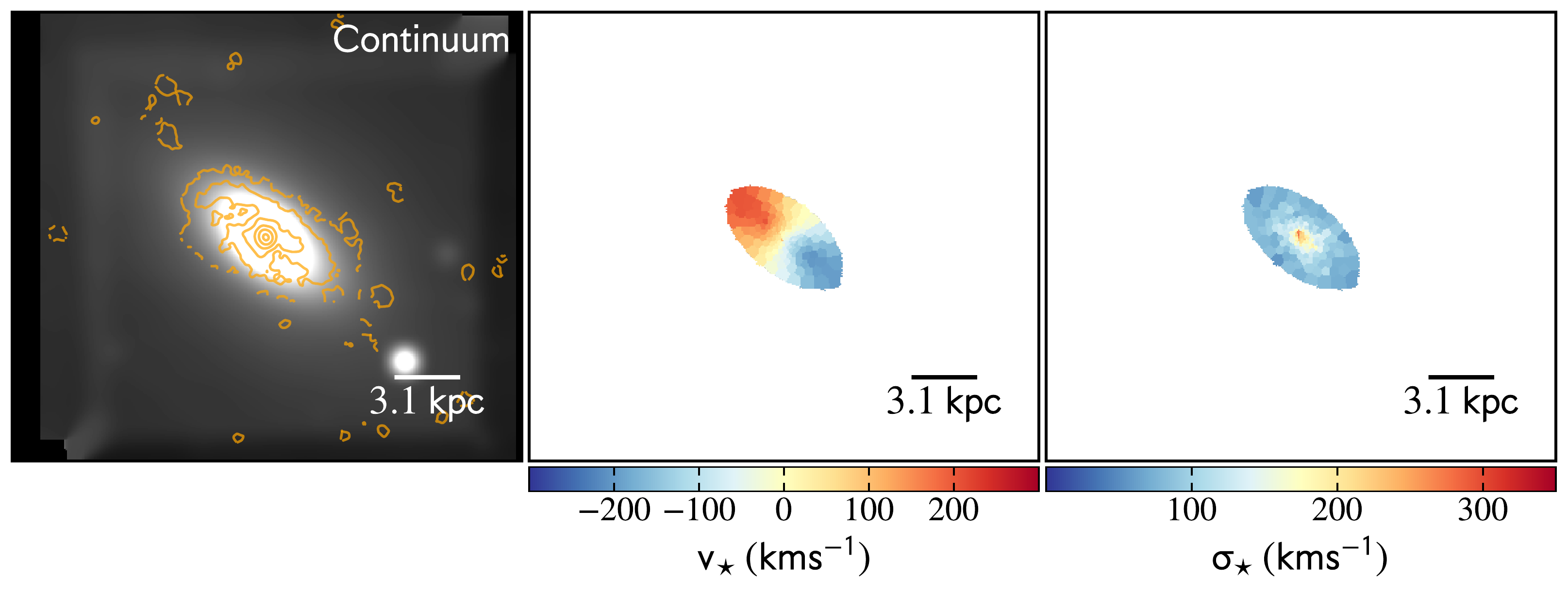}}
            \subfigure{\includegraphics[width=\setwidthsmall\textwidth]{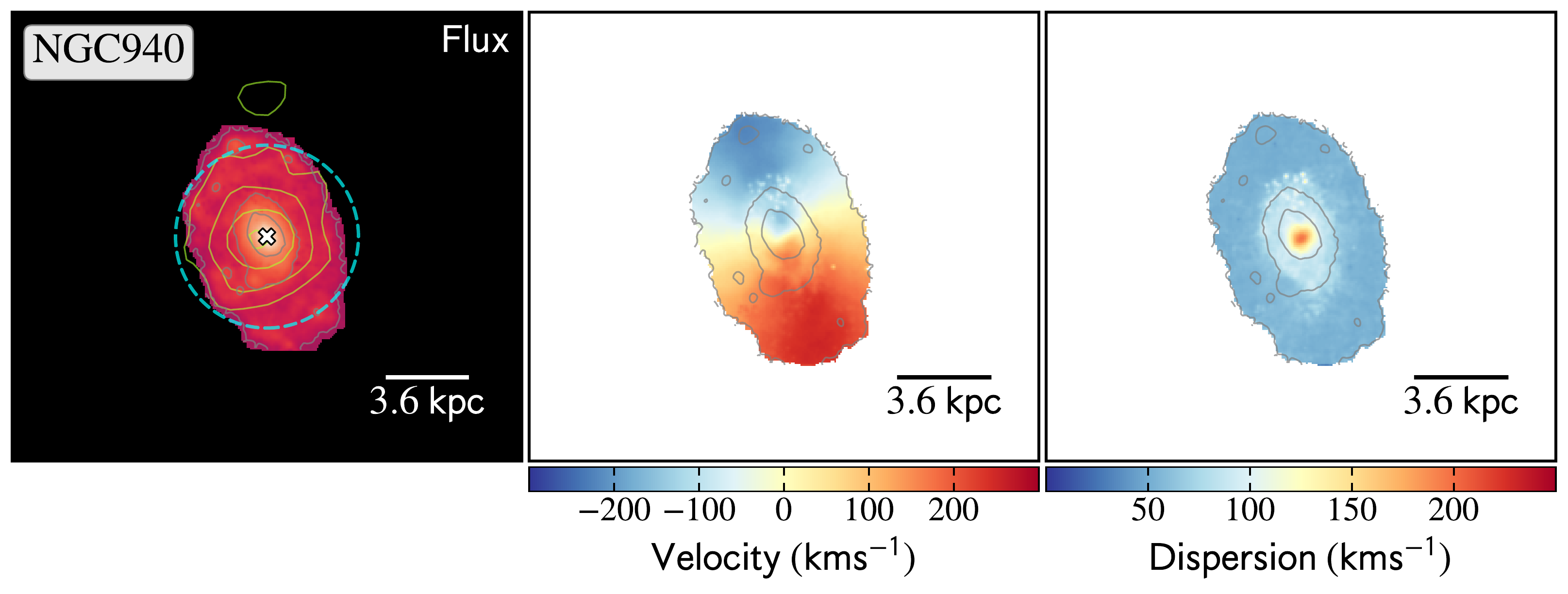}}\\
        \vspace{-0.5cm}
        \subfigure{\includegraphics[width=\setwidthsmall\textwidth]{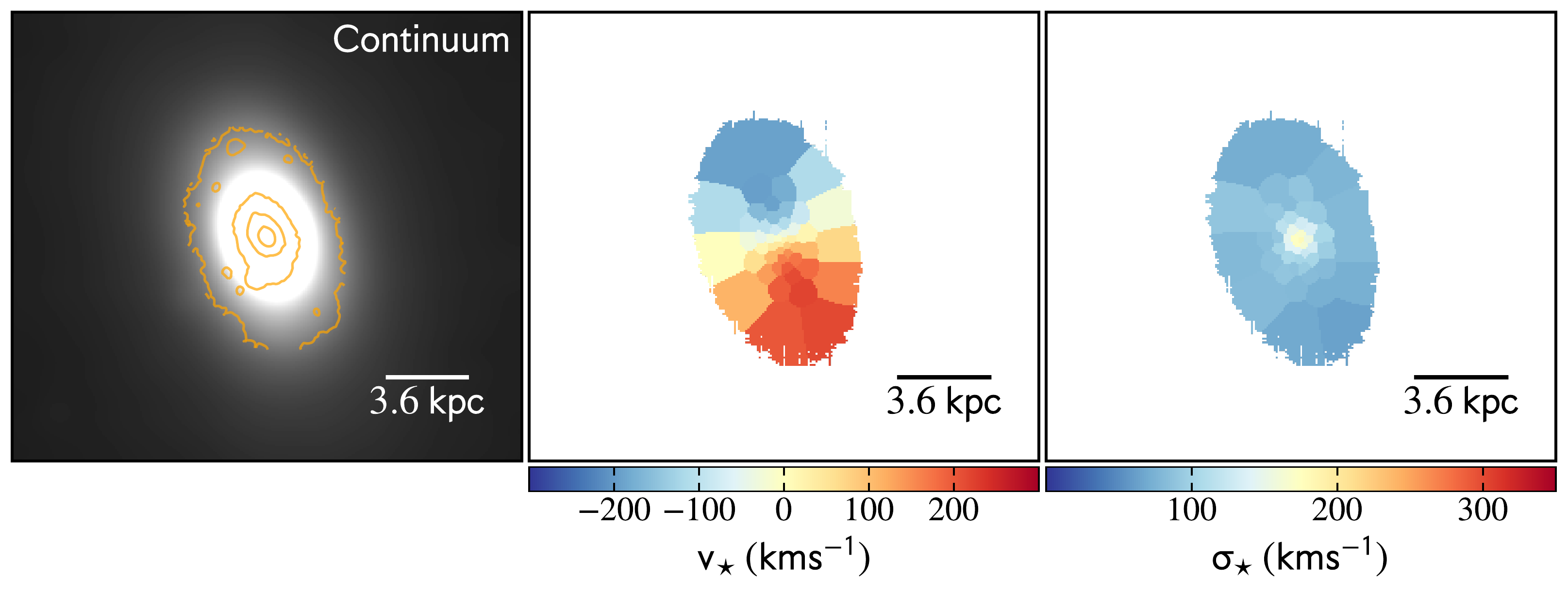}}\\
      
        \caption{Continuation of Fig.~\ref{fig:nii_maps_examples} (see text for details).} \label{fig:nii_maps_examples_app}
\end{figure*}
    
\begin{figure*}[htbp!]
    \ContinuedFloat
    \captionsetup{list=off,format=cont}
    \centering
        \subfigure{\includegraphics[width=\setwidthsmall\textwidth]{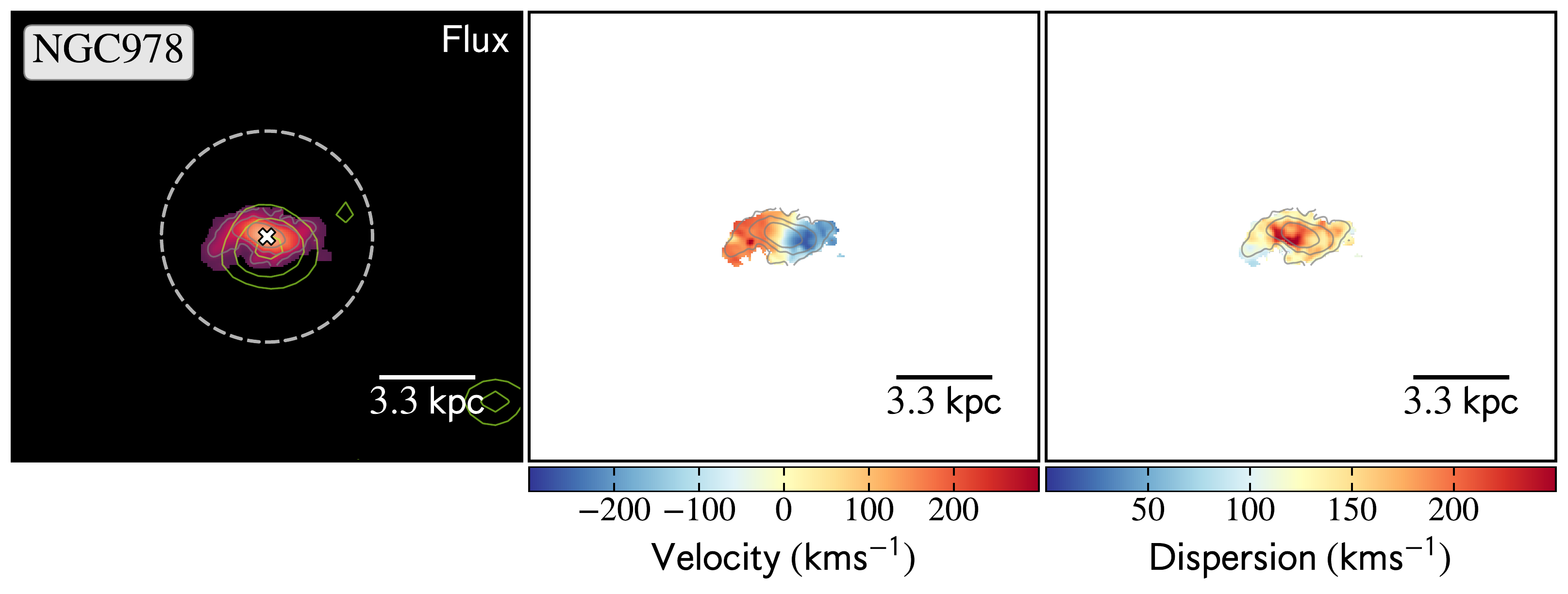}}\\
        \vspace{-0.5cm}
        \subfigure{\includegraphics[width=\setwidthsmall\textwidth]{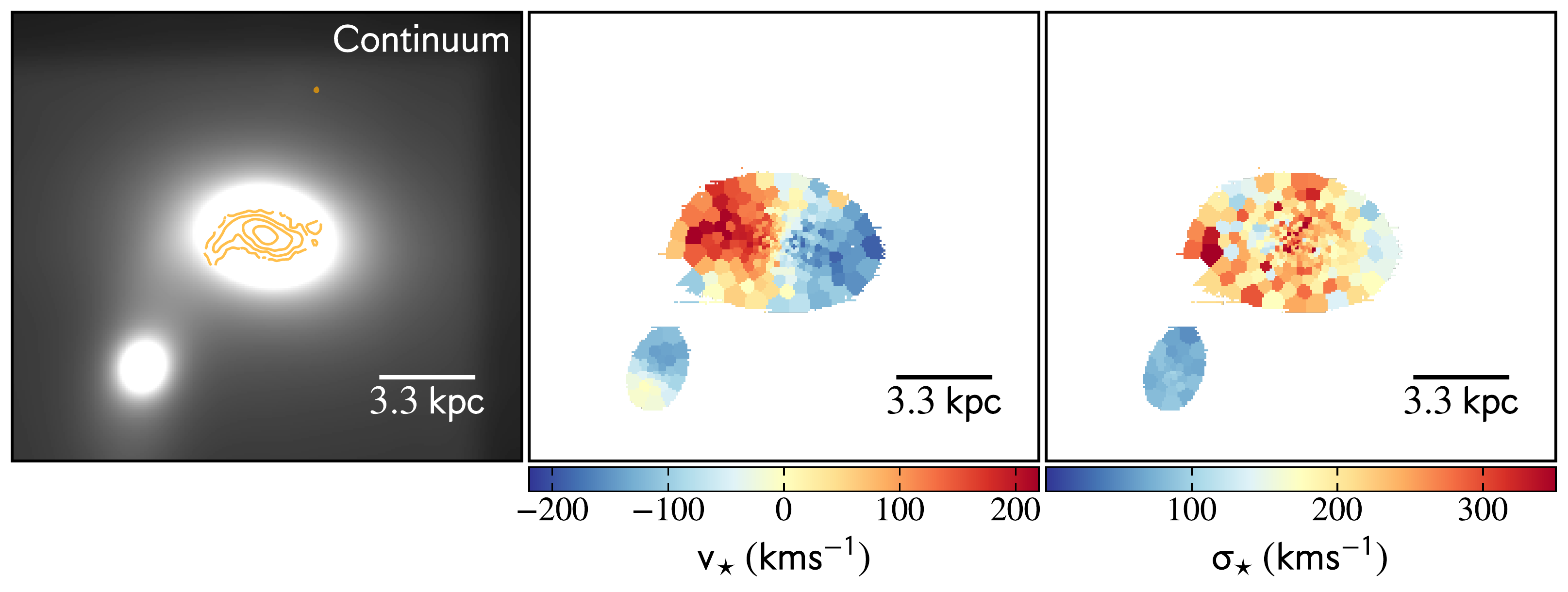}}\\
        \subfigure{\includegraphics[width=\setwidthsmall\textwidth]{Figures/Optical_maps/NGC1453_new.pdf}}\\
        \vspace{-0.5cm}
        \subfigure{\includegraphics[width=\setwidthsmall\textwidth]{Figures/Optical_maps/NGC1453_stellar_new.pdf}}\\
        \caption{Continuation of Fig.~\ref{fig:nii_maps_examples} (see text for details).}
       % \vfill        
      %  \vspace{2cm}
\end{figure*}

\begin{figure*}[htbp!]
\ContinuedFloat
\captionsetup{list=off,format=cont}
\centering
        \subfigure{\includegraphics[width=\setwidthsmall\textwidth]{Figures/Optical_maps/NGC4261_new.pdf}}\\
        \vspace{-0.5cm}
        \subfigure{\includegraphics[width=\setwidthsmall\textwidth]{Figures/Optical_maps/NGC4261_stellar_new.pdf}}\\

        \subfigure{\includegraphics[width=\setwidthsmall\textwidth]{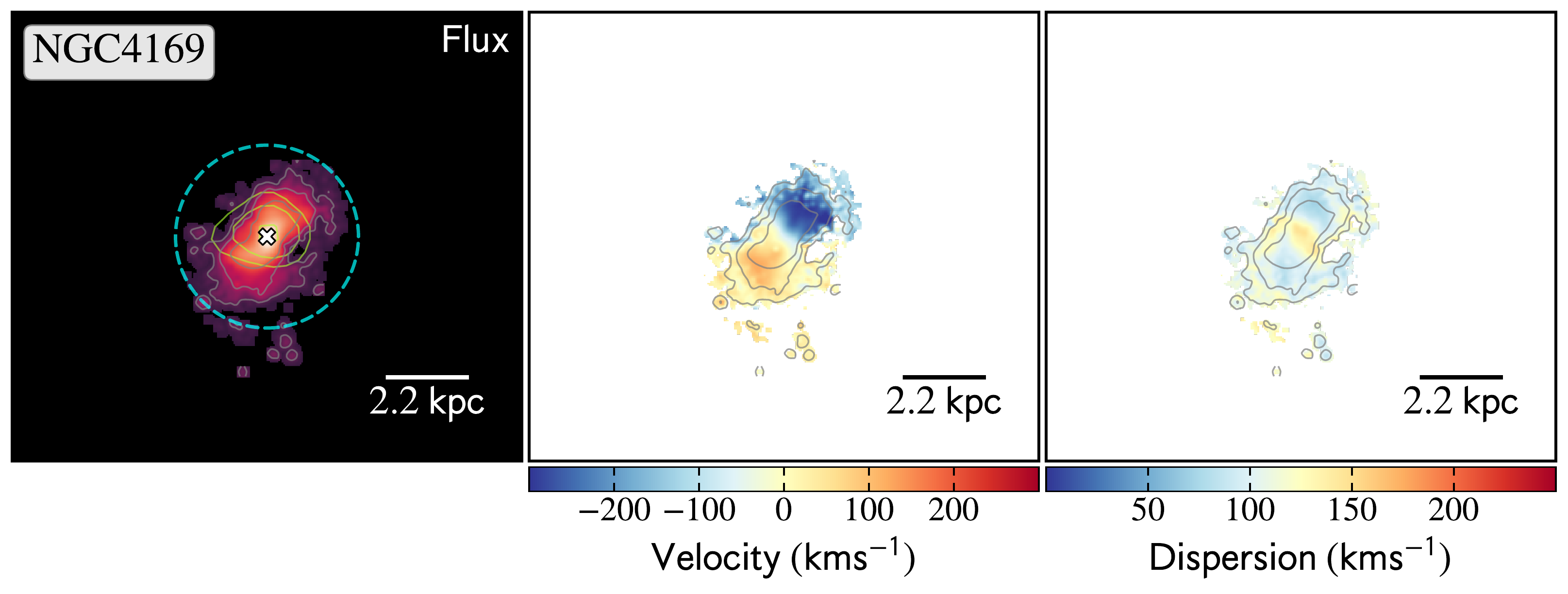}}\\
        \vspace{-0.5cm}
        \subfigure{\includegraphics[width=\setwidthsmall\textwidth]{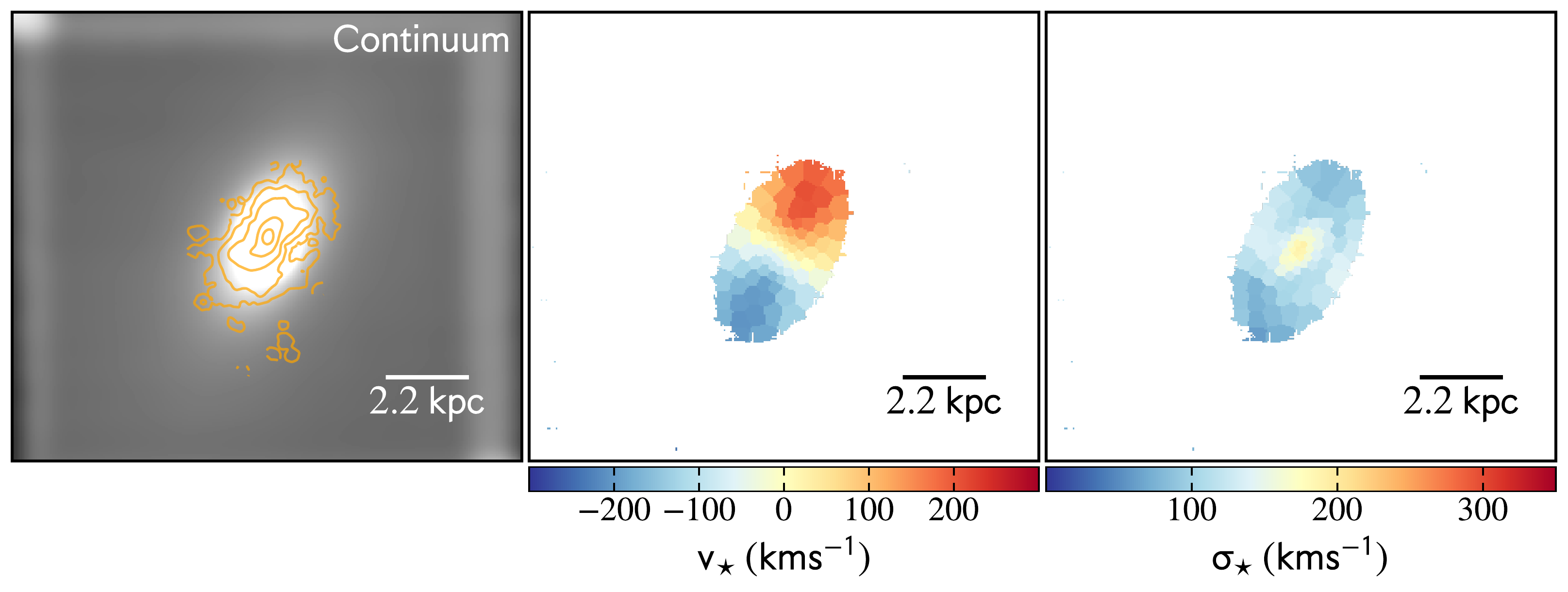}}\\
        \caption{Continuation of Fig.~\ref{fig:nii_maps_examples} (see text for details).}
\end{figure*}

\begin{figure*}[htbp!]
\ContinuedFloat
\captionsetup{list=off,format=cont}
\centering

        \subfigure{\includegraphics[width=\setwidthsmall\textwidth]{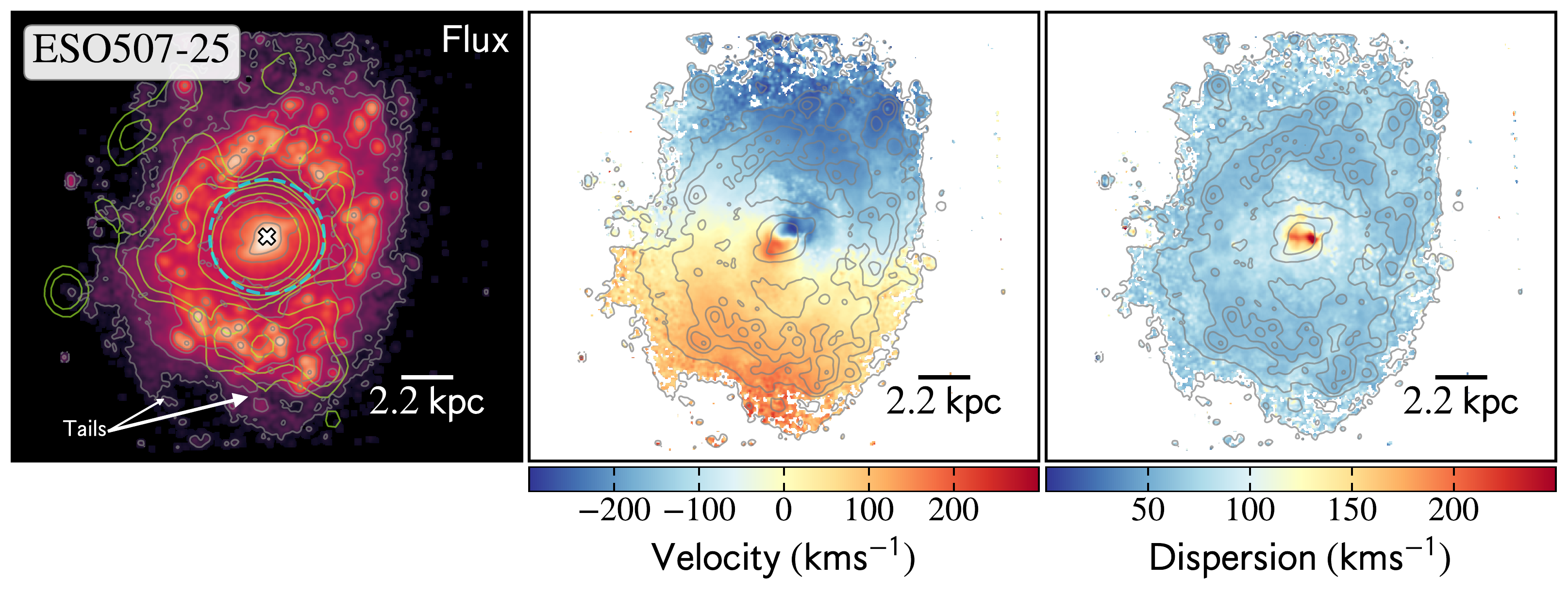}}\\
      \vspace{-0.5cm}
        \subfigure{\includegraphics[width=\setwidthsmall\textwidth]{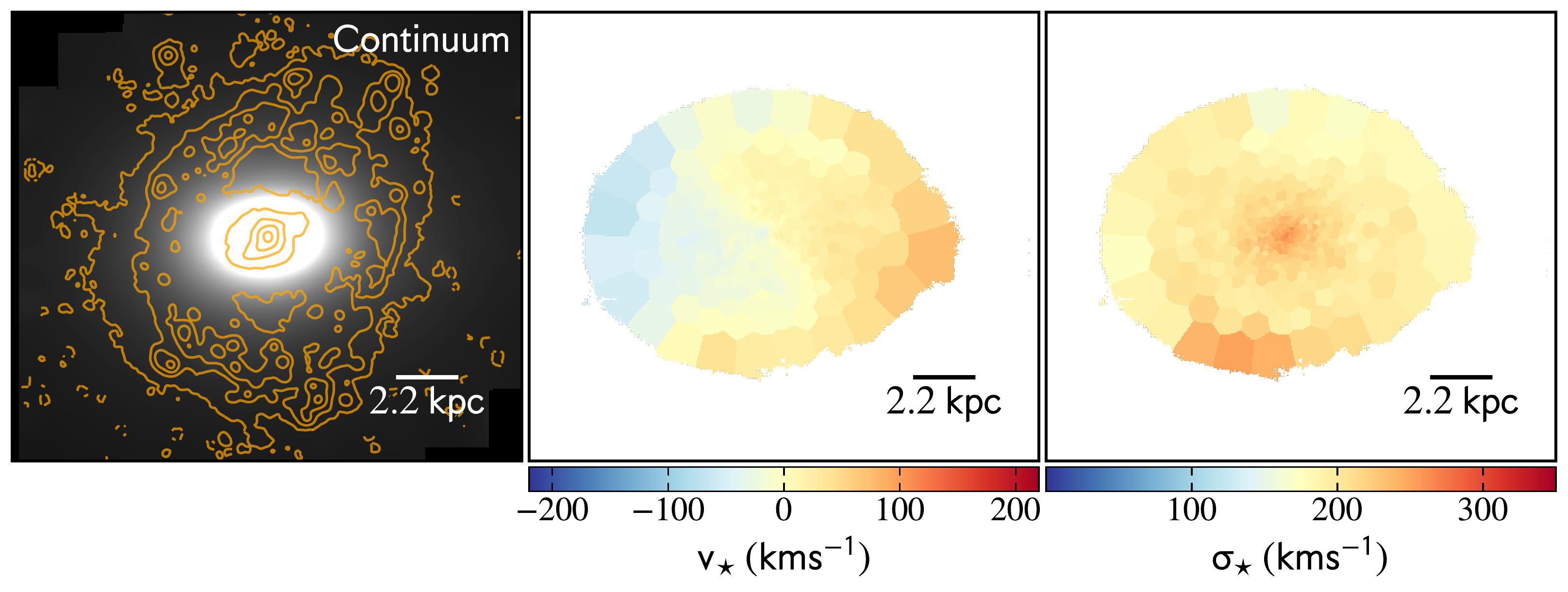}}\\

        \caption{Continuation of Fig.~\ref{fig:nii_maps_examples} (see text for details).}
\end{figure*}

% North right, east left 
{\noindent\it NGC\,924 --} NGC\,924 shows a more complex kinematics. Ordered velocity structures are found along the inner 9~kpc ($\sim$30$\arcsec$) of the galaxy with a velocity range from $-$260 to $+$110~km~s$^{-1}$ from SE to NW. Around the rotating disk, a clumpy outer ring of gas is moving from $+$170 to $-$280~km~s$^{-1}$ located at $\sim$7.5~kpc (25$\arcsec$) of the central disk. A few clumps located SE and NW of the disk are inflowing/outflowing of the inner disk with a velocity of $-$270~km~s$^{-1}$ and $+$120~km~s$^{-1}$, respectively.\\

{\noindent\it NGC\,940 --} A smooth rotating disk of gas is detected centered on the AGN in NGC\,940, with the gas moving from $-$300 to $+$170~km~s$^{-1}$ from NW to SE across $\sim$11~kpc. An inner nuclear disk of $\sim$2.5~kpc (7$\arcsec$) is located in the inner region of the galaxy with a slightly different PA that the main extended rotating disk. \\

{\noindent\it NGC\,978 --} The gas distribution in NGC\,978 is a very compact. NGC\,978 has a smooth velocity gradient from $-$210 to $+$180~km~s$^{-1}$ in the EW direction along 3.6~kpc ($\sim$11$\arcsec$). Small tails of ionized gas are detected to the SW and NE of the central disk following the kinematics of the disk. \\% (see Fig.~\ref{fig:ngc4261_hst}).\\ checked

{\noindent\it NGC\,1453 --} Similarly, NGC\,1453 shows a smooth velocity gradient from $+$300 to $-$360~km~s$^{-1}$ from North to South across 21~kpc. An inner compact disk of $\sim$3~kpc (9.5$\arcsec$) of diameter, with a slightly different PA from that of the major disk, is found at the very center of the galaxy. Extended structures are seen in NGC\,1453 located $\sim$9~kpc (30$\arcsec$) at the {NW} the center of the galaxy, moving at $+$250~km~s$^{-1}$.\\ %

{\noindent\it NGC\,4261 --} NGC\,4261 reveals a very compact gas distributions. A compact rotating disk is detected in NGC\,4261, with a velocity gradient from $-$150 up to $+$70~km~s$^{-1}$ across 0.5~kpc ($\sim$3$\arcsec$) along the SN axis (minor-axis). Additionally, some blueshifted gas is moving at $-$80~km~s$^{-1}$, towards the E and W directions from the galaxy center across $\sim$1~kpc. HST images reveal a dusty torus around the inner rotating disk of ionized gas (see Fig.~\ref{fig:ngc4261_hst}).\\ % (see Fig.~\ref{fig:ngc4261_hst}).\\

{\noindent\it NGC\,4169 --} The velocity field of NGC\,4169 is moving from $+$150 to $-$280~km~s$^{-1}$ going from {SE} to {NW} across $\sim$4.8~kpc of extension, with some clumpy structure at 3~kpc (15$\arcsec$) of the SE center. It also hosts some extended structures located $\sim$2.8~kpc (13$\arcsec$) to the {SW} of the galaxy center, moving at $+$30~km~s$^{-1}$.\\

{\noindent\it ESO\,507-25 --} This galaxy shows a nuclear rotating disk centered on the galaxy, with a velocity gradient from $+$150 to $-$250~km~s$^{-1}$ from the SE to the NW extending along $\sim$2.2~kpc ($\sim$10.2$\arcsec$). A ring structure is found misaligned with the inner disk, which exhibits a smooth velocity gradient from $-$260 to $+$210~km~s$^{-1}$ from the SE to the NW across $\sim$15~kpc of extension. Also, two tails of ionized gas are extending out of the ring to the {SW} moving at $\sim$160~km~s$^{-1}$ and 90~km~s$^{-1}$.\\

\FloatBarrier

\subsection{Compact and Extended Filaments}
\begin{figure*}[hbtp!]
\ContinuedFloat
\captionsetup{list=off,format=cont}
        {\bf Filaments}\\
        \centering
        \subfigure{\includegraphics[width=\setwidthsmall\textwidth]{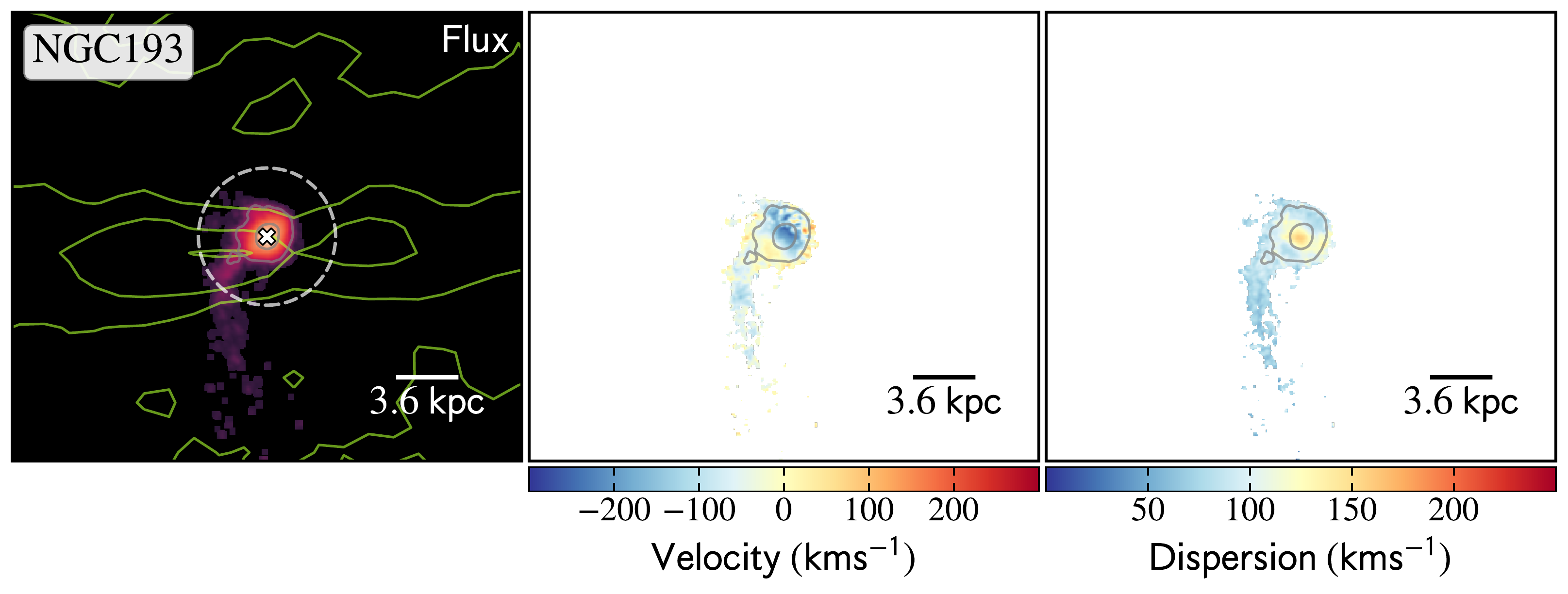}}\\
        \vspace{-0.5cm}
        \subfigure{\includegraphics[width=\setwidthsmall\textwidth]{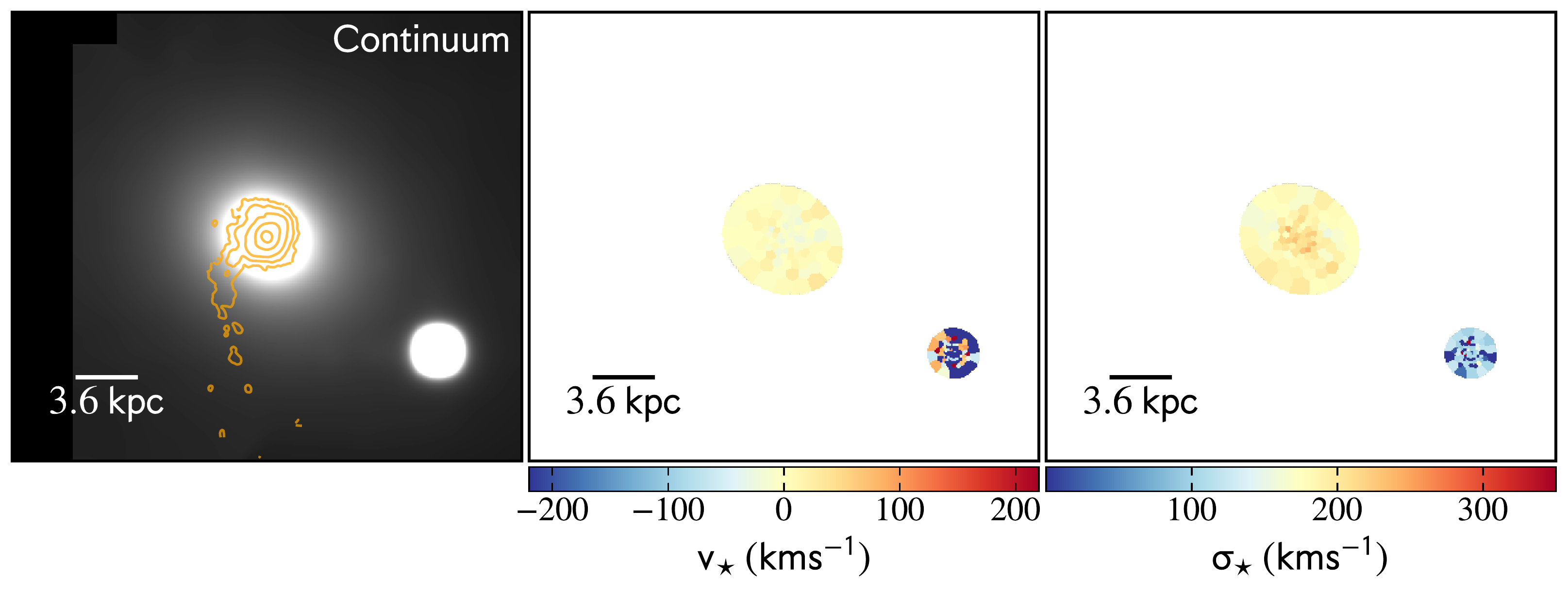}}\\
        \subfigure{\includegraphics[width=\setwidthsmall\textwidth]{Figures/Optical_maps/NGC410_new.pdf}}\\
        \vspace{-0.5cm}
        \subfigure{\includegraphics[width=\setwidthsmall\textwidth]{Figures/Optical_maps/NGC410_stellar_new.pdf}}\\
        \caption{Continuation of Fig.~\ref{fig:nii_maps_examples_app} (see text for details).}
    %    \vspace{1.5cm}
\end{figure*}

\begin{figure*}[hbtp!]
\ContinuedFloat
\captionsetup{list=off,format=cont}
      %  {\bf Filaments}\\
        \centering
        \subfigure{\includegraphics[width=\setwidthsmall\textwidth]{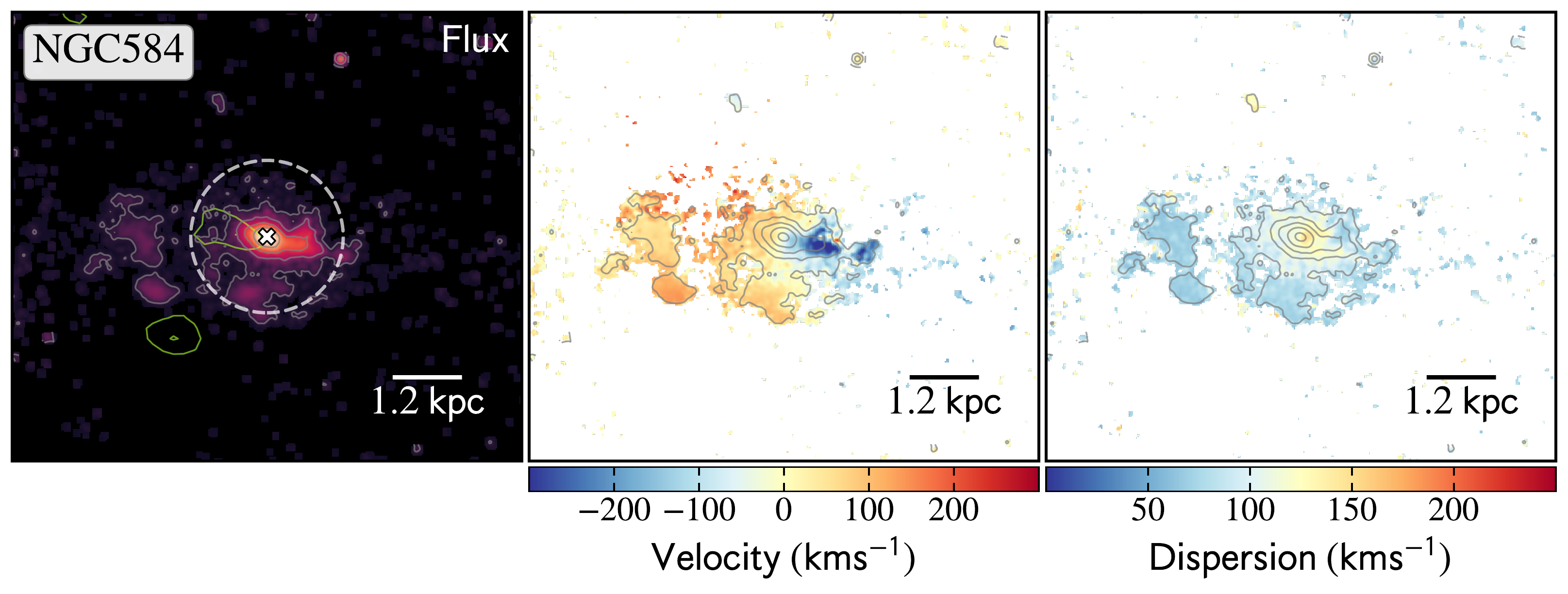}}\\
        \vspace{-0.5cm}
        \subfigure{\includegraphics[width=\setwidthsmall\textwidth]{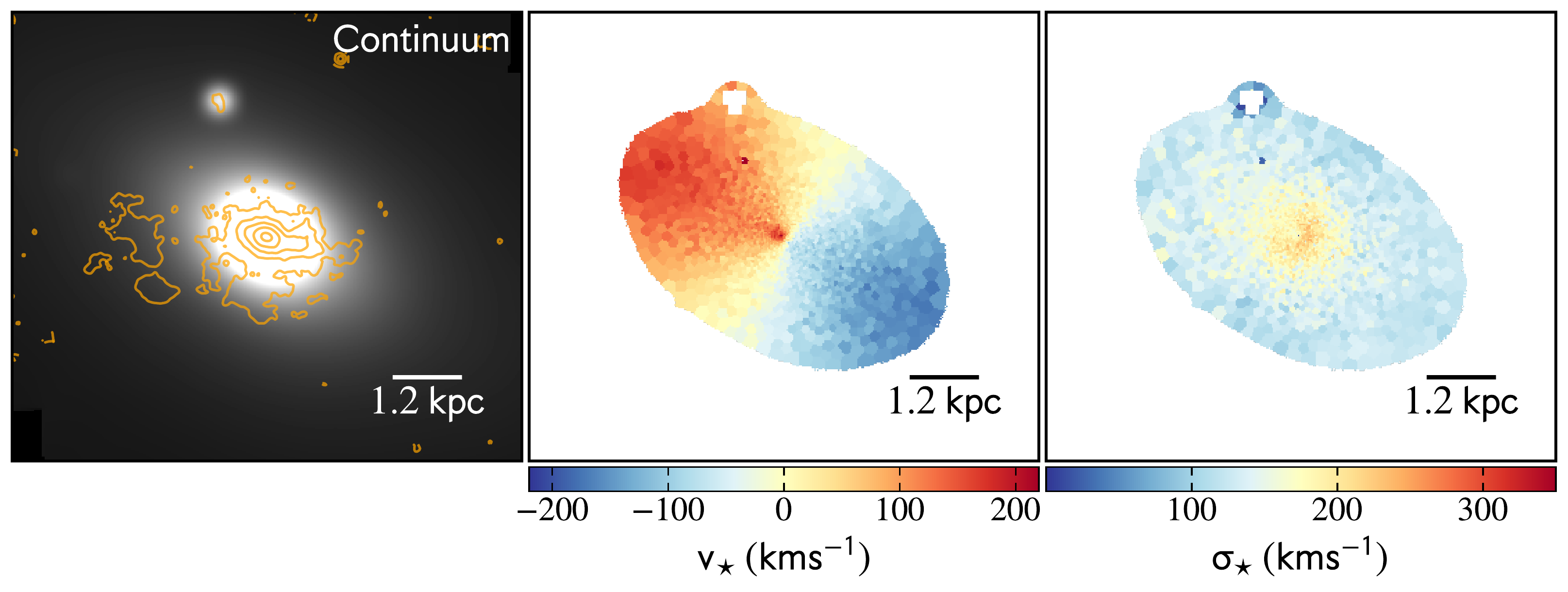}}\\

        \subfigure{\includegraphics[width=\setwidthsmall\textwidth]{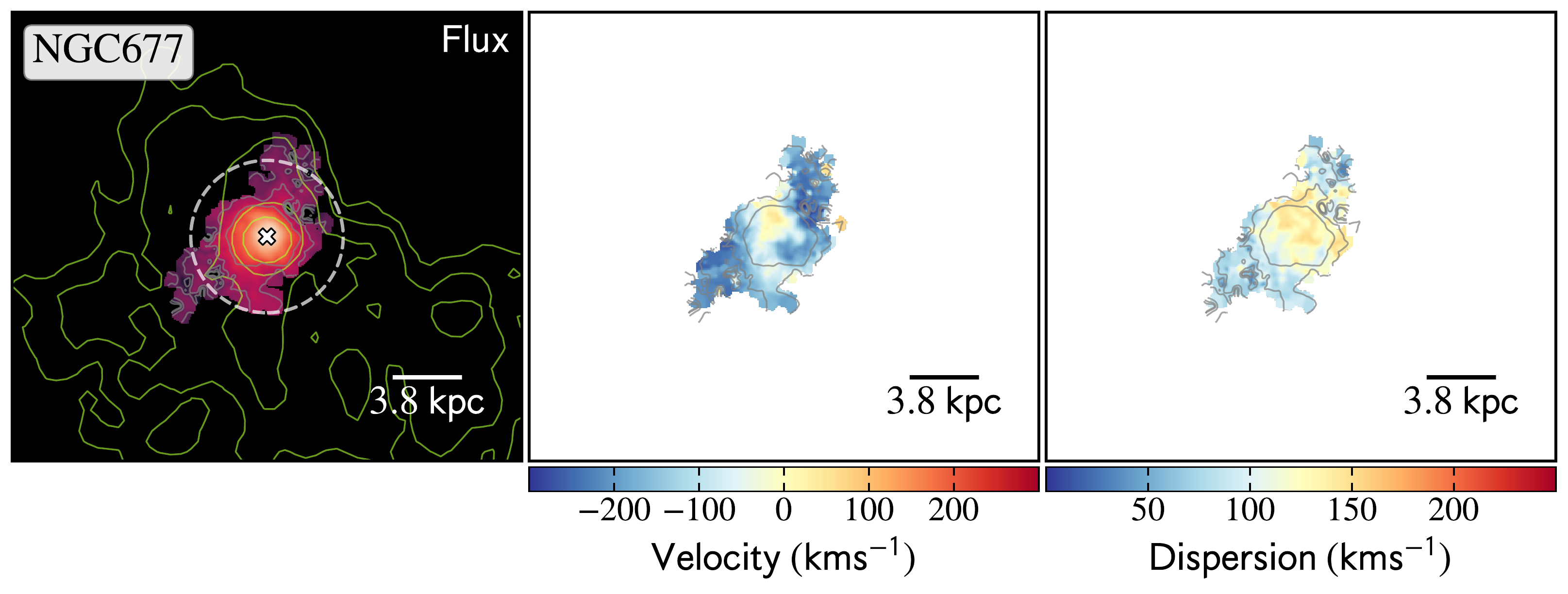}}\\
        \vspace{-0.5cm}
        \subfigure{\includegraphics[width=\setwidthsmall\textwidth]{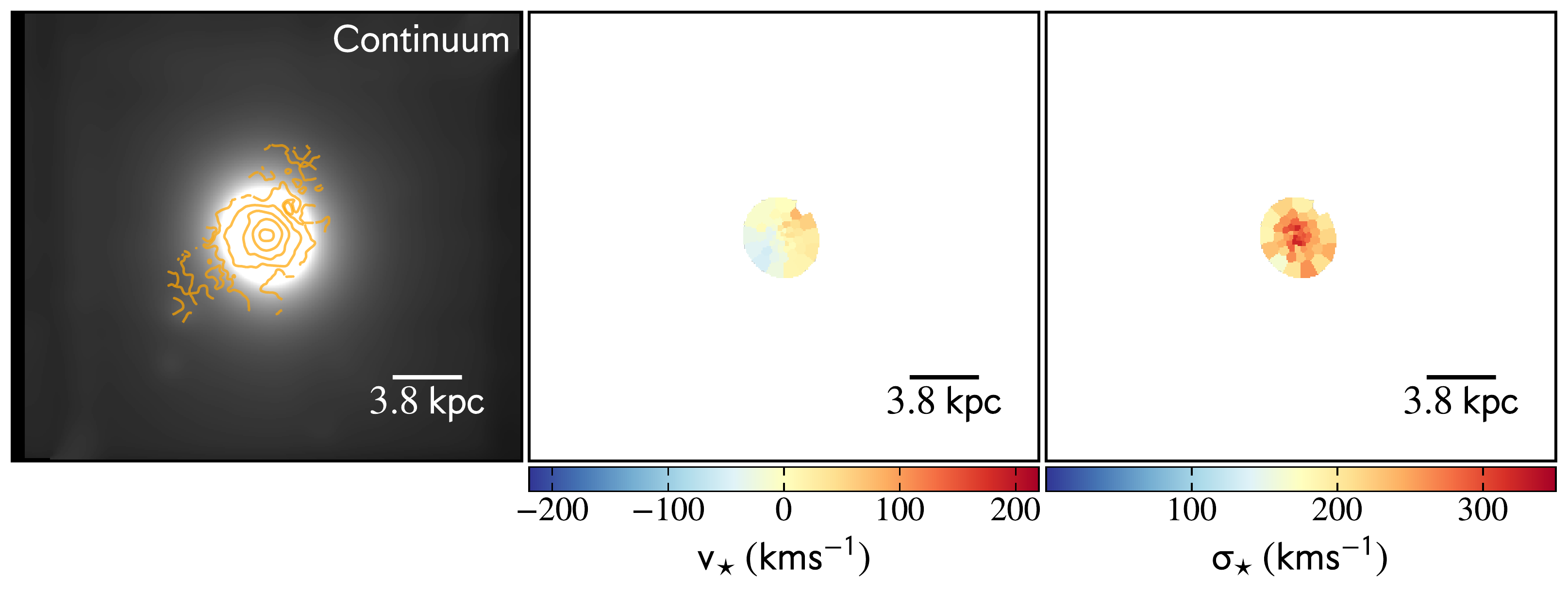}}\\
        \caption{Continuation of Fig.~\ref{fig:nii_maps_examples_app} (see text for details).}
\end{figure*}

\begin{figure*}[hbtp!]
\ContinuedFloat
\captionsetup{list=off,format=cont}
        \centering
        \subfigure{\includegraphics[width=\setwidthsmall\textwidth]{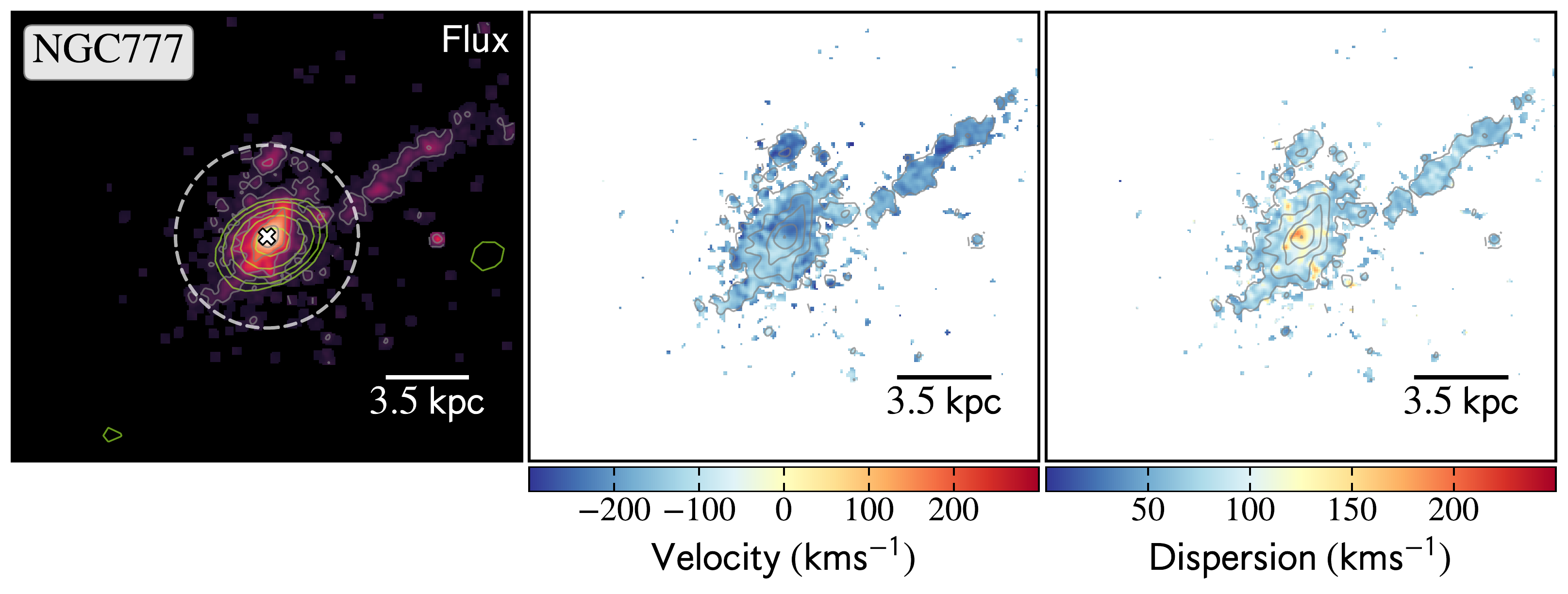}}\\
        \vspace{-0.5cm}
        \subfigure{\includegraphics[width=\setwidthsmall\textwidth]{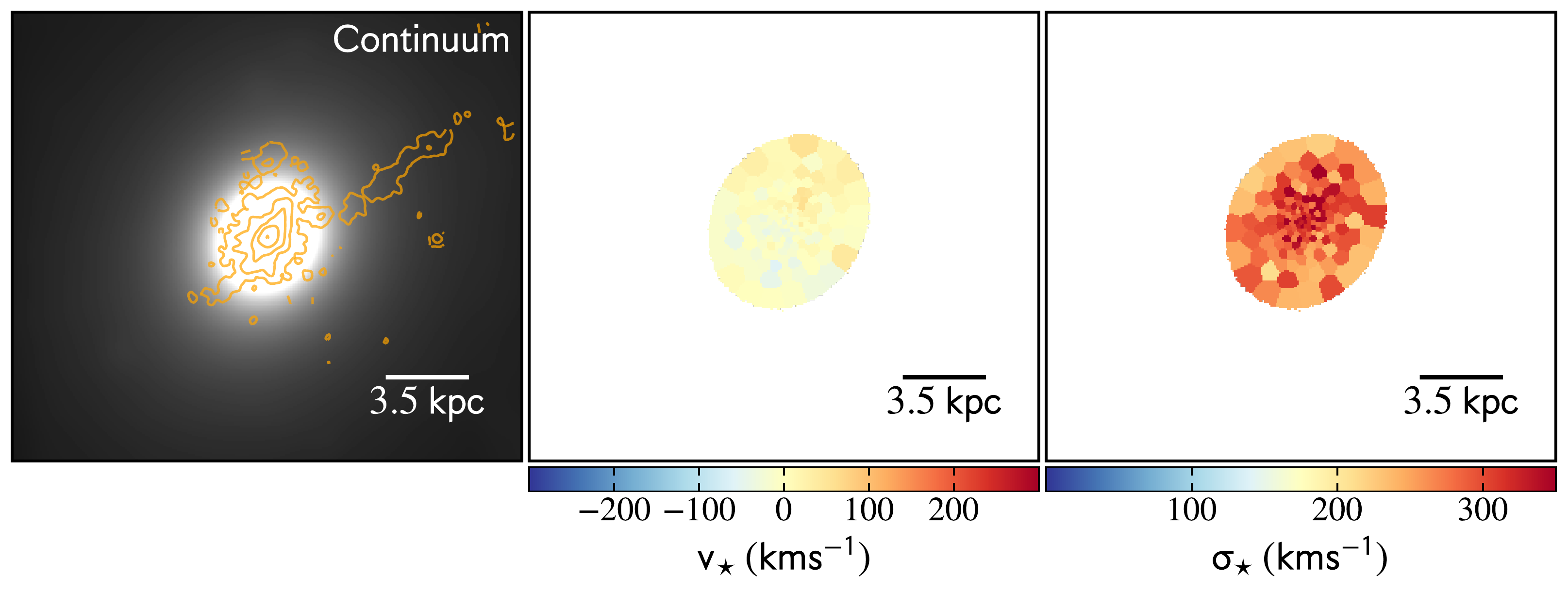}}\\

        \subfigure{\includegraphics[width=\setwidthsmall\textwidth]{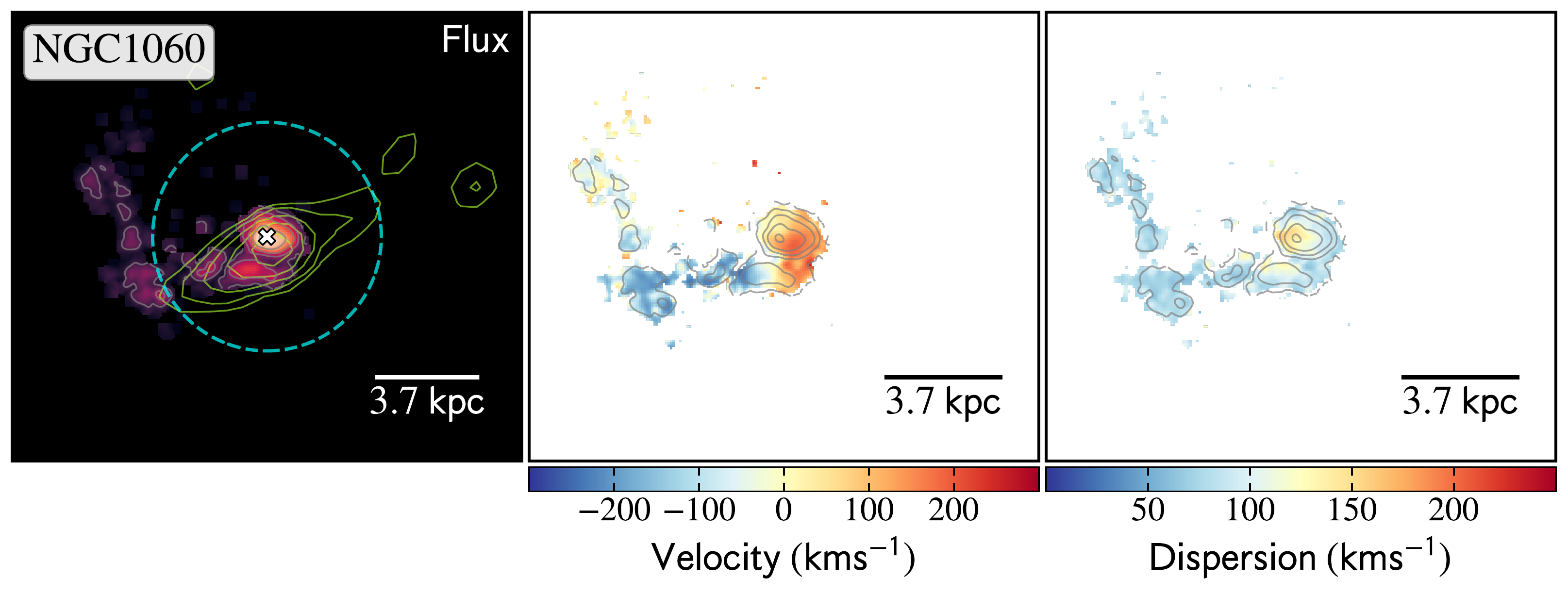}}\\
        \vspace{-0.5cm}
        \subfigure{\includegraphics[width=\setwidthsmall\textwidth]{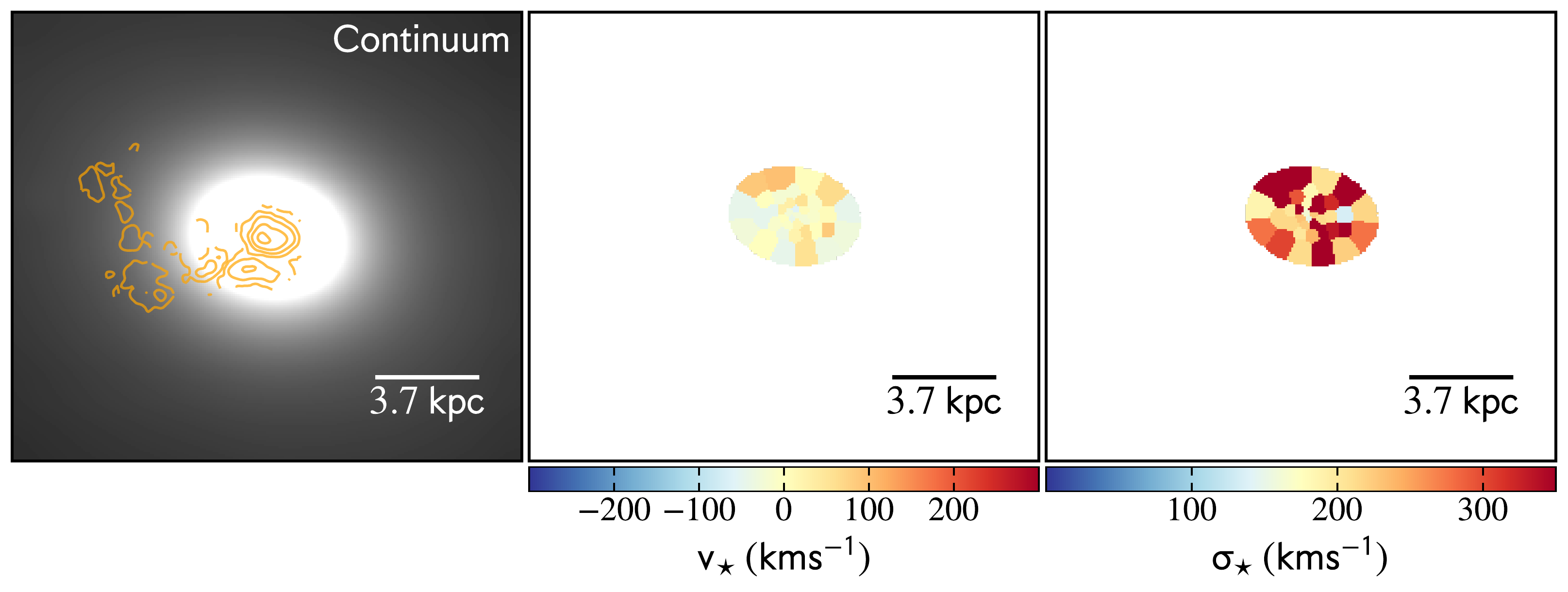}}\\

        \caption{Continuation of Fig.~\ref{fig:nii_maps_examples} (see text for details).}
      %  \vspace{1.5cm}
\end{figure*}

\begin{figure*}[hbtp!]
\ContinuedFloat
\captionsetup{list=off,format=cont}
        \centering
         \subfigure{\includegraphics[width=\setwidthsmall\textwidth]{Figures/Optical_maps/NGC1587_new.pdf}}\\
        \vspace{-0.5cm}
        \subfigure{\includegraphics[width=\setwidthsmall\textwidth]{Figures/Optical_maps/NGC1587_stellar_new.pdf}}\\
         \subfigure{\includegraphics[width=\setwidthsmall\textwidth]{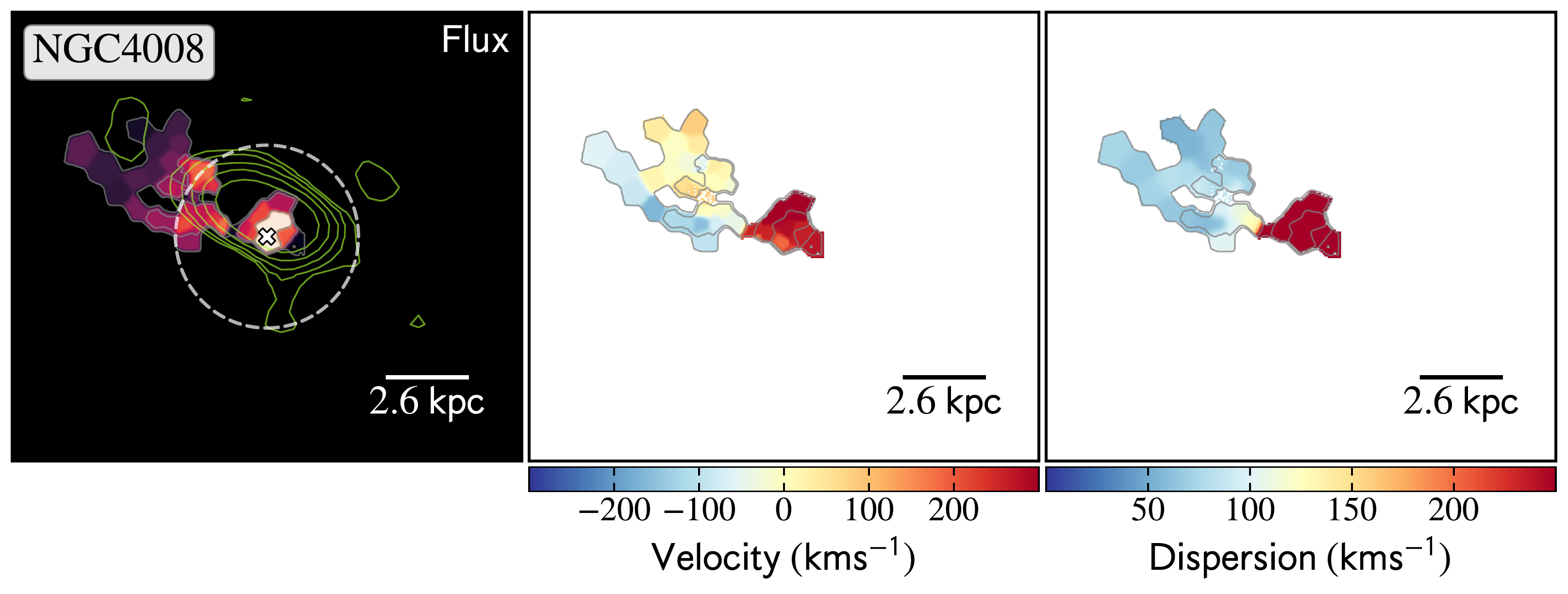}}\\
        \vspace{-0.5cm}
        \subfigure{\includegraphics[width=\setwidthsmall\textwidth]{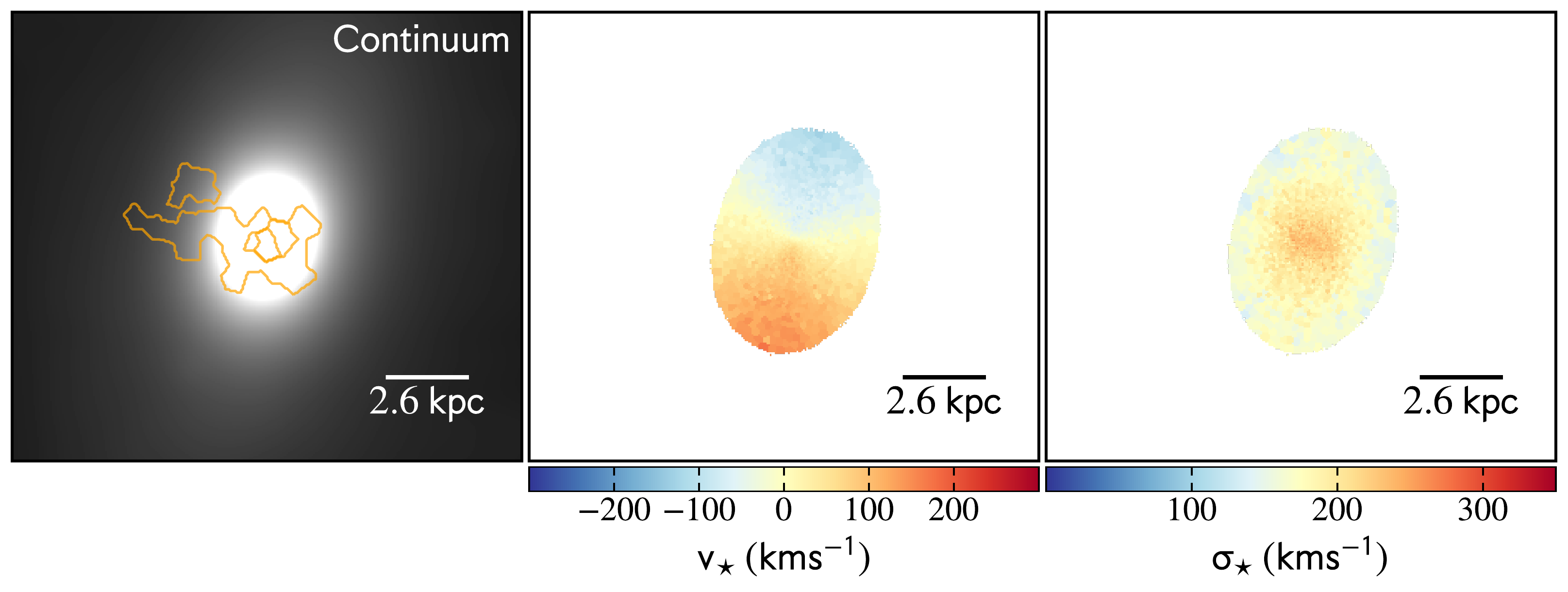}}\\
        \caption{Continuation of Fig.~\ref{fig:nii_maps_examples} (see text for details).}
\end{figure*}

\begin{figure*}[hbtp!]
\ContinuedFloat
\captionsetup{list=off,format=cont}
    \centering
        \subfigure{\includegraphics[width=\setwidthsmall\textwidth]{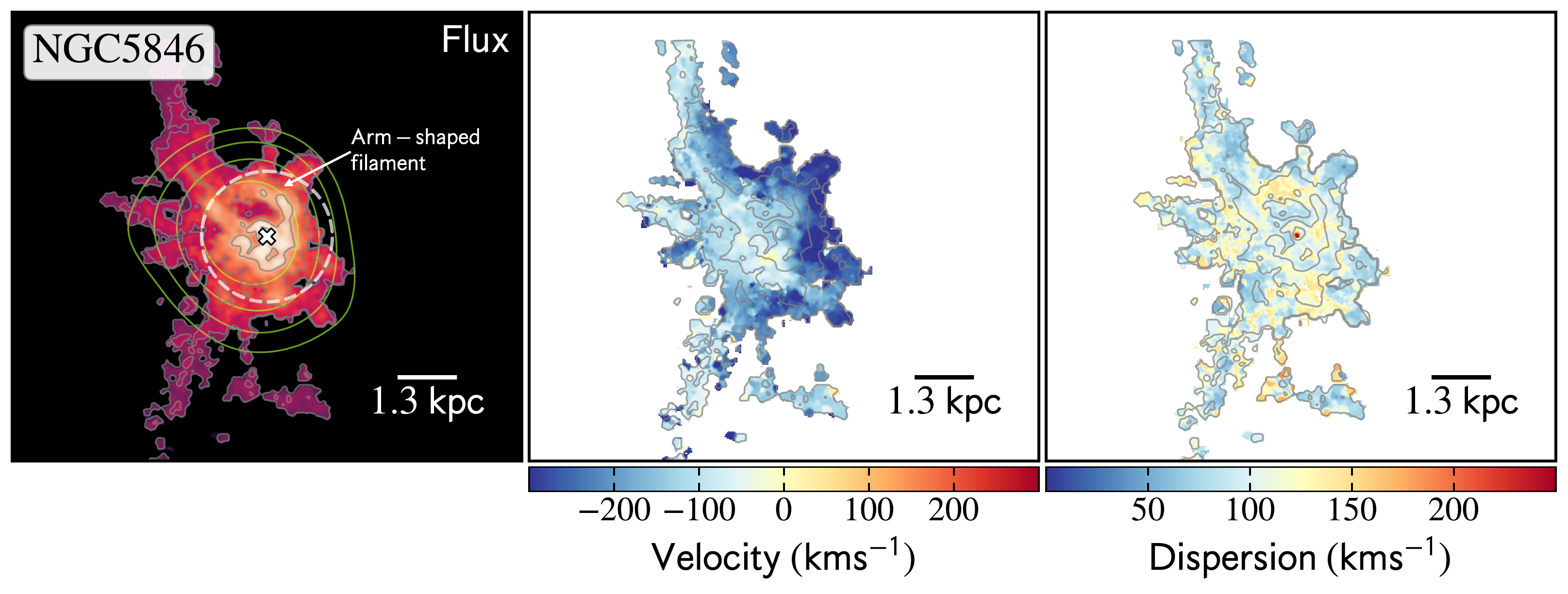}}\\

        \subfigure{\includegraphics[width=\setwidthsmall\textwidth]{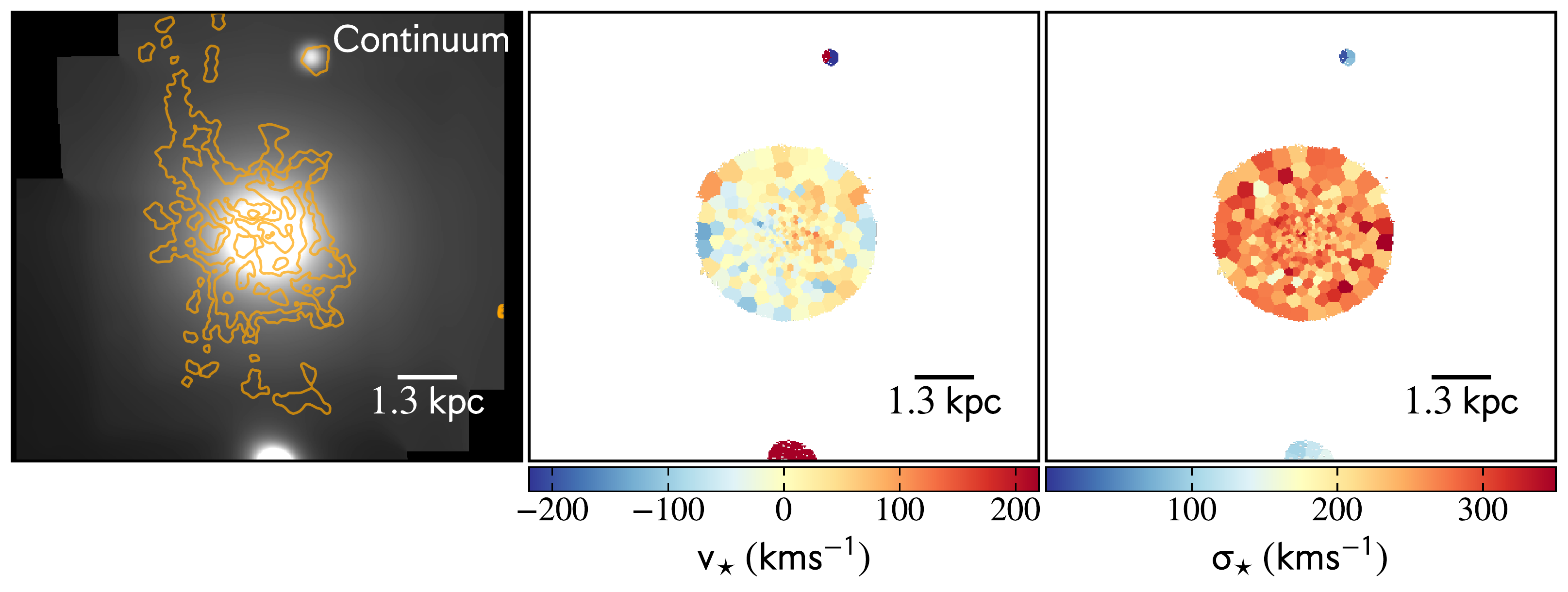}}\\

        \subfigure{\includegraphics[width=\setwidthsmall\textwidth]{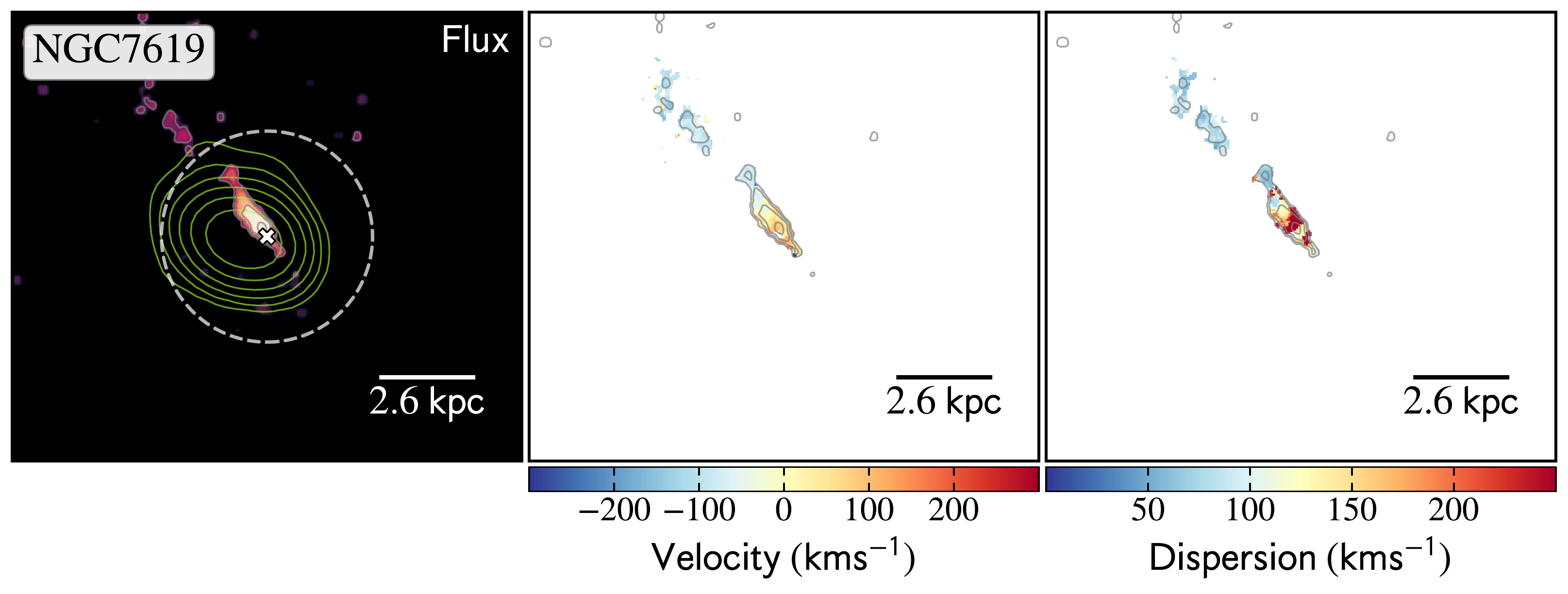}}\\
        \vspace{-0.5cm}
        \subfigure{\includegraphics[width=\setwidthsmall\textwidth]{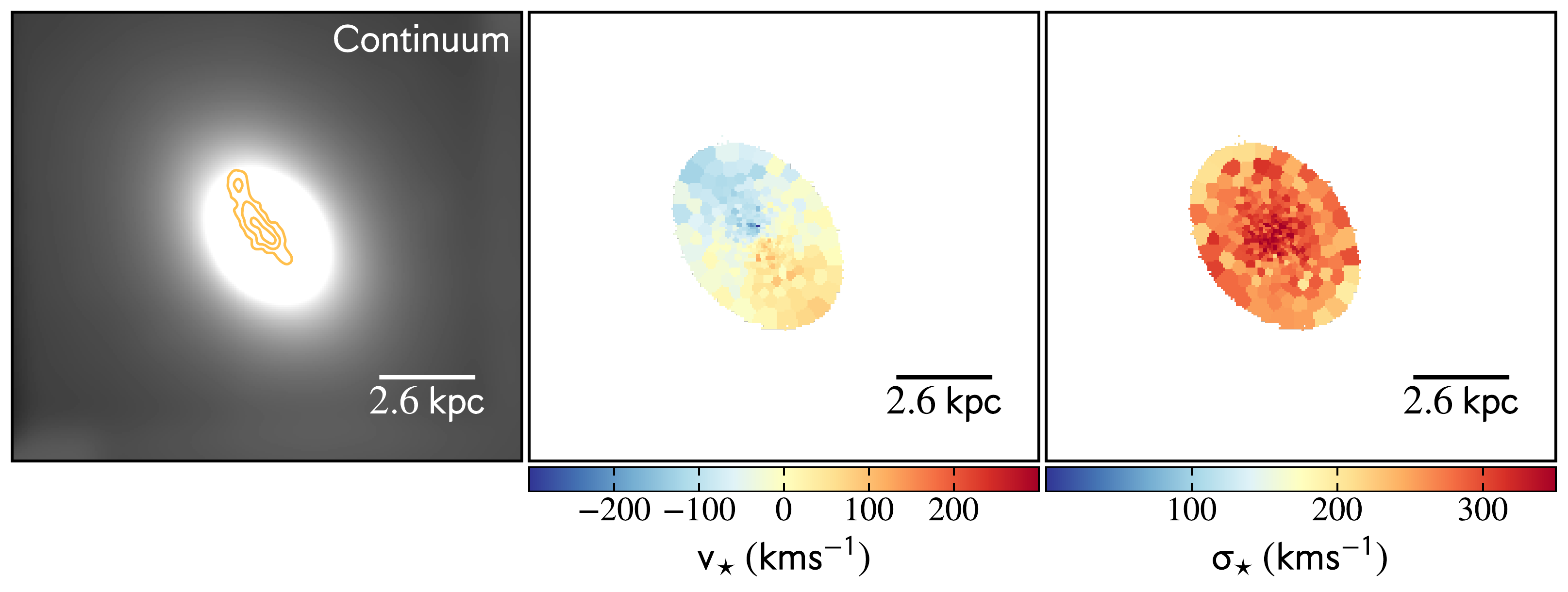}}

        \caption{Continuation of Fig.~\ref{fig:nii_maps_examples} (see text for details).}
       % \vspace{2cm}
\end{figure*}

{\noindent\it NGC\,193 --} In the central region of NGC\,193, we note signs of gas rotation, with a velocity gradient going from $-$300 to $+$30~km~s$^{-1}$ across 3.5~kpc ($\sim$10$\arcsec$). This inner region is followed by a blueshifted velocity stream of ionized gas extending from $-$60 to $-$120~km~s$^{-1}$ towards the south direction of the galaxy. An X-ray cavity has been detected at the location of the SW filament from X-ray observations \citep{osullivan17}. \\

{\noindent\it NGC\,410 --} hosts a stream of ionized gas with a smooth gradient going from {NE to SW} with a velocity from $+$50 to $-$400~km~s$^{-1}$. \\

{\noindent\it NGC\,584 --} has a velocity field slightly dominated by rotation, with a velocity gradient from $+$150 to $-$300~km~s$^{-1}$ in the WE direction along with $\sim$5~kpc (50$\arcsec$) in extension. However, the distribution of the ionized gas is quite perturbed and clumpy, with a few radial filaments.\\

{\noindent\it NGC\,677 --} host complex set of filaments, with several clumpy structures at the center. A set of 5.6~kpc (15$\arcsec$) overlapping filaments is found at southwest of the center that appears to be spiraling moving with a velocity of $-$180~km~s$^{-1}$ and $-$270~km~s$^{-1}$, and to the southeast with velocities of $-$150~km~s$^{-1}$. A faint 9.5~kpc(25$\arcsec$) clumpy structures are seen at northeast of the galaxy center with similar velocities. \\

{\noindent\it NGC\,777 --} there is indication of rotation along the {SE--NW} axis at the center, with a velocity gradient between $-$90~km~s$^{-1}$ to $-250$~km~s$^{-1}$. The peak is followed by a 7.5~kpc (21$\arcsec$) NW clumpy filament moving at about $-$170~km~s$^{-1}$, and a NE filament of 1.8~kpc (5$\arcsec$) on size filament moving with a velocity of $-$270~km~s$^{-1}$.\\

{\noindent\it NGC\,1060 --} Similarly, NGC\,1060 have a smooth velocity gradient along its filament length. This clumpy ionized filament is moving to the EW with a velocity from $+$180~km~s$^{-1}$ at the peak of the emission to $-$130~km~s$^{-1}$, followed by another NE clumpy filament moving from $-$180 to $+$40~km~s$^{-1}$ along 3.5~kpc ($\sim$10$\arcsec$).\\

{\noindent\it NGC\,1587 --} shows indications of rotation, but this is probably due to several thin filaments that are inflowing or outflowing toward the center of the galaxy. In this source, the ionized gas is moving from $-$280 to $+$80~km~s$^{-1}$ from NW to SE across $\sim$11~kpc (44$\arcsec$) in a spiral form.\\

{\noindent\it NGC\,4008 --} reveal three faint almost parallel NE filaments. The southern filament has a size of (9$\arcsec$) and a velocity of $-$80~km~s$^{-1}$, the middle (3.5$\arcsec$) and velocity of 20~km~s$^{-1}$, finally the northern filament has a length (5$\arcsec$) and a velocity of $-100$~km~s$^{-1}$.\\

{\noindent\it NGC\,5846 --} NGC hosts a very complex net of clumpy filaments, and most of them are connected to the center. The clumpy ``arm-shaped'' filament follows the distribution of the knotty-rimmed X-ray cavities with a velocity $-$120--$-$140~km~s$^{-1}$ along its (11$\arcsec$) length. A very elongated (40$\arcsec$) parallel net of clumpy filament is detected to the Northeast of the center moving at different velocities between $+$20~km~s$^{-1}$~km~s$^{-1}$, $-$100~km~s$^{-1}$, and $-$170~km~s$^{-1}$. There are some filaments coming out to the southeast of the galaxy center, in particular a very elongated filament (40$\arcsec$) has a velocity of $-$80~km~s$^{-1}$ to $+$20~km~s$^{-1}$.\\

{\noindent\it NGC\,7619 --} host an elongated (12$\arcsec$) filament along the NE--SW axis. It has a shallow velocity gradient of $+$20~km~s$^{-1}$--40~km~s$^{-1}$. There is a fainter filament located 3.6~kpc (14$arcsec$) NE of the center moving with a projected velocity of roughly $-$90~km~s$^{-1}$. \\

%\begin{figure*}[hbtp!]
%\ContinuedFloat
%\captionsetup{list=off,format=cont}
%\centering
        %\subfigure{\includegraphics[width=\setwidthsmall\textwidth]{Figures/Optical_maps/NGC1060_new.pdf}}\\
        %\subfigure{\includegraphics[width=\setwidthsmall\textwidth]{Figures/Optical_maps/NGC1060_new.pdf}}\\
%        \caption{Continuation of Fig.~\ref{fig:nii_maps_examples} (see text for details).}
%\end{figure*}

%\section{NGC\,193 and NGC\,5846 Chandra X-ray images}

%Figure~\ref{fig:ngc193_ngc5846_chandra} shows the X-ray Chandra images of NGC\,193 and NGC\,5846 and overlaid by with optical emitting gas, respectively.

%\begin{figure*}
%    \centering
%    \includegraphics[width=0.49\textwidth]{Figures/NGC193-CC_chandra.png}
%    \includegraphics[width=0.49\textwidth]{Figures/NGC5846-CC_chandra.png}
 %   \caption{Left panel: \textit{Chandra} image of NGC\,193 overlaid by with [NII] optical emission in green. Right panel: \textit{Chandra} image of NGC\,5846 overlaid with [NII] optical emission in green}
 %   \label{fig:ngc193_ngc5846_chandra}
%\end{figure*}

\FloatBarrier

\def\setwithhalf{0.45}

{\section{Double components}
\def\setwithhalf{0.48}
% Double Halpha emission lines from those regions.
%\section{Double components}
\begin{figure}
    \centering
    \subfigure{\includegraphics[width=\setwithhalf\textwidth]{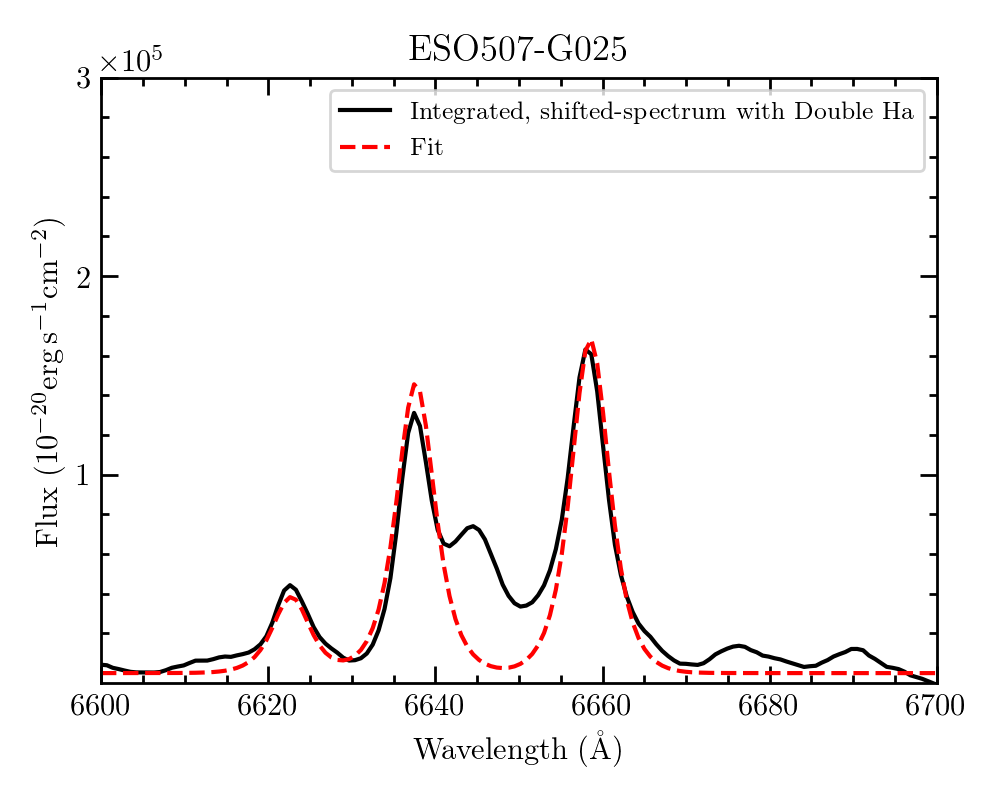}}
    \subfigure{\includegraphics[width=\setwithhalf\textwidth]{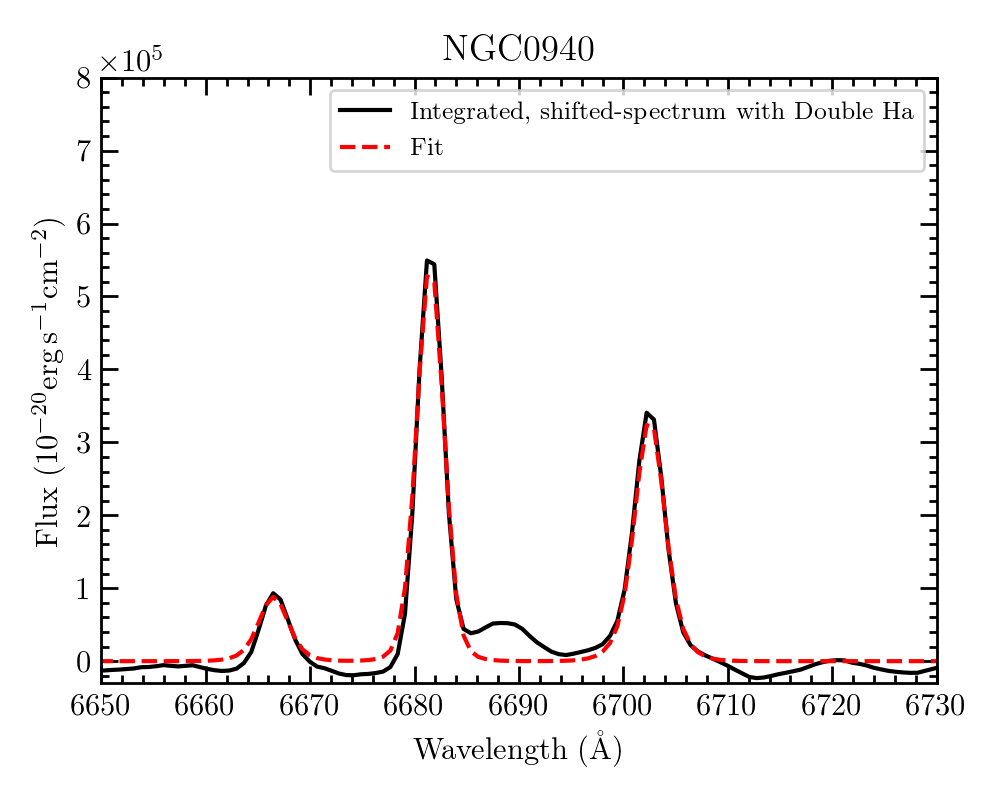}}\\
    \caption[Integrated spectrum of the double velocity component region detected in some BGGs.]{Integrated spectrum of the region that has double H$\alpha$ emission lines. {The spectra have been shifted to $v=0$ at the respective redshift of each source with the purpose of showing the double H$\alpha$ emission line.}}
    \label{fig:double_components}
\end{figure}
}

\section{Velocity dispersion of gas and stars}
\begin{table}[htb]
\caption{Velocity dispersion of the gas structures and stars.}
\centering
\scalebox{0.8}{
\begin{tabular}{lcccccc}
    \noalign{\smallskip} \hline \hline \noalign{\smallskip}
    BGG & $\sigma_{\rm gas, 0}$ & $\sigma_{\rm gas}$ & $\sigma_{\rm \star,0}$ & $\sigma_{\rm \star,mean}$ & $V_{\rm max}$ &  $V_{\rm max}/\sigma$\\
     &  ($\rm km~s^{-1}$) & ($\rm km~s^{-1}$) & ($\rm km~s^{-1}$) & ($\rm km~s^{-1}$) & ($\rm km~s^{-1}$) \\
     (1) &  (2) & (3) & (4) & (5) & (6) & (7)\\

    \noalign{\smallskip} \hline \noalign{\smallskip}
    {\it Disks}    &  &  & \\
    NGC924&330$\pm$1.5&91$\pm$2.0&288&94&203&2.17\\
    NGC940&190$\pm$1.4&55$\pm$1.0&179&81&219&2.71\\
    NGC978&250$\pm$2.0&82$\pm$4.0&401&210&227&1.08\\
    NGC1453&249$\pm$1.5&111$\pm$1.4&328&263&125&0.47\\
    NGC4169&148$\pm$1.5&104$\pm$0.6&190&112&208&1.86\\
    NGC4261&195$\pm$1.1&109$\pm$0.3&415&256&130&0.51\\
    ESO507-25&278$\pm$1.5&55$\pm$1.0&264&197&73&0.37\\

    {\it Filaments}    &  &  & \\ 
    NGC193&159$\pm$1.5&89$\pm$3.0&517&183&34&0.18\\
    NGC410&173$\pm$1.4&87$\pm$7.0&362&258&150&0.58\\
    NGC584&162$\pm$1.6&85$\pm$5.0&246&135&298&2.21\\
    NGC677&130$\pm$1.4&118$\pm$4.0&322&228&79&0.35\\
    NGC777&190$\pm$2.0&77$\pm$3.0&375&269&59&0.22\\
    NGC1060&161$\pm$2.0&84$\pm$4.0&452&299&98&0.33\\
    NGC1587&153$\pm$1.5&73$\pm$2.0&295&207&158&0.76\\
    NGC4008&249$\pm$3.0&66$\pm$10.0&258&131&172&1.31\\
    NGC5846&307$\pm$1.5&100$\pm$2.0&361&257&119&0.46\\
    NGC7619&249$\pm$3.0&76$\pm$7.0&401&266&124&0.47\\ 
    \noalign{\smallskip} \hline \noalign{\smallskip}
\end{tabular}
}
\tablefoot{
(1) Source name. \\
(2) Gas velocity dispersion measure at the central region of each source.\\
(3) Gas dispersion measure at one of the structures (e.g. filaments, disk and clumps).\\
(4) Central stellar velocity dispersion.\\
(5) Mean stellar dispersion measure from the dispersion map.\\
(6) Maximum stellar velocity measured from the stellar velocity map.\\
(7) Ratio of the maximum stellar velocity and the mean stellar dispersion.
}\\ 
\end{table}

\begin{figure}
    \centering
        \includegraphics[scale=0.7]{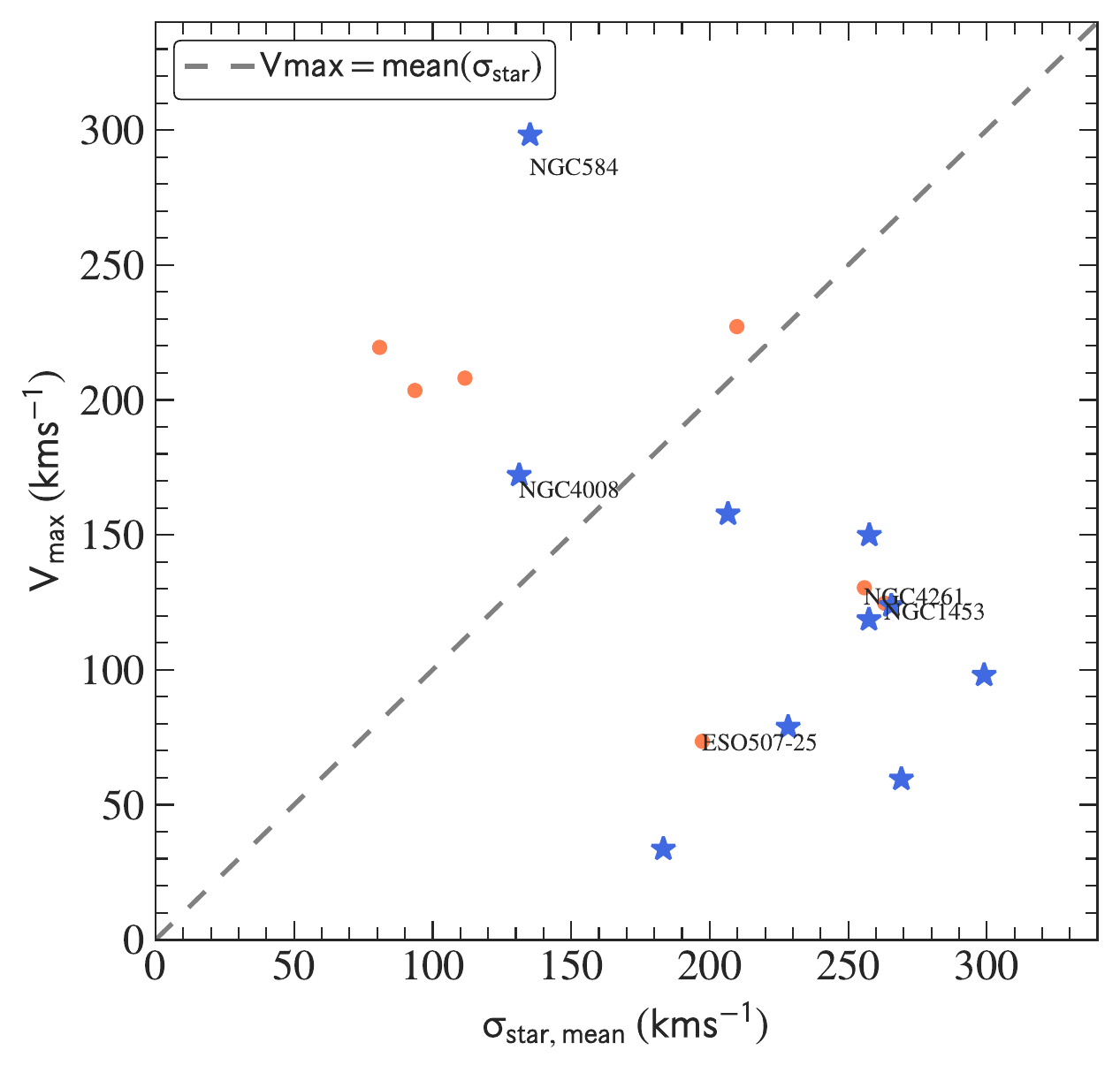}
    \caption{Maximum stellar velocity versus mean stellar velocity for our BGG sample. The dashed gray line mark the $V_{\rm max}/\sigma=1$ value. Sources with filaments and disks are shown with blue stars and orange circles, respectively.}
    \label{fig:vmax_sigma_class}
\end{figure}

%\FloatBarrier

\section{Surface brightness profiles}
\def\setwidthsmall{0.32}
\begin{figure*}
    \centering
    \vspace{-0.2cm}
    \subfigure{\includegraphics[width=\setwidthsmall\textwidth]{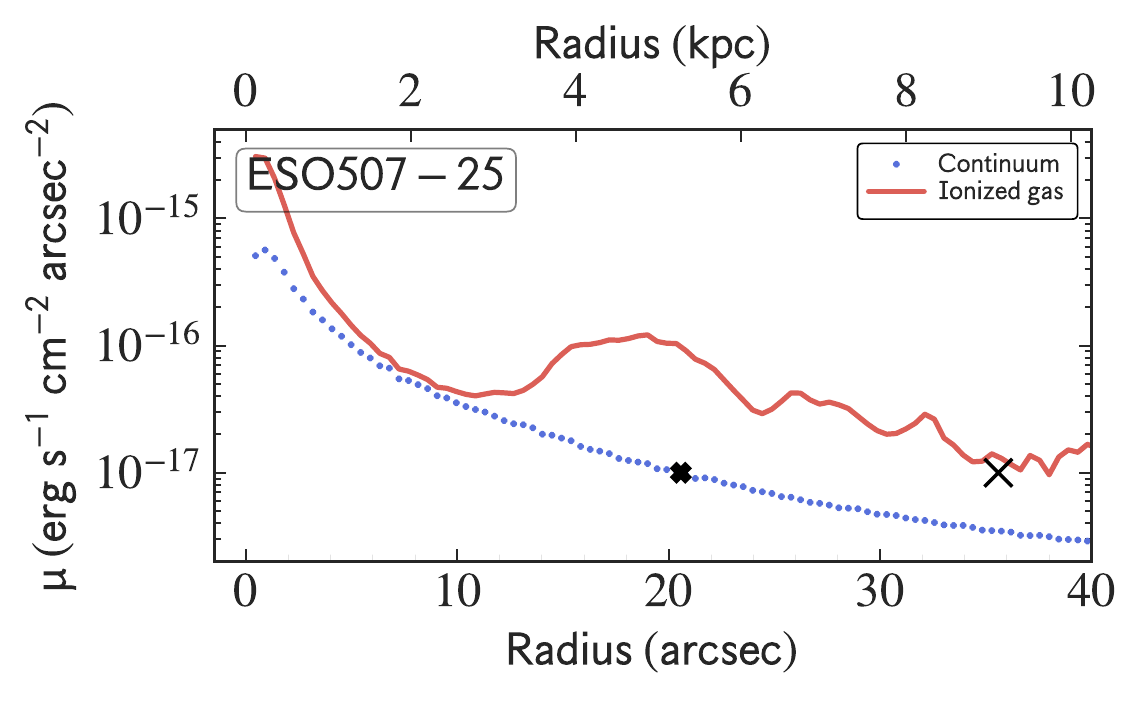}}
    \subfigure{\includegraphics[width=\setwidthsmall\textwidth]{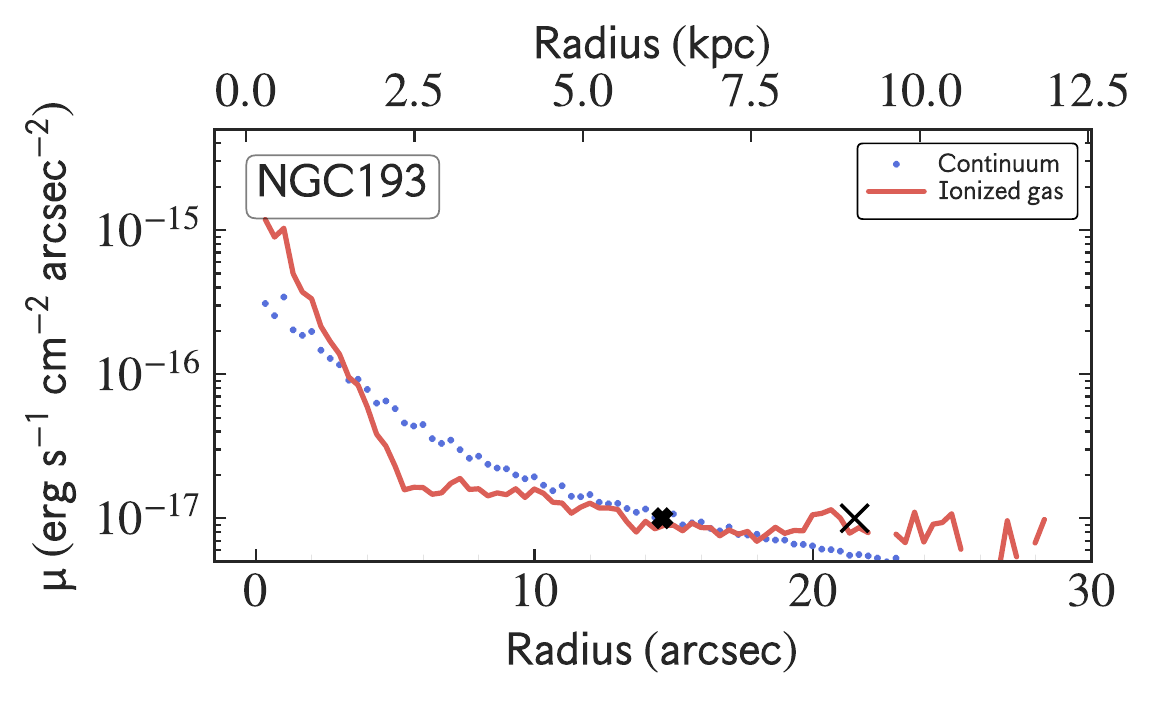}}
    \subfigure{\includegraphics[width=\setwidthsmall\textwidth]{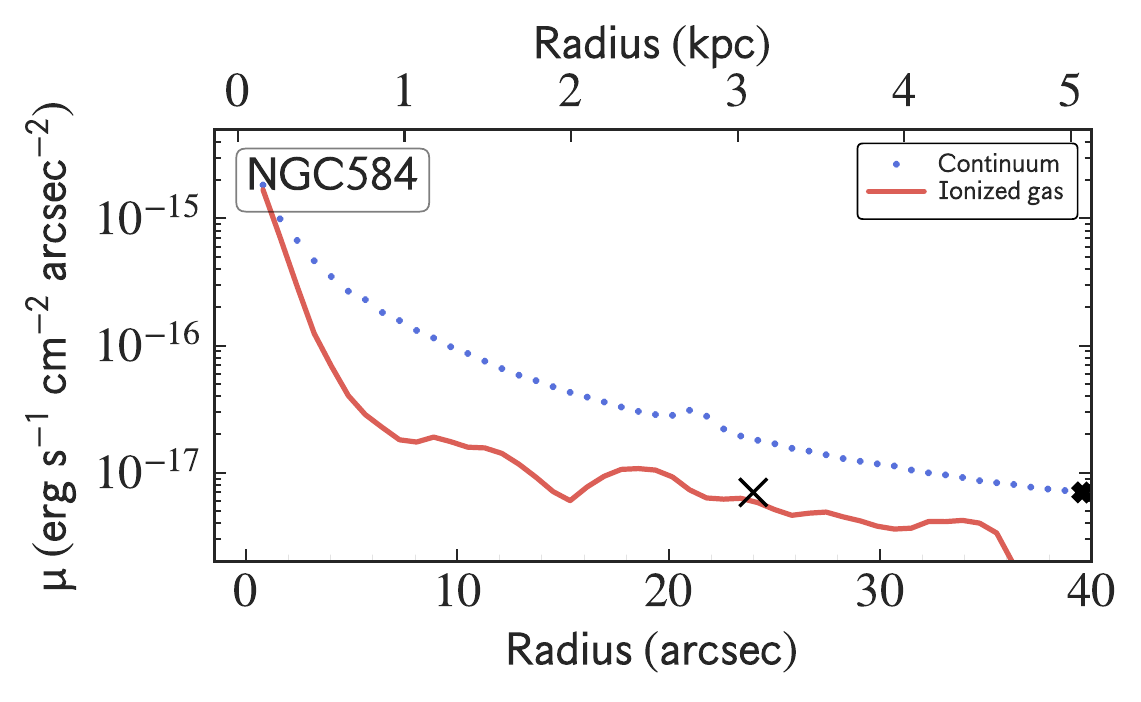}}\\
    \vspace{-0.2cm}
    \subfigure{\includegraphics[width=\setwidthsmall\textwidth]{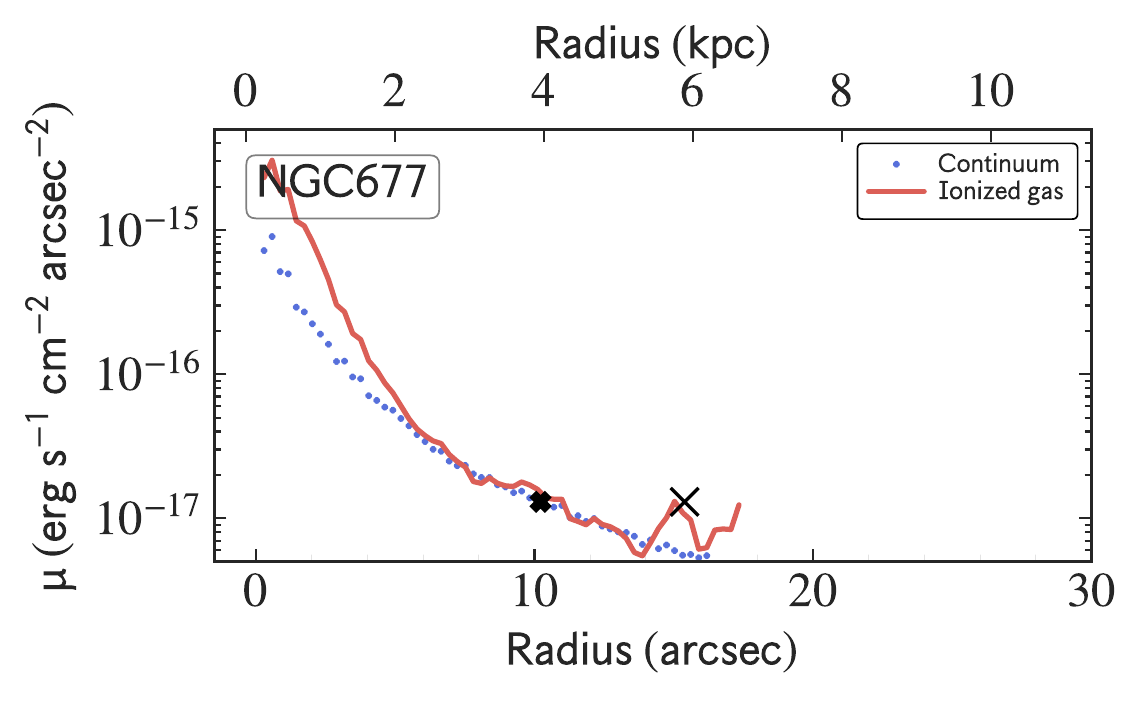}}
    \subfigure{\includegraphics[width=\setwidthsmall\textwidth]{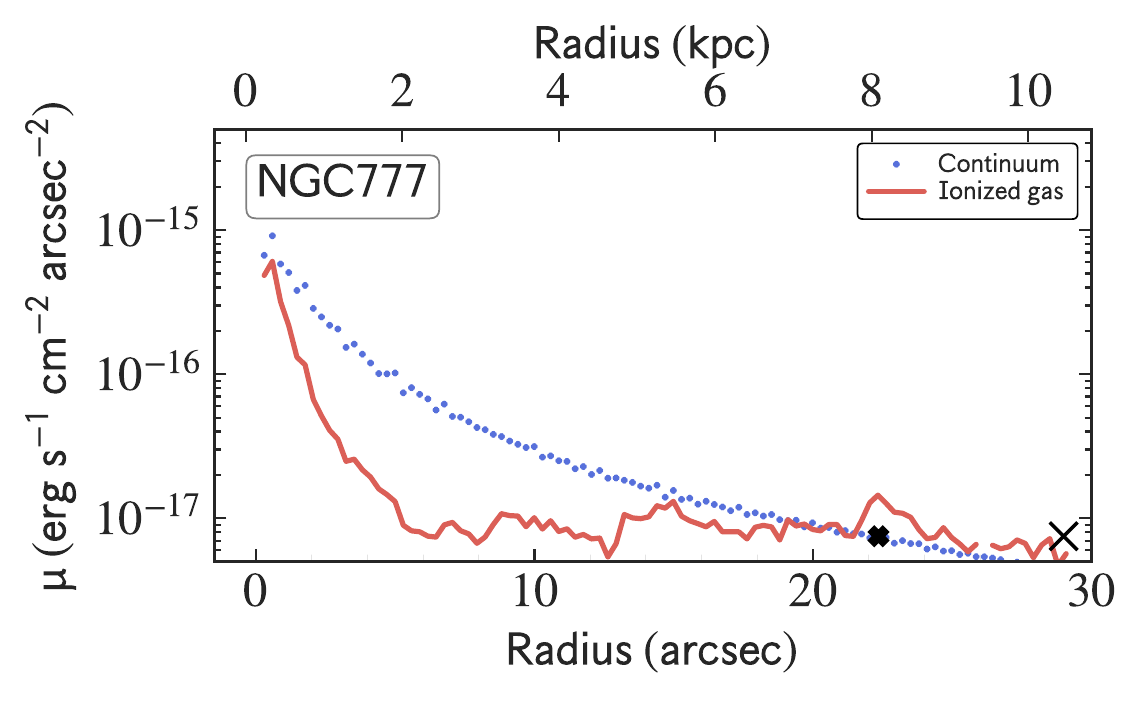}}
    \subfigure{\includegraphics[width=\setwidthsmall\textwidth]{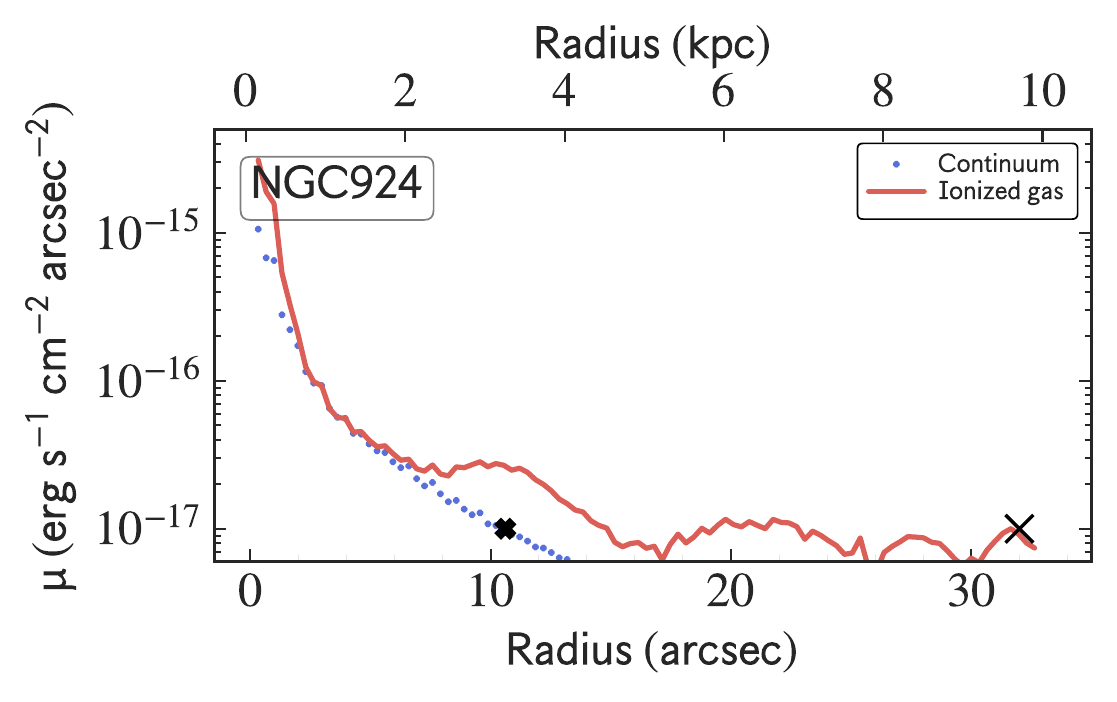}}\\
    \vspace{-0.2cm}
    \subfigure{\includegraphics[width=\setwidthsmall\textwidth]{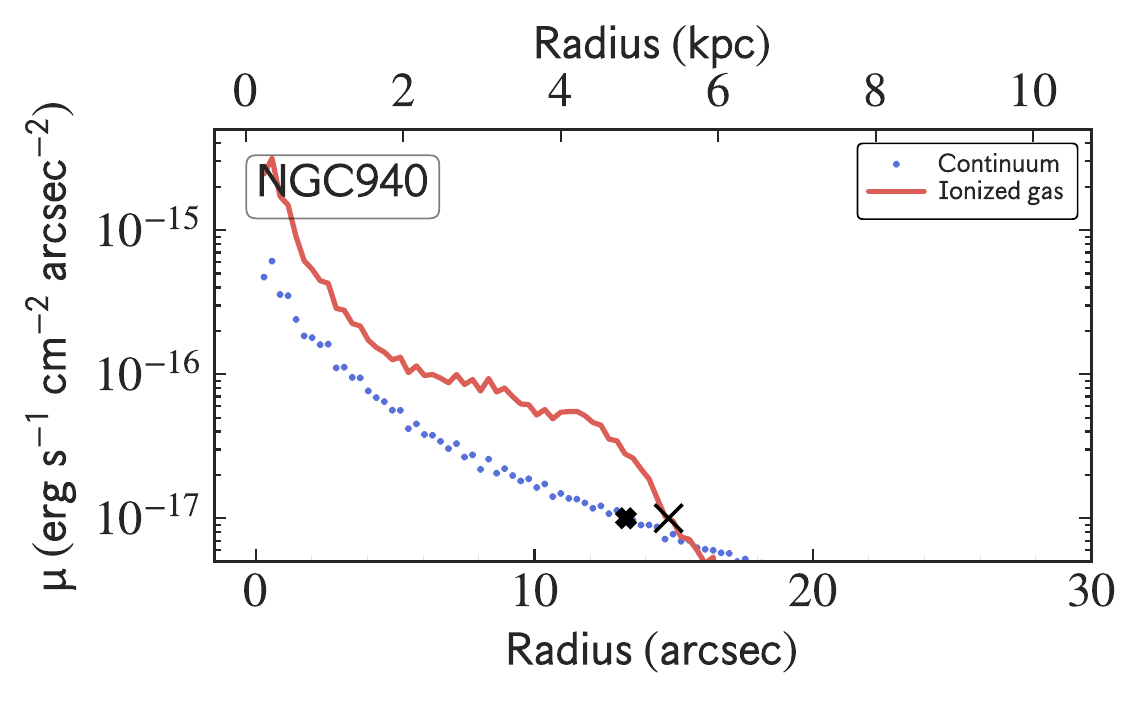}}
    \subfigure{\includegraphics[width=\setwidthsmall\textwidth]{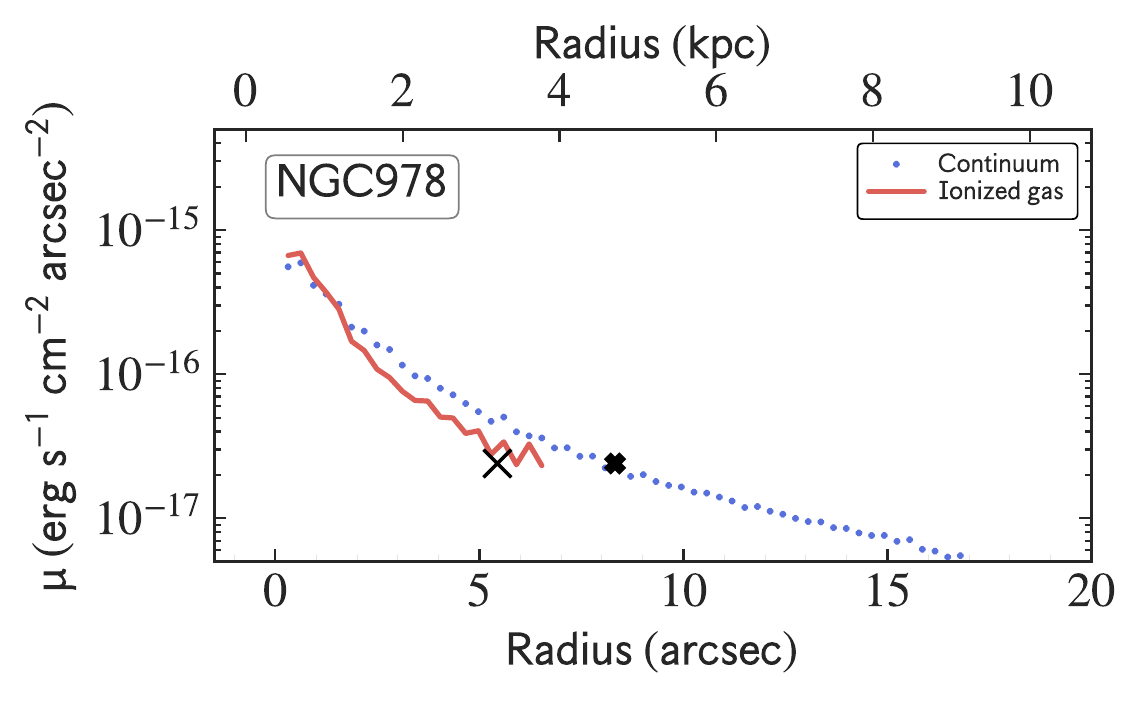}}
    \subfigure{\includegraphics[width=\setwidthsmall\textwidth]{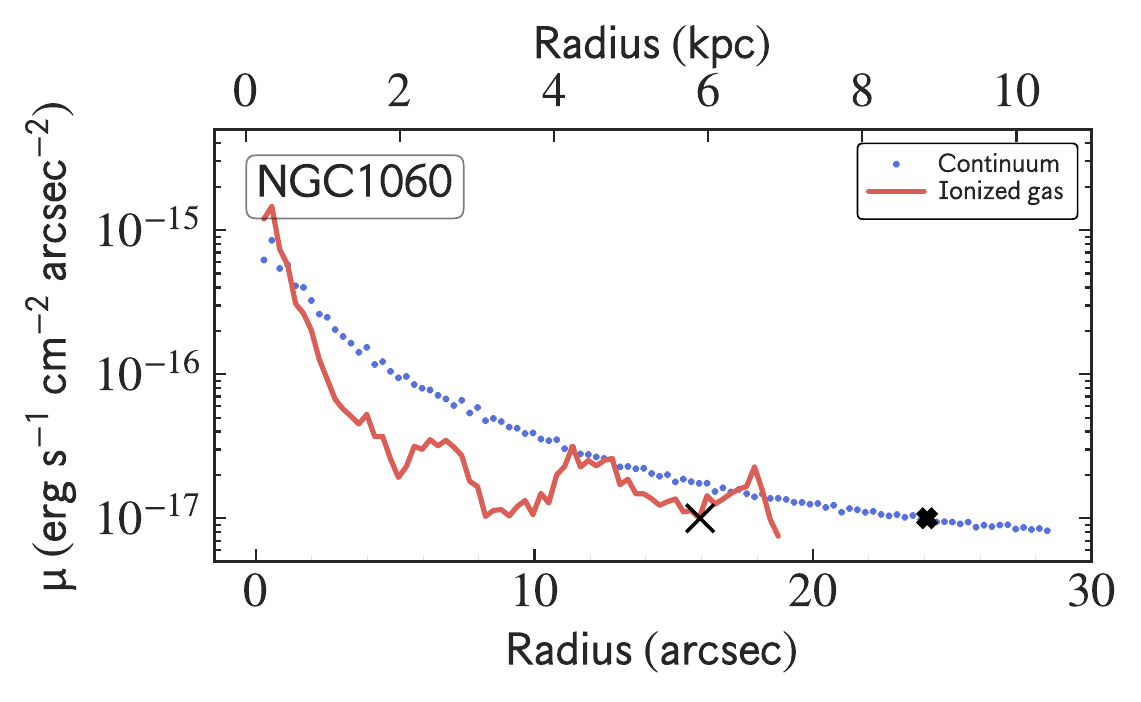}}\\
    \vspace{-0.2cm}
    
    \subfigure{\includegraphics[width=\setwidthsmall\textwidth]{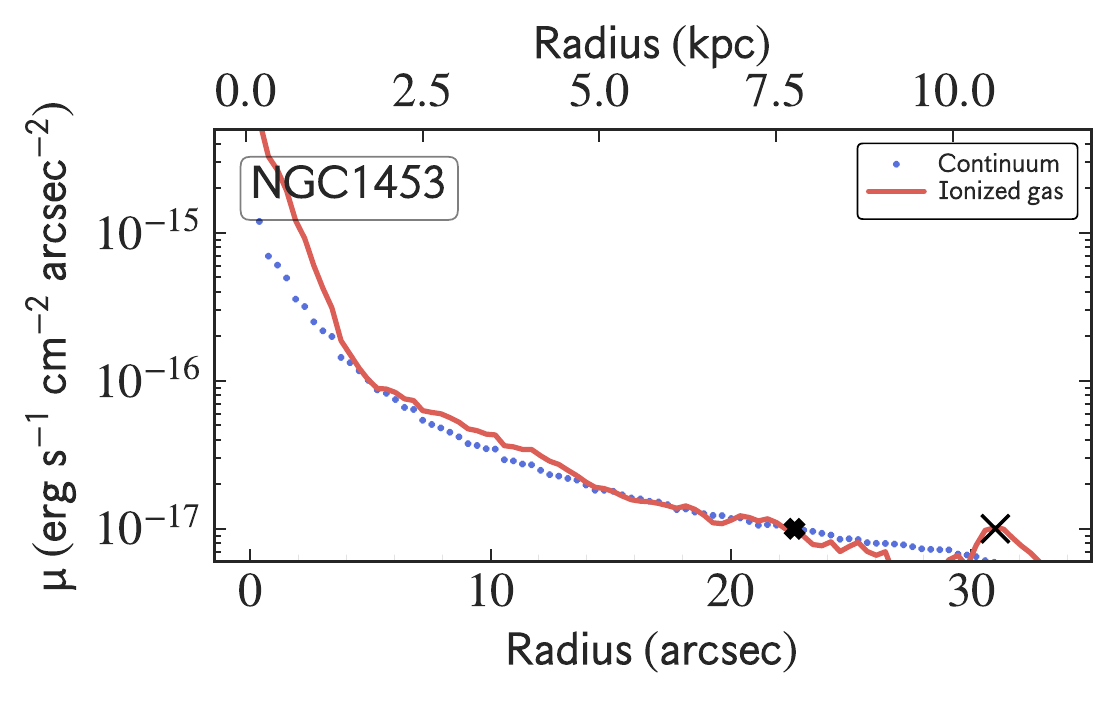}}
     \subfigure{\includegraphics[width=\setwidthsmall\textwidth]{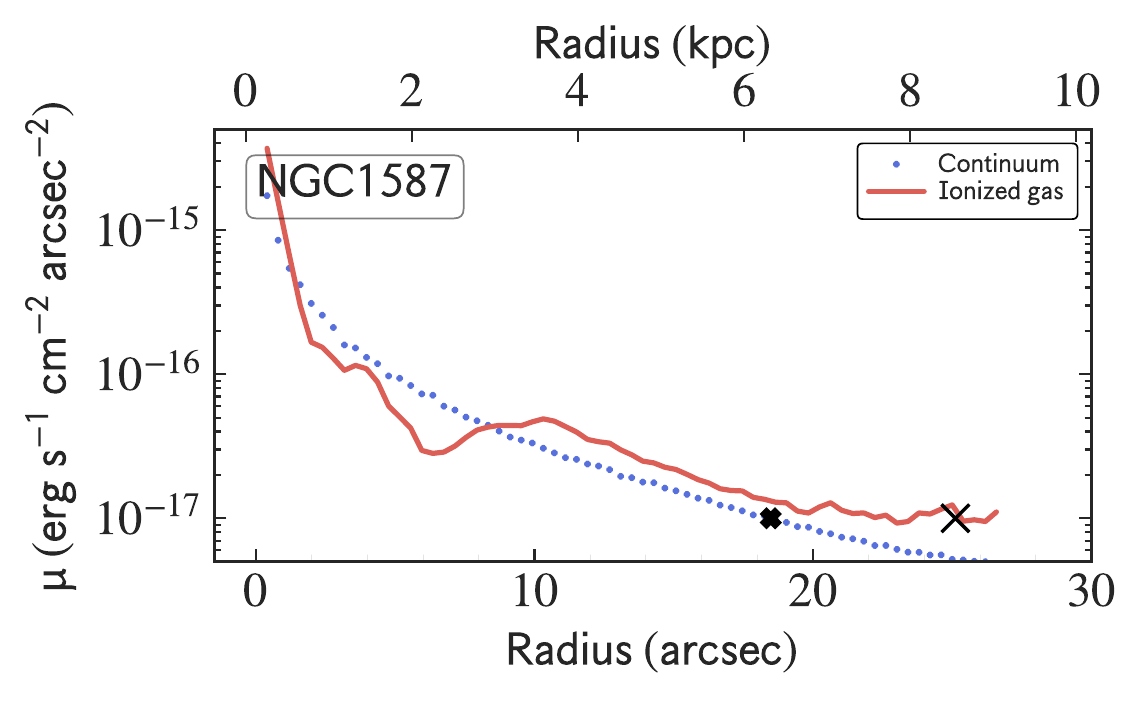}}
    \subfigure{\includegraphics[width=\setwidthsmall\textwidth]{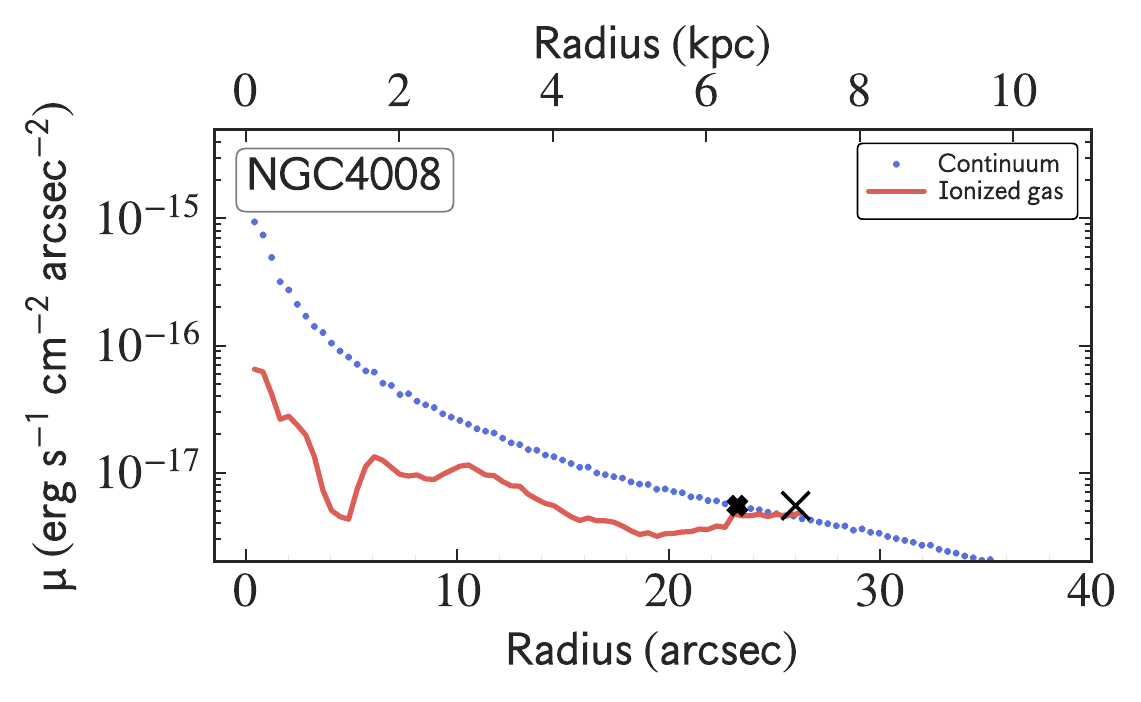}}\\
    \vspace{-0.2cm}

     \subfigure{\includegraphics[width=\setwidthsmall\textwidth]{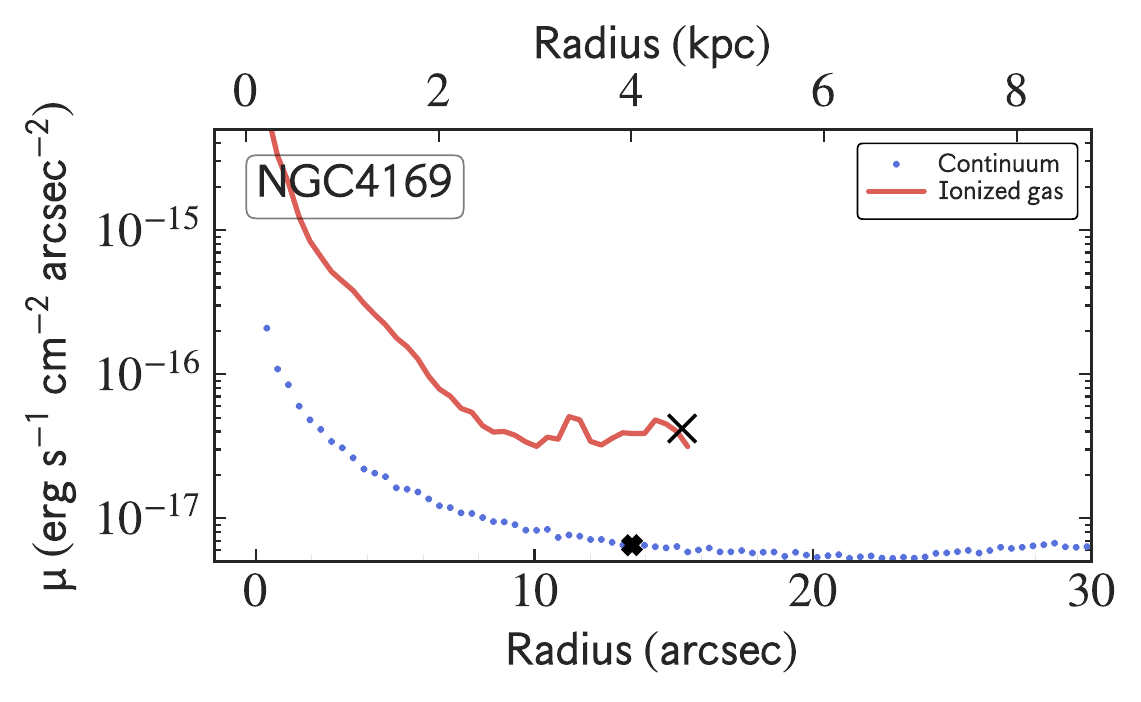}}
    \subfigure{\includegraphics[width=\setwidthsmall\textwidth]{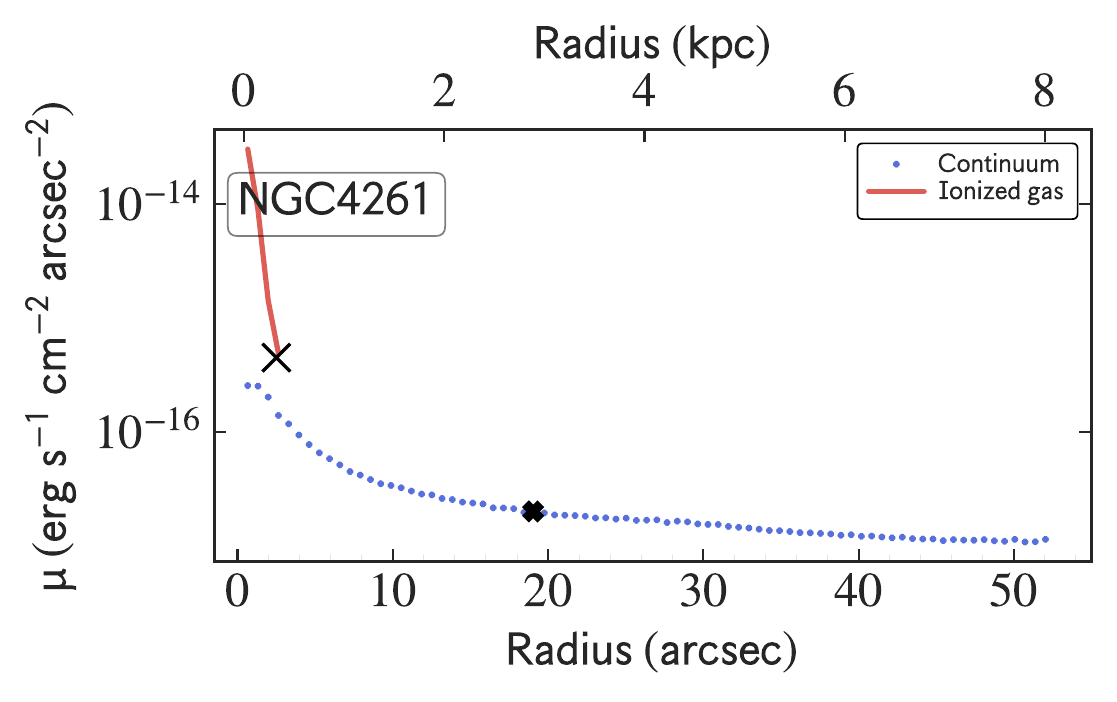}}
     \subfigure{\includegraphics[width=\setwidthsmall\textwidth]{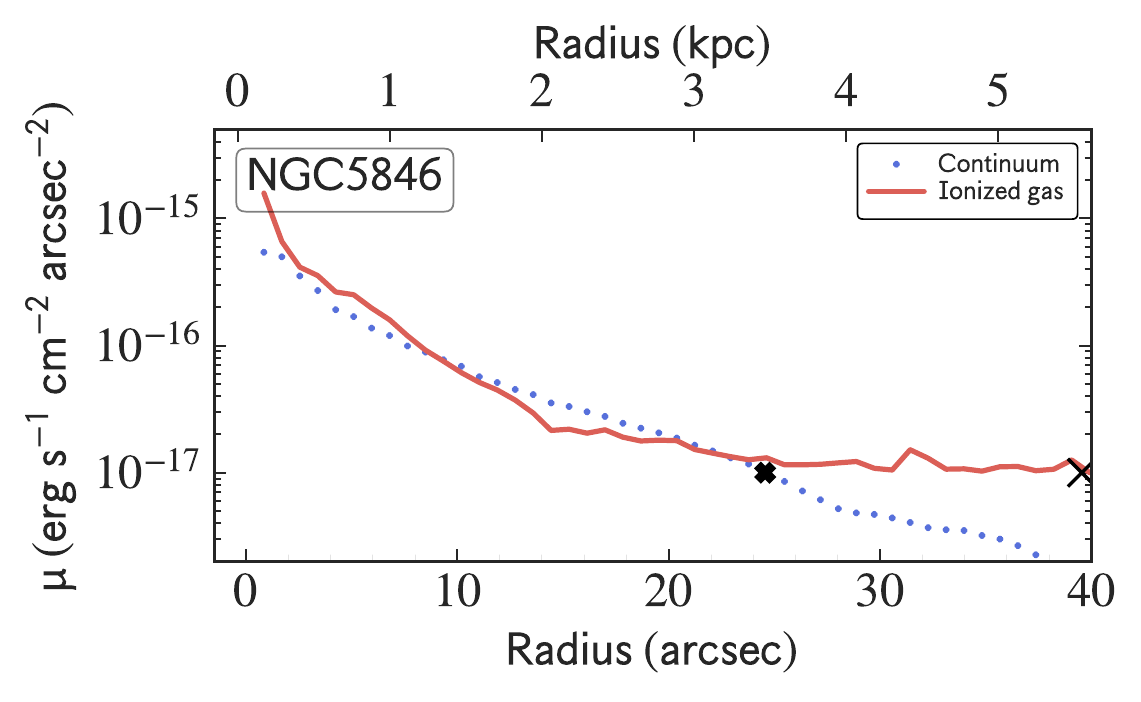}}\\    
        \vspace{-0.2cm}
 
    \subfigure{\includegraphics[width=\setwidthsmall\textwidth]{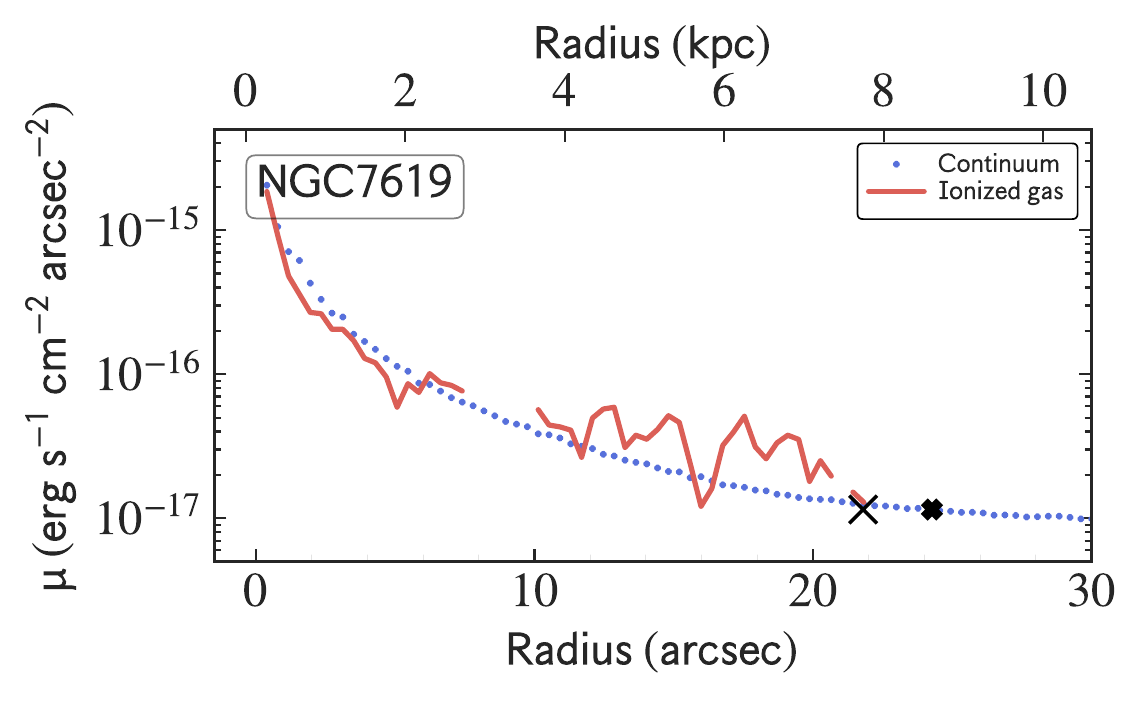}}
    \caption{Surface brightness profiles of the continuum (blue dots) and [NII] emission line (red line). The black crosses correspond to the location where we measured the projected radius of the ionized gas and the continuum roughly at $\rm \sim10^{-17}~erg~s^{-1}~cm^{-2}~arcsec^{-2}$. No optical emitting gas is found at $\rm \sim10^{-17}~erg~s^{-1}~cm^{-2}~arcsec^{-2}$ for NGC\,4169 and NGC\,4261.}
    \label{fig:SB_radial_profile}
\end{figure*}

\end{document}